%%%%%%%%%%%%%%%%%%%%%%
\def\yyy{}
\def\Ehet{E_{het}}
\def\Kv{\tx Z_V}\def\Kh{Z_H}\def\g{\ux G}

\def\mn{\the\secno.\the\subsecno}
%%%%%%%%%%%%%%%%%%%%%%
% packages
%%%%%%%%%%%%%%%%%%%%%%
\input amssym
\input harvmac
\input xymatrix
\input xyarrow
%\input epsf
%\input tables
%\draftmode
\noblackbox

%%%%%%%%%%%%%%%%%%%%%%
% hypertex
%%%%%%%%%%%%%%%%%%%%%%
\newif\ifhypertex
\ifx\hyperdef\UnDeFiNeD
\hypertexfalse
\message{[HYPERTEX MODE OFF]}

\def\hyperdef#1#2#3#4{#4}

\def\e@tf@ur#1{}

\else
\hypertextrue
\message{[HYPERTEX MODE ON]}

\fi

%%%%%%%%%%%%%%%%%%%%%%
% Sizes
%%%%%%%%%%%%%%%%%%%%%%
\def\mn{\the\secno.\the\subsecno}
\baselineskip=16pt plus 2pt minus 1pt
\parskip=2pt plus 16pt minus 1pt

%%%%%%%%%%%%%%%%%%%%%%
% macros
%%%%%%%%%%%%%%%%%%%%%%
\def\opt{\cx O(-3)_{\IP^2}}
\def\-{\hphantom{-}}

\def\cf{{\it cf.\ }}

\def\cxH{D}\def\cxS{\Si}
\def\mus{\mu^\star}

\def\la{\lambda}\def\Si{\Sigma}\def\Im{{\rm Im\ }}\def\La{\Lambda}
\def\Vol{{\rm Vol}}
\def\zh{\hat{z}}
\def\that{{\hat{t}}}

\def\br{\hfill\break}
\def\noi{\noindent}
\def\cx#1{{\cal #1}}
\def\IP{{\bf P}}
\def\IZ{{\bf Z}}
\def\IC{{\bf C}}
\def\IR{{\bf R}}
\def\tx#1{{\tilde{#1}}}\def\wtx#1{{\widetilde{#1}}}
\def\hx#1{{\hat{#1}}}
\def\bx#1{{\bf #1}}
\def\bb#1{{\bar{#1}}}
\def\ux#1{\underline{#1}}
\def\fc#1#2{{#1 \over #2}}
\def\p{\partial}

\def\al{\alpha}
\def\be{\beta}
\def\ga{\gamma}\def\Ga{\Gamma}
\def\de{\delta}

\def\vphi{\varphi}
\def\om{\omega}
\def\Om{\Omega}
\def\th{\theta}
\def\subsubsec#1{\ \br \noindent {\it #1} \br}
\def\zh{\hx z}

%%%%
%%%%

%%
\def\Xa{X}
\def\Za{Z}
\def\D{D}
\def\S{S}\def\DS{D_S}

\def\hD{{\hat D}}
\def\DH{H}
\def\I{{\cx K}}\def\SI{{\cx K}'}

\def\Xabms{\check X_{nc}}
\def\Ds{\Delta^*}\def\D{\Delta}\def\nus{\nu^\star}
\def\S{\Si}\def\tn{\check{t}}\def\dcl{\delta}
\def\XXb{X_{B}^{nc}}\def\XXa{X_{A}^{nc}}\def\Zb{Z_B}\def\Za{Z_A}\def\Xb{X_B}\def\Xa{X_A}
\def\ZG{{\tilde\Zb}}

%%%%%%%%%%%%%%%%%%%%%%
% references
%%%%%%%%%%%%%%%%%%%%%%

%\DonagiXE
\lref\DonagiXE{
R.~Donagi, A.~Lukas, B.~A.~Ovrut and D.~Waldram,
``Non-perturbative vacua and particle physics in M-theory,''
JHEP {\bf 9905}, 018 (1999)
[arXiv:hep-th/9811168].
%%CITATION = JHEPA,9905,018;%%
}
\lref\RajeshIK{
G.~Rajesh,
``Toric geometry and F-theory/heterotic duality in four dimensions,''
JHEP {\bf 9812}, 018 (1998)
[arXiv:hep-th/9811240].
%%CITATION = JHEPA,9812,018;%%
}
\lref\refhm{
M.~Ro\v cek, C.~Vafa and S.~Vandoren,
``Hypermultiplets and topological strings,''
JHEP {\bf 0602}, 062 (2006)
[arXiv:hep-th/0512206];\br
D.~Robles-Llana, F.~Saueressig and S.~Vandoren,
``String loop corrected hypermultiplet moduli spaces,''
JHEP {\bf 0603}, 081 (2006)
[arXiv:hep-th/0602164];\br
D.~Robles-Llana, M.~Ro\v cek, F.~Saueressig, U.~Theis and S.~Vandoren,
``Nonperturbative corrections to 4D string theory effective actions from
SL(2,Z) duality and supersymmetry,''
Phys.\ Rev.\ Lett.\ {\bf 98}, 211602 (2007)
[arXiv:hep-th/0612027];\br
S.~Alexandrov, B.~Pioline, F.~Saueressig and S.~Vandoren,
``D-instantons and twistors,''
JHEP {\bf 0903}, 044 (2009)
[arXiv:0812.4219 [hep-th]].
}
\lref\refdil{
B.~de Wit, V.~Kaplunovsky, J.~Louis and D.~L\"ust,
``Perturbative couplings of vector multiplets in N=2 heterotic string
vacua,''
Nucl.\ Phys.\ B {\bf 451}, 53 (1995)
[arXiv:hep-th/9504006];\br
%%CITATION = NUPHA,B451,53;%%
I.~Antoniadis, S.~Ferrara, E.~Gava, K.~S.~Narain and T.~R.~Taylor,
``Perturbative Prepotential And Monodromies In N=2 Heterotic Superstring,''
Nucl.\ Phys.\ B {\bf 447}, 35 (1995)
[arXiv:hep-th/9504034].
%%CITATION = NUPHA,B447,35;%%
}
\lref\AspinwallII{
P.~S.~Aspinwall and M.~R.~Plesser,
``T-duality can fail,''
JHEP {\bf 9908}, 001 (1999)
[arXiv:hep-th/9905036].
%%CITATION = JHEPA,9908,001;%%
}
\lref\SING{G.-M. Greuel, G.~Pfister, and H.~{Sch\"onemann}, {{Singular} 3.0.1 --
{A} {C}omputer {A}lgebra {S}ystem for {P}olynomial {C}omputations};
Center for Computer Algebra, University of Kaiserslautern, 2006; {\tt http://www.singular.uni-kl.de}.
}
\lref\CandelasHW{
P.~Candelas, A.~Font, S.~H.~Katz and D.~R.~Morrison,
``Mirror symmetry for two parameter models. 2,''
Nucl.\ Phys.\ å B {\bf 429}, 626 (1994)
[arXiv:hep-th/9403187].
%%CITATION = NUPHA,B429,626;%%
}
%\GranaHR
\lref\GranaHR{
M.~Grana, J.~Louis and D.~Waldram,
``SU(3) x SU(3) compactification and mirror duals of magnetic fluxes,''
JHEP {\bf 0704}, 101 (2007)
[arXiv:hep-th/0612237].
%%CITATION = JHEPA,0704,101;%%
}
%\DAuriaTR
\lref\DAuriaTR{
R.~D'Auria, S.~Ferrara, M.~Trigiante and S.~Vaula,
``Gauging the Heisenberg algebra of special quaternionic manifolds,''
Phys.\ Lett.\ å B {\bf 610}, 147 (2005)
[arXiv:hep-th/0410290].
%%CITATION = PHLTA,B610,147;%%
}

%\BershadskyNH
\lref\BershadskyNH{
M.~Bershadsky, K.~A.~Intriligator, S.~Kachru, D.~R.~Morrison, V.~Sadov and C.~Vafa,
``Geometric singularities and enhanced gauge symmetries,''
Nucl.\ Phys.\ å B {\bf 481}, 215 (1996)
[arXiv:hep-th/9605200].
%%CITATION = NUPHA,B481,215;%%
}
%\BershadskyVN
\lref\Fflux{
M.~Bershadsky, T.~Pantev and V.~Sadov,
``F-theory with quantized fluxes,''
Adv.\ Theor.\ Math.\ Phys.\ å {\bf 3}, 727 (1999)
[arXiv:hep-th/9805056];\br
%%CITATION = 00203,3,727;%%
}
\lref\Witsi{N.~Seiberg and E.~Witten,
``Comments on String Dynamics in Six Dimensions,''
Nucl.\ Phys.\ å B {\bf 471}, 121 (1996)
[arXiv:hep-th/9603003].
%%CITATION = NUPHA,B471,121;%%
}
%\HoriZJ
\lref\HoriZJ{
K.~Hori, H.~Ooguri and C.~Vafa,
``Non-Abelian conifold transitions and N = 4 dualities in three
dimensions,''
Nucl.\ Phys.\ å B {\bf 504}, 147 (1997)
[arXiv:hep-th/9705220].
%%CITATION = NUPHA,B504,147;%%
}
%\MayrBK
\lref\MayrBK{
P.~Mayr,
``Conformal field theories on K3 and three-dimensional gauge theories,''
JHEP {\bf 0008}, 042 (2000)
[arXiv:hep-th/9910268].
%%CITATION = JHEPA,0008,042;%%
}
%\NekrasovJS
\lref\NekrasovJS{
N.~Nekrasov, H.~Ooguri and C.~Vafa,
``S-duality and topological strings,''
JHEP {\bf 0410}, 009 (2004)
[arXiv:hep-th/0403167].
%%CITATION = JHEPA,0410,009;%%
}
%\BerglundVA
\lref\BerglundVA{P.~Berglund, A.~Klemm, P.~Mayr and S.~Theisen,
``On type IIB vacua with varying coupling constant,''
Nucl.\ Phys.\ å B {\bf 558}, 178 (1999)
[arXiv:hep-th/9805189].
%%CITATION = NUPHA,B558,178;%%
}
%\GraberDW
\lref\GraberDW{
T.~Graber and E.~Zaslow,
``Open string Gromov-Witten invariants: Calculations and a mirror
'theorem',''
arXiv:hep-th/0109075.
%%CITATION = HEP-TH/0109075;%%
}
%\PerevalovHT
\lref\PerevalovHT{
E.~Perevalov and G.~Rajesh,
``Mirror symmetry via deformation of bundles on K3 surfaces,''
Phys.\ Rev.\ Lett.\ å {\bf 79}, 2931 (1997)
[arXiv:hep-th/9706005].
%%CITATION = PRLTA,79,2931;%%
}
%\KlemmTJ
\lref\KlemmTJ{
A.~Klemm, W.~Lerche and P.~Mayr,
``K3 Fibrations and heterotic type II string duality,''
Phys.\ Lett.\ å B {\bf 357}, 313 (1995)
[arXiv:hep-th/9506112].
%%CITATION = PHLTA,B357,313;%%
}
%\AspinwallVK
\lref\AspinwallVK{
P.~S.~Aspinwall and J.~Louis,
``On the Ubiquity of K3 Fibrations in String Duality,''
Phys.\ Lett.\ å B {\bf 369}, 233 (1996)
[arXiv:hep-th/9510234].
%%CITATION = PHLTA,B369,233;%%
}
\lref\VafOS{ C.~Vafa,
``Extending mirror conjecture to Calabi-Yau with bundles,''
arXiv:hep-th/9804131.}
\lref\CandF{P.~Candelas and A.~Font,
``Duality between the webs of heterotic and type II vacua,''
Nucl.\ Phys.\ å B {\bf 511}, 295 (1998)
[arXiv:hep-th/9603170].
%%CITATION = NUPHA,B511,295;%%
}
%\CandelasEH
\lref\CandelasEH{
P.~Candelas, E.~Perevalov and G.~Rajesh,
``Toric geometry and enhanced gauge symmetry of F-theory/heterotic å vacua,''
Nucl.\ Phys.\ å B {\bf 507}, 445 (1997)
[arXiv:hep-th/9704097].
%%CITATION = NUPHA,B507,445;%%
}
\lref\Avram{
A.~C.~Avram, M.~Kreuzer, M.~Mandelberg and H.~Skarke,
``Searching for K3 fibrations,''
Nucl.\ Phys.\ å B {\bf 494}, 567 (1997)
[arXiv:hep-th/9610154];\br
%%CITATION = NUPHA,B494,567;%%
M.~Kreuzer and H.~Skarke,
``Calabi-Yau 4-folds and toric fibrations,''
J.\ Geom.\ Phys.\ å {\bf 26}, 272 (1998)
[arXiv:hep-th/9701175].
%%CITATION = JGPHE,26,272;%%
}
\lref\ferrara{
L.~Andrianopoli, R.~D'Auria, S.~Ferrara and M.~A.~Lledo,
%``4-D gauged supergravity analysis of type IIB vacua on K3 x T**2/Z(2),''
JHEP {\bf 0303}, 044 (2003)
[arXiv:hep-th/0302174];\br
%%CITATION = JHEPA,0303,044;%%
C.~Angelantonj, R.~D'Auria, S.~Ferrara and M.~Trigiante,
``$K3 x T^2/Z_2$ orientifolds with fluxes, open string moduli and critical
points,''
Phys.\ Lett.\ å B {\bf 583}, 331 (2004)
[arXiv:hep-th/0312019].
%%CITATION = PHLTA,B583,331;%%
}
\lref\AKV{M.~Aganagic, A.~Klemm and C.~Vafa,
``Disk instantons, mirror symmetry and the duality web,''
Z.\ Naturforsch.\ å A {\bf 57}, 1 (2002)
[arXiv:hep-th/0105045].
%%CITATION = ZNTFA,A57,1;%%
}
\lref\Luest{
D.~L\"ust, P.~Mayr, S.~Reffert and S.~Stieberger,
``F-theory flux, destabilization of orientifolds and soft terms on
D7-branes,''
Nucl.\ Phys.\ å B {\bf 732}, 243 (2006)
[arXiv:hep-th/0501139].
%%CITATION = NUPHA,B732,243;%%
}
\lref\CurioBVA{
G.~Curio and R.~Y.~Donagi,
``Moduli in N = 1 heterotic/F-theory duality,''
Nucl.\ Phys.\ å B {\bf 518}, 603 (1998)
[arXiv:hep-th/9801057].
%%CITATION = NUPHA,B518,603;%%
}
\lref\VafaGM{
C.~Vafa and E.~Witten,
``Dual string pairs with N = 1 and N = 2 supersymmetry in four å dimensions,''
Nucl.\ Phys.\ Proc.\ Suppl.\ å {\bf 46}, 225 (1996)
[arXiv:hep-th/9507050].
%%CITATION = NUPHZ,46,225;%%
}
\lref\HV{
K.~Hori and C.~Vafa,
``Mirror symmetry,''
arXiv:hep-th/0002222.
%%CITATION = HEP-TH/0002222;%%
}
\lref\Yau{S.~Li, B.~H.~Lian and S.~T.~Yau,
``Picard-Fuchs Equations for Relative Periods and Abel-Jacobi Map for
Calabi-Yau Hypersurfaces,''
arXiv:0910.4215 [math.AG].
%%CITATION = ARXIV:0910.4215;%%
}
\lref\OV{
H.~Ooguri and C.~Vafa,
``Knot invariants and topological strings,''
Nucl.\ Phys.\ å B {\bf 577}, 419 (2000)
[arXiv:hep-th/9912123].
%%CITATION = NUPHA,B577,419;%%
}
\lref\JSii{H.~Jockers and M.~Soroush,
``Relative periods and open-string integer invariants for a compact
Calabi-Yau hypersurface,''
arXiv:0904.4674 [hep-th].}
\lref\MW{
D.~R.~Morrison and J.~Walcher,
``D-branes and Normal Functions,''
arXiv:0709.4028 [hep-th].}
\lref\Wa{
J.~Walcher,
``Opening mirror symmetry on the quintic,''
Commun.\ Math.\ Phys.\ å {\bf 276}, 671 (2007)
[arXiv:hep-th/0605162].
%%CITATION = CMPHA,276,671;%%
}
\lref\PaWa{R.~Pandharipande, J.~Solomon and J.~Walcher, 
``Disk enumeration on the quintic 3-fold,'' arXiv.org:math/0610901.}

\lref\SW{SW}
\lref\Katz{S.~Kachru, S.~H.~Katz, A.~E.~Lawrence and J.~McGreevy,
``Open string instantons and superpotentials,''
Phys.\ Rev.\ å D {\bf 62}, 026001 (2000)
[arXiv:hep-th/9912151]; ``Mirror symmetry for open strings,''
Phys.\ Rev.\ å D {\bf 62}, 126005 (2000)
[arXiv:hep-th/0006047].}
\lref\BJPS{M.~Bershadsky, A.~Johansen, T.~Pantev and V.~Sadov,
``On four-dimensional compactifications of F-theory,''
Nucl.\ Phys.\ å B {\bf 505}, 165 (1997)
[arXiv:hep-th/9701165].
%%CITATION = NUPHA,B505,165;%%
}
\lref\Bat{ V.~V.~Batyrev,
``Dual polyhedra and mirror symmetry for Calabi-Yau hypersurfaces in toric
varieties,''
J.\ Alg.\ Geom.\ å {\bf 3}, 493 (1994). 
}
\lref\KMV{
S.~Katz, P.~Mayr and C.~Vafa,
``Mirror symmetry and exact solution of 4D N = 2 gauge theories. I,''
Adv.\ Theor.\ Math.\ Phys.\ å {\bf 1}, 53 (1998)
[arXiv:hep-th/9706110].
%%CITATION = 00203,1,53;%%
}
\lref\GMPh{B.~R.~Greene, D.~R.~Morrison and M.~R.~Plesser,
``Mirror manifolds in higher dimension,''
Commun.\ Math.\ Phys.\ å {\bf 173}, 559 (1995)[arXiv:hep-th/9402119].
%%CITATION = CMPHA,173,559;%%
}
\lref\PMff{P.~Mayr,
``Mirror symmetry, N = 1 superpotentials and tensionless strings on
Calabi-Yau four-folds,''
Nucl.\ Phys.\ å B {\bf 494}, 489 (1997)
[arXiv:hep-th/9610162].
%%CITATION = NUPHA,B494,489;%%
}
\lref\WitCS{
E.~Witten,
``Chern-Simons Gauge Theory As A String Theory,''
Prog.\ Math.\ å {\bf 133}, 637 (1995)
[arXiv:hep-th/9207094].
%%CITATION = PMTMA,133,637;%%
}
\lref\KLRY{
A.~Klemm, B.~Lian, S.~S.~Roan and S.~T.~Yau,
``Calabi-Yau fourfolds for M- and F-theory compactifications,''
Nucl.\ Phys.\ å B {\bf 518}, 515 (1998)
[arXiv:hep-th/9701023].
%%CITATION = NUPHA,B518,515;%%
}
\lref\TVsp{
T.~R.~Taylor and C.~Vafa,
``RR flux on Calabi-Yau and partial supersymmetry breaking,''
Phys.\ Lett.\ å B {\bf 474}, 130 (2000)
[arXiv:hep-th/9912152].
}
\lref\WitCh{E.~Witten,
``Branes and the dynamics of {QCD},
Nucl.\ Phys.\ å B {\bf 507}, 658 (1997)
[arXiv:hep-th/9706109].}
\lref\mibo{K. Hori et. al, ``Mirror Symmetry'', Clay Mathematics Monographs v.1., 2003}
\lref\vafaf{
C.~Vafa,
``Evidence for F-Theory,''
Nucl.\ Phys.\ å B {\bf 469}, 403 (1996)
[arXiv:hep-th/9602022].
%%CITATION = NUPHA,B469,403;%%
}
\lref\GVW{
S.~Gukov, C.~Vafa and E.~Witten,
``CFT's from Calabi-Yau four-folds,''
Nucl.\ Phys.\ å B {\bf 584}, 69 (2000)
[Erratum-ibid.\ å B {\bf 608}, 477 (2001)]
[arXiv:hep-th/9906070].
}
\lref\MV{
D.~R.~Morrison and C.~Vafa,
``Compactifications of F-Theory on Calabi--Yau Threefolds -- I,''
Nucl.\ Phys.\ å B {\bf 473}, 74 (1996)
[arXiv:hep-th/9602114];
%%CITATION = NUPHA,B473,74;%%
%D.~R.~Morrison and C.~Vafa,
``Compactifications of F-Theory on Calabi--Yau Threefolds -- II,''
Nucl.\ Phys.\ å B {\bf 476}, 437 (1996)
[arXiv:hep-th/9603161].
%%CITATION = NUPHA,B476,437;%%
}
\lref\FMW{
R.~Friedman, J.~Morgan and E.~Witten,
``Vector bundles and F theory,''
Commun.\ Math.\ Phys.\ å {\bf 187}, 679 (1997)
[arXiv:hep-th/9701162].
%%CITATION = CMPHA,187,679;%%
}
%\AlimBX
\lref\AlimBX{
M.~Alim, M.~Hecht, H.~Jockers, P.~Mayr, A.~Mertens and M.~Soroush,
``Hints for Off-Shell Mirror Symmetry in type II/F-theory Compactifications,''
arXiv:0909.1842 [hep-th].
%%CITATION = ARXIV:0909.1842;%%
}
%\AlimRF
\lref\AlimRF{
M.~Alim, M.~Hecht, P.~Mayr and A.~Mertens,
``Mirror Symmetry for Toric Branes on Compact Hypersurfaces,''
arXiv:0901.2937 [hep-th].
%%CITATION = ARXIV:0901.2937;%%
}
%\AganagicGS
\lref\AV{
M.~Aganagic and C.~Vafa,
``Mirror symmetry, D-branes and counting holomorphic discs,''
arXiv:hep-th/0012041.
%%CITATION = HEP-TH/0012041;%%
}
%\AganagicJQ
\lref\AganagicJQ{
M.~Aganagic and C.~Beem,
``The Geometry of D-Brane Superpotentials,''
arXiv:0909.2245 [hep-th].
%%CITATION = ARXIV:0909.2245;%%
}
\lref\AMsd{
P.~S.~Aspinwall and D.~R.~Morrison,
``Point-like instantons on K3 orbifolds,''
Nucl.\ Phys.\ å B {\bf 503}, 533 (1997)
[arXiv:hep-th/9705104].
%%CITATION = NUPHA,B503,533;%%
}
%\AndreasVE
\lref\AndreasVE{
B.~Andreas, G.~Curio, D.~Hernandez Ruiperez and S.~T.~Yau,
``Fibrewise T-duality for D-branes on elliptic Calabi-Yau,''
JHEP {\bf 0103}, 020 (2001)
[arXiv:hep-th/0101129].
%%CITATION = JHEPA,0103,020;%%
}
%\AspinwallBW
\lref\AspinwallBW{
P.~S.~Aspinwall,
``Aspects of the hypermultiplet moduli space in string duality,''
JHEP {\bf 9804}, 019 (1998)
[arXiv:hep-th/9802194].
%%CITATION = JHEPA,9804,019;%%
}
%\AspinwallXS
\lref\AspinwallXS{
P.~S.~Aspinwall and M.~R.~Plesser,
``Heterotic string corrections from the dual type II string,''
JHEP {\bf 0004}, 025 (2000)
[arXiv:hep-th/9910248].
%%CITATION = JHEPA,0004,025;%%
}
%\AspinwallVK
\lref\AspinwallVK{
P.~S.~Aspinwall and J.~Louis,
``On the Ubiquity of K3 Fibrations in String Duality,''
Phys.\ Lett.\ å B {\bf 369}, 233 (1996)
[arXiv:hep-th/9510234].
%%CITATION = PHLTA,B369,233;%%
}
%\BenmachicheMA
\lref\BenmachicheMA{
I.~Benmachiche, J.~Louis and D.~Martinez-Pedrera,
``The effective action of the heterotic string compactified on manifolds with
SU(3) structure,''
Class.\ Quant.\ Grav.\ å {\bf 25}, 135006 (2008)
[arXiv:0802.0410 [hep-th]].
%%CITATION = CQGRD,25,135006;%%
}
%\BerglundDM
\lref\BMof{
P.~Berglund and P.~Mayr,
``Non-perturbative superpotentials in F-theory and string duality,''
arXiv:hep-th/0504058.
%%CITATION = HEP-TH/0504058;%%
}
%\BerglundQK
\lref\BerglundQK{
P.~Berglund and P.~Mayr,
``Stability of vector bundles from F-theory,''
JHEP {\bf 9912}, 009 (1999)
[arXiv:hep-th/9904114].
%%CITATION = JHEPA,9912,009;%%
}
%\BerglundEJ
\lref\BMff{
P.~Berglund and P.~Mayr,
``Heterotic string/F-theory duality from mirror symmetry,''
Adv.\ Theor.\ Math.\ Phys.\ å {\bf 2}, 1307 (1999)
[arXiv:hep-th/9811217].
%%CITATION = 00203,2,1307;%%
}
%\CandelasRM
\lref\CandelasRM{
P.~Candelas, X.~C.~De La Ossa, P.~S.~Green and L.~Parkes,
``A pair of Calabi-Yau manifolds as an exactly soluble superconformal
theory,''
Nucl.\ Phys.\ å B {\bf 359}, 21 (1991).
%%CITATION = NUPHA,B359,21;%%
}
%\CurioAE
\lref\CurioAE{
G.~Curio, A.~Klemm, B.~Kors and D.~Lust,
``Fluxes in heterotic and type II string compactifications,''
Nucl.\ Phys.\ å B {\bf 620}, 237 (2002)
[arXiv:hep-th/0106155].
%%CITATION = NUPHA,B620,237;%%
}
%\DasguptaSS
\lref\DasguptaSS{
K.~Dasgupta, G.~Rajesh and S.~Sethi,
``M theory, orientifolds and G-flux,''
JHEP {\bf 9908}, 023 (1999)
[arXiv:hep-th/9908088].
%%CITATION = JHEPA,9908,023;%%
}
%\GatesDU
\lref\GatesDU{
S.~J.~Gates, M.~T.~Grisaru and M.~E.~Wehlau,
``A Study of General 2D, N=2 Matter Coupled to Supergravity in Superspace,''
Nucl.\ Phys.\ å B {\bf 460}, 579 (1996)
[arXiv:hep-th/9509021].
%%CITATION = NUPHA,B460,579;%%
}
%\GukovIQ
\lref\GukovIQ{
S.~Gukov and M.~Haack,
``IIA string theory on Calabi-Yau fourfolds with background fluxes,''
Nucl.\ Phys.\ å B {\bf 639}, 95 (2002)
[arXiv:hep-th/0203267].
%%CITATION = NUPHA,B639,95;%%
}
%\GrimmEF
\lref\GrimmEF{
T.~W.~Grimm, T.~W.~Ha, A.~Klemm and D.~Klevers,
``Computing Brane and Flux Superpotentials in F-theory Compactifications,''
arXiv:0909.2025 [hep-th].
%%CITATION = ARXIV:0909.2025;%%
}
%\HaackDI
\lref\HaackDI{
M.~Haack, J.~Louis and M.~Marquart,
``Type IIA and heterotic string vacua in D = 2,''
Nucl.\ Phys.\ å B {\bf 598}, 30 (2001)
[arXiv:hep-th/0011075].
%%CITATION = NUPHA,B598,30;%%
}
\lref\HalmagyiWI{
N.~Halmagyi, I.~V.~Melnikov and S.~Sethi,
``Instantons, Hypermultiplets and the Heterotic String,''
JHEP {\bf 0707}, 086 (2007)
[arXiv:0704.3308 [hep-th]].
%%CITATION = JHEPA,0707,086;%%
}
%\HullJV
\lref\HullJV{
C.~M.~Hull and E.~Witten,
``Supersymmetric Sigma Models And The Heterotic String,''
Phys.\ Lett.\ å B {\bf 160}, 398 (1985).
%%CITATION = PHLTA,B160,398;%%
}
%\HullYS
\lref\HullYS{
C.~M.~Hull and P.~K.~Townsend,
``Unity of superstring dualities,''
Nucl.\ Phys.\ å B {\bf 438}, 109 (1995)
[arXiv:hep-th/9410167].
%%CITATION = NUPHA,B438,109;%%
}
%\JockersZY
\lref\JockersZY{
H.~Jockers and J.~Louis,
``D-terms and F-terms from D7-brane fluxes,''
Nucl.\ Phys.\ å B {\bf 718}, 203 (2005)
[arXiv:hep-th/0502059].
%%CITATION = NUPHA,B718,203;%%
}
%\JockersYJ
\lref\JockersYJ{
H.~Jockers and J.~Louis,
``The effective action of D7-branes in N = 1 Calabi-Yau orientifolds,''
Nucl.\ Phys.\ å B {\bf 705}, 167 (2005)
[arXiv:hep-th/0409098].
%%CITATION = NUPHA,B705,167;%%
}
%\JockersPE
\lref\JockersPE{
H.~Jockers and M.~Soroush,
``Effective superpotentials for compact D5-brane Calabi-Yau geometries,''
Commun.\ Math.\ Phys.\ å {\bf 290}, 249 (2009)
[arXiv:0808.0761 [hep-th]].
%%CITATION = CMPHA,290,249;%%
} 
\lref\LM{
W.~Lerche and P.~Mayr,
``On N = 1 mirror symmetry for open type II strings,''
arXiv:hep-th/0111113.
%%CITATION = HEP-TH/0111113;%%
}
%\LercheZB
\lref\LercheZB{
W.~Lerche,
``Fayet-Iliopoulos potentials from four-folds,''
JHEP {\bf 9711}, 004 (1997)
[arXiv:hep-th/9709146].
%%CITATION = JHEPA,9711,004;%%
}
%\LercheYW
\lref\LMW{
W.~Lerche, P.~Mayr and N.~Warner,
``N = 1 special geometry, mixed Hodge variations and toric geometry,''
arXiv:hep-th/0208039;
%%CITATION = HEP-TH/0208039;%%
``Holomorphic N = 1 special geometry of open-closed type II strings,''
arXiv:hep-th/0207259.
%%CITATION = HEP-TH/0207259;%%
}
%\LopesCardosoAF
\lref\LopesCardosoAF{
G.~Lopes Cardoso, G.~Curio, G.~Dall'Agata and D.~Lust,
``BPS action and superpotential for heterotic string compactifications å with
fluxes,''
JHEP {\bf 0310}, 004 (2003)
[arXiv:hep-th/0306088].
%%CITATION = JHEPA,0310,004;%%
}
%\LouisUY
\lref\LouisUY{
J.~Louis and A.~Micu,
``Heterotic string theory with background fluxes,''
Nucl.\ Phys.\ å B {\bf 626}, 26 (2002)
[arXiv:hep-th/0110187].
%%CITATION = NUPHA,B626,26;%%
}
%\LouisKB
\lref\LouisKB{
J.~Louis and A.~Micu,
``Heterotic-type IIA duality with fluxes,''
JHEP {\bf 0703}, 026 (2007)
[arXiv:hep-th/0608171].
%%CITATION = JHEPA,0703,026;%%
}
%\MayrXK
\lref\MayrXK{
P.~Mayr,
``N = 1 mirror symmetry and open/closed string duality,''
Adv.\ Theor.\ Math.\ Phys.\ å {\bf 5}, 213 (2002)
[arXiv:hep-th/0108229].
%%CITATION = 00203,5,213;%%
}
%\PolchinskiRR
\lref\PolchinskiRR{
J.~Polchinski,
``String theory. Vol. 2: Superstring theory and beyond,''
%\href{http://www.slac.stanford.edu/spires/find/hep/www?irn=4634802}{SPIRES entry}
{\it å Cambridge, UK: University Press (1998).}
}
%\SenBP
\lref\SenBP{
A.~Sen,
``Orientifold limit of F-theory vacua,''
Nucl.\ Phys.\ Proc.\ Suppl.\ å {\bf 68}, 92 (1998)
[Nucl.\ Phys.\ Proc.\ Suppl.\ å {\bf 67}, 81 (1998)]
[arXiv:hep-th/9709159].
%%CITATION = NUPHZ,67,81;%%
}
%\SenPM
\lref\SenPM{
A.~Sen and S.~Sethi,
``The mirror transform of type I vacua in six dimensions,''
Nucl.\ Phys.\ å B {\bf 499}, 45 (1997)
[arXiv:hep-th/9703157].
%%CITATION = NUPHA,B499,45;%%
}
%\SenTQ
\lref\SenTQ{
A.~Sen,
``Local Gauge And Lorentz Invariance Of The Heterotic String Theory,''
Phys.\ Lett.\ å B {\bf 166}, 300 (1986).
%%CITATION = PHLTA,B166,300;%%
}
%\SethiES
\lref\SethiES{
S.~Sethi, C.~Vafa and E.~Witten,
``Constraints on low-dimensional string compactifications,''
Nucl.\ Phys.\ å B {\bf 480}, 213 (1996)
[arXiv:hep-th/9606122].
%%CITATION = NUPHA,B480,213;%%
}
%\StromingerUH
\lref\StromingerUH{
A.~Strominger,
``Superstrings with Torsion,''
Nucl.\ Phys.\ å B {\bf 274}, 253 (1986).
%%CITATION = NUPHA,B274,253;%%
}

\lref\grothen{
A.~Grothendieck,
``La th\'eorie des classes de {C}hern,'' 
Bull.\ Soc.\ Math.\ France {\bf 86}, 137--154 (1958)}

\lref\ThomasAA{
R.~P.~Thomas,
``A holomorphic Casson invariant for Calabi-Yau 3-folds, and bundles on K3
fibrations,''
Jour.\ Diff.\ Geom. {\bf 54}, no. 2, 367 (2000)
[arXiv:math.AG/9806111]}
%\deWitXR
\lref\deWitXR{
B.~de Wit, M.~T.~Grisaru, E.~Rabinovici and H.~Nicolai,
``Two Loop Finiteness Of D = 2 Supergravity,''
Phys.\ Lett.\ å B {\bf 286}, 78 (1992)
[arXiv:hep-th/9205012].
%%CITATION = PHLTA,B286,78;%%
}
%\WittenBZ
\lref\WittenBZ{
E.~Witten,
``New Issues In Manifolds Of SU(3) Holonomy,''
Nucl.\ Phys.\ å B {\bf 268}, 79 (1986).
%%CITATION = NUPHA,B268,79;%%
}
%\WittenEX
\lref\WittenEX{
E.~Witten,
``String theory dynamics in various dimensions,''
Nucl.\ Phys.\ å B {\bf 443}, 85 (1995)
[arXiv:hep-th/9503124].
%%CITATION = NUPHA,B443,85;%%
}
%\WittenYC
\lref\WittenYC{
E.~Witten,
``Phases of N = 2 theories in two dimensions,''
Nucl.\ Phys.\ å B {\bf 403}, 159 (1993)
[arXiv:hep-th/9301042].
%%CITATION = NUPHA,B403,159;%%
}
\lref\WittenFQ{
E.~Witten,
``Heterotic string conformal field theory and A-D-E singularities,''
JHEP {\bf 0002}, 025 (2000)
[arXiv:hep-th/9909229].
%%CITATION = JHEPA,0002,025;%%
}
\lref\WittenBN{
E.~Witten,
``Non-Perturbative Superpotentials In String Theory,''
Nucl.\ Phys.\ å B {\bf 474}, 343 (1996)
[arXiv:hep-th/9604030].
%%CITATION = NUPHA,B474,343;%%
}
%\WittenHC
\lref\WittenHC{
E.~Witten,
``Five-brane effective action in M-theory,''
J.\ Geom.\ Phys.\ å {\bf 22}, 103 (1997)
[arXiv:hep-th/9610234].
%%CITATION = JGPHE,22,103;%%
}
\lref\OoguriWJ{
H.~Ooguri and C.~Vafa,
``Two-Dimensional Black Hole and Singularities of CY Manifolds,''
Nucl.\ Phys.\ å B {\bf 463}, 55 (1996)
[arXiv:hep-th/9511164].
%%CITATION = NUPHA,B463,55;%%
}
%\AspinwallQW
\lref\AspinwallQW{
P.~S.~Aspinwall,
``An analysis of fluxes by duality,''
arXiv:hep-th/0504036.
%%CITATION = HEP-TH/0504036;%%
}
%%%%%%%%%%%%%%%%%
\lref\BottTu{
R.~Bott and L.~W.~Tu,
``Differential forms in algebraic topology,''
Springer-Verlag.
}
\lref\BBDG{
K.~Becker, M.~Becker, K.~Dasgupta and P.~S.~Green,
``Compactifications of heterotic theory on non-Kaehler complex manifolds. I,''
JHEP {\bf 0304}, 007 (2003)
[arXiv:hep-th/0301161];
K.~Becker, M.~Becker, P.~S.~Green, K.~Dasgupta and E.~Sharpe,
``Compactifications of heterotic strings on non-Kaehler complex å manifolds. II,''
Nucl.\ Phys.\ å B {\bf 678}, 19 (2004)
[arXiv:hep-th/0310058].
}
%\AndriotFP
\lref\AndriotFP{
D.~Andriot, R.~Minasian and M.~Petrini,
``Flux backgrounds from Twist duality,''
arXiv:0903.0633 [hep-th].
%%CITATION = ARXIV:0903.0633;%%
}
%\BarsQQ
\lref\BarsQQ{
I.~Bars, D.~Nemeschansky and S.~Yankielowicz,
``Compactified Superstrings And Torsion,''
Nucl.\ Phys.\ å B {\bf 278}, 632 (1986).
%%CITATION = NUPHA,B278,632;%%
}
%\BeckerET
\lref\BeckerET{
K.~Becker, M.~Becker, J.~X.~Fu, L.~S.~Tseng and S.~T.~Yau,
``Anomaly cancellation and smooth non-Kaehler solutions in heterotic string
theory,''
Nucl.\ Phys.\ å B {\bf 751}, 108 (2006)
[arXiv:hep-th/0604137].
%%CITATION = NUPHA,B751,108;%%
}
%\BeckerGQ
\lref\BeckerGQ{
K.~Becker, M.~Becker, K.~Dasgupta and S.~Prokushkin,
``Properties of heterotic vacua from superpotentials,''
Nucl.\ Phys.\ å B {\bf 666}, 144 (2003)
[arXiv:hep-th/0304001].
%%CITATION = NUPHA,B666,144;%%
}
%\CurioAE
\lref\CurioAE{
G.~Curio, A.~Klemm, B.~K\"ors and D.~L\"ust,
``Fluxes in heterotic and type II string compactifications,''
Nucl.\ Phys.\ å B {\bf 620}, 237 (2002)
[arXiv:hep-th/0106155].
%%CITATION = NUPHA,B620,237;%%
}
%\FuSM
\lref\FuSM{
J.~X.~Fu and S.~T.~Yau,
``Existence of supersymmetric Hermitian metrics with torsion on å non-Kaehler
manifolds,''
arXiv:hep-th/0509028.
%%CITATION = HEP-TH/0509028;%%
}
\lref\GurrieriDTJG{
S.~Gurrieri, A.~Lukas and A.~Micu,
``Heterotic on half-flat,''
Phys.\ Rev.\ å D {\bf 70}, 126009 (2004)
[arXiv:hep-th/0408121];
``Heterotic String Compactifications on Half-flat Manifolds II,''
JHEP {\bf 0712}, 081 (2007)
[arXiv:0709.1932 [hep-th]].
}
%\HullKZ
\lref\HullKZ{
C.~M.~Hull,
``Compactifications of the heterotic superstring,''
Phys.\ Lett.\ å B {\bf 178}, 357 (1986).
%%CITATION = PHLTA,B178,357;%%
}
%\LopesCardosoSP
\lref\LopesCardosoSP{
G.~Lopes Cardoso, G.~Curio, G.~Dall'Agata and D.~L\"ust,
``Heterotic string theory on non-K\"ahler manifolds with H-flux and gaugino
condensate,''
Fortsch.\ Phys.\ å {\bf 52}, 483 (2004)
[arXiv:hep-th/0310021].
%%CITATION = FPYKA,52,483;%%
}
\lref\KaSu{
å Y.~Kazama and H.~Suzuki,
å ``New N=2 Superconformal Field Theories and Superstring Compactification,''
å Nucl.\ Phys.\ å B {\bf 321}, 232 (1989).
å %%CITATION = NUPHA,B321,232;%%
}
\lref\LVW{
å W.~Lerche, C.~Vafa and N.~P.~Warner,
å ``Chiral Rings in N=2 Superconformal Theories,''
å Nucl.\ Phys.\ å B {\bf 324}, 427 (1989).
å %%CITATION = NUPHA,B324,427;%%
}
\lref\KnOm{
å J.~Knapp and H.~Omer,
å ``Matrix factorizations, minimal models and Massey products,''
å JHEP {\bf 0605}, 064 (2006)
å [arXiv:hep-th/0604189].
å %%CITATION = JHEPA,0605,064;%%
}
\lref\EguchiFM{
å T.~Eguchi, N.~P.~Warner and S.~K.~Yang,
å ``ADE singularities and coset models,''
å Nucl.\ Phys.\ å B {\bf 607}, 3 (2001)
å [arXiv:hep-th/0105194].
å %%CITATION = NUPHA,B607,3;%%
}
\lref\GovindarajanJS{
å S.~Govindarajan, T.~Jayaraman and T.~Sarkar,
å ``Worldsheet approaches to D-branes on supersymmetric cycles,''
å Nucl.\ Phys.\ å B {\bf 580}, 519 (2000)
å [arXiv:hep-th/9907131].
å %%CITATION = NUPHA,B580,519;%%
}
%\HerbstJP
\lref\HerbstJP{
M.~Herbst, C.~I.~Lazaroiu and W.~Lerche,
``Superpotentials, A(infinity) relations and WDVV equations for open
topological strings,''
JHEP {\bf 0502}, 071 (2005)
[arXiv:hep-th/0402110].
%%CITATION = JHEPA,0502,071;%%
}
\lref\MoCu{
C.~Curto, D.~R.~Morrison,
``Threefold flops via Matrix Factorizations,''
arXiv:math.AG/0611014}
\lref\DistlerMK{
å J.~Distler and S.~Kachru,
å ``(0,2) Landau-Ginzburg theory,''
å Nucl.\ Phys.\ å B {\bf 413}, 213 (1994)
å [arXiv:hep-th/9309110].
å %%CITATION = NUPHA,B413,213;%%
}
\lref\DouglasZN{
 M.~R.~Douglas,
 ``Effective potential and warp factor dynamics,''
 arXiv:0911.3378 [hep-th].
 %%CITATION = ARXIV:0911.3378;%%
}
\lref\GiddingsYU{
 S.~B.~Giddings, S.~Kachru and J.~Polchinski,
 ``Hierarchies from fluxes in string compactifications,''
 Phys.\ Rev.\  D {\bf 66}, 106006 (2002)
 [arXiv:hep-th/0105097].
 %%CITATION = PHRVA,D66,106006;%%
}
\lref\BeckerKS{
 K.~Becker, M.~Becker, C.~Vafa and J.~Walcher,
 ``Moduli stabilization in non-geometric backgrounds,''
 Nucl.\ Phys.\  B {\bf 770}, 1 (2007)
 [arXiv:hep-th/0611001].
 %%CITATION = NUPHA,B770,1;%%
}

\lref\VafaWI{
 C.~Vafa,
 ``Superstrings and topological strings at large N,''
 J.\ Math.\ Phys.\  {\bf 42}, 2798 (2001)
 [arXiv:hep-th/0008142].
 %%CITATION = JMAPA,42,2798;%%
}

%\DonagiCA
\lref\DonagiCA{
R.~Donagi and M.~Wijnholt,
``Model Building with F-Theory,''
arXiv:0802.2969 [hep-th].
%%CITATION = ARXIV:0802.2969;%%
}

%%%%%%%%%%%%%%%%%
%%%%%%%%%%%%%%%%%

%%%%%%%%%%%%%%%%%%%%%%
% end references
%%%%%%%%%%%%%%%%%%%%%%

%%%%%%%%%%%%%%% Title Page  %%%%%%%%%%%%%
\Title{\vbox{
\hbox{\tt CERN-PH-TH/2009-250}
\hbox{\tt LMU-ASC 59/09}
\hbox{\tt {SU-ITP-09/52} }} }
{\vbox{
\vskip -1cm
\centerline{\hbox{On $\cx N=1$ 4d å Effective Couplings for}}
\vskip 0.5cm
\centerline{\hbox{F-theory and Heterotic Vacua}}
\vskip -0.1cm
}}
\centerline{{\bf Hans Jockers${}^{\,a}$, Peter Mayr${}^{\,b}$, Johannes Walcher${}^{\,c}$}}
\bigskip
\bigskip
\centerline{${}^a${\it Department of Physics, Stanford University}}
\centerline{{\it å Stanford, CA 94305-4060, USA}}
\vskip1.5ex
\centerline{\it ${}^b$Arnold Sommerfeld Center for Theoretical Physics}
\centerline{\it LMU, Theresienstr. 37, D-80333 Munich, Germany}
\vskip1.5ex
\centerline{\it ${}^c$PH-TH Division, CERN}
\centerline{\it CH-1211 Geneva 23, Switzerland}
\bigskip
\vskip 1.cm
\centerline{\bf Abstract}
We show that certain superpotential and K\"ahler potential couplings of $\cx N=1$ supersymmetric
compactifications with branes or bundles can be computed from Hodge theory and
mirror symmetry. This applies to F-theory on a Calabi--Yau four-fold and 
three-fold compactifications of type II and heterotic strings with branes. The heterotic case includes a
class of bundles on elliptic manifolds constructed by Friedmann, Morgan and 
Witten. Mirror symmetry of the four-fold computes non-perturbative
corrections to mirror symmetry on the three-folds, including D-instanton corrections.
We also propose a physical interpretation for the observation by Warner that relates
the deformation spaces of certain matrix factorizations and the periods of non-compact 
4-folds that are ALE fibrations.

\bigskip
\Date{\sl {December 2009}}
\vfill\eject

\def\mrm{\rm}
\noi{\bf Contents}\vskip-1cm
$$
\hskip-0.5cm{\vbox{\offinterlineskip\halign{\strut 
\hskip-12pt \ninerm #~ \hfil&
%$\phantom{\oplus\atop\oplus^3}$\hskip-12pt
~{\bf #}~\hfil&
~{\bf #}~\hfil&
~{#}~\hfil&
~{#}~\hfil&
~#\hfil \cr
& & & å \cr
&1.&Introduction& & &\cr
&2.&Hodge theoretic data and $\cx N=1$ superpotentials& &&\cr
&&\mrm 2.1. Hodge variations in open-closed duality&&\cr
&&\mrm 2.2. Hodge variations for heterotic superpotentials&&\cr
&&\mrm 2.3. Holomorphic-Chern Simons functional for heterotic bundles&&\cr
&&\mrm 2.4. Chern-Simons vs.\ F-theory/heterotic duality&&\cr
&3.&Quantum corrected superpotentials in F-theory from mirror symmetry of 4-folds&&&\cr
&&\mrm 3.1. Four-fold superpotentials: a first look at the quantum corrections&&\cr
&&\mrm 3.2. $\cx N=1$ Duality chain&&\cr
&&\mrm 3.3. The decoupling limit as a stable degeneration&&\cr
&&\mrm 3.4. Open-closed duality as a limit of F-theory/heterotic duality&&\cr
&&\mrm 3.5. Instanton corrections and mirror symmetry in F-theory&&\cr
&4.&\bf Heterotic superpotential from F-theory/heterotic duality\cr
&&\mrm 4.1. Generalized Calabi--Yau contribution to $W_F(\Xb)$&&\cr
&&\mrm 4.2. The Chern-Simons contribution to $W_F(\Xb)$&&\cr
&&\mrm 4.3. Type II / heterotic map&&\cr
&5.&Type II/heterotic duality in two space-time dimensions\cr
&&\mrm 5.1. Type IIA on Calabi-Yau fourfolds&&\cr
&&\mrm 5.2. Type IIA on the Calabi-Yau 4-folds $\Xa$ and $\Xb$&&\cr
&&\mrm 5.3. Heterotic string on $T^2\times\Zb$&&\cr
&6.&A heterotic bundle on the mirror of the quintic\cr
&&\mrm 6.1. Heterotic string on the threefold in the decoupling limit&&\cr
&&\mrm 6.2. F-theory superpotential on the four-fold $\Xb$&&\cr
&&\mrm 6.3. Finite $S$ corrections: perturbative contributions&&\cr
&&\mrm 6.4. D-instanton corrections and Gromov--Witten invariants on the 4-fold
&&\cr
&7.&Heterotic five-branes and non-trivial Jacobians\cr
&&\mrm 7.1. Structure group $SU(1)$: Heterotic five-branes&&\cr
&&\mrm 7.2. Non-trivial Jacobians: $SU(2)$ bundle on a degree 9 hypersurface&&\cr
&8.&ADE Singularities, Kazama-Suzuki models and matrix factorizations\cr
&9.&Conclusions\cr
\phantom{${X\over X}\over X$}&&Appendix A: Some toric data for the examples\cr
&&\mrm A.1. The quintic in $P^4(1,1,1,1,1)$\cr
&&\mrm A.2. Heterotic 5-branes\cr
&&\mrm A.3. $SU(2)$ bundle of the degree 9 hypersurface in $\IP^4(1,1,1,3,3)$\cr
}}}
$$
\vfill\eject
%%%%%%%%%%%%%%%%%%%%%%%%%%%%%%%%%%
\newsec{Introduction}
Let $\Zb$ be a Calabi--Yau (CY) three-fold and $E$ a holomorphic bundle or
sheaf on it. In a certain decoupling limit, 
where one neglects the backreaction of the
full string theory to the degrees of freedom of the bundle,
$E$ can describe either a (sub-)bundle of a heterotic string compactification
on $\Zb$, a heterotic 5-brane å or a $B$-type brane in a type II å compactification 
on $\Zb$. In the latter case we will also be interested in the 
geometry $(\Za,L)$ associated to $(\Zb,E)$ by open string mirror symmetry, 
which consists of an $A$-type brane $L$ on the mirror three-fold $\Za$ of $\Zb$. å 
The contribution of the bundle to the space-time superpotential of 
a string compactification on $\Zb$ is, in a 
certain approximation, given by the holomorphic Chern-Simons functional
for both the heterotic bundle \WittenBZ\ and the $B$-type brane \WitCS 
\eqn\ecs{
W_{CS} = \int_{\Zb} \Om \wedge \tr (\fc{1}{2}A\wedge \bb \p A+\fc{1}{3}A\wedge A \wedge A)\ .
}
Here $\Om$ is the holomorphic (3,0) form on $\Zb$ and $A$ is the $(0,1)$ 
part of the å connection on $E$. There is another superpotential proportional
to the periods of $\Om$, which, 
again in a certain approximation, is of the form 
\eqn\ecsp{
W_{G} = \int_{\Zb} \Om \wedge G= (N_\Si + S \hx N_\Si)\int_{\ga_\Si} \Om,\qquad \gamma_\Si \in H_3(\Zb,\IZ) \ å .
}
In the type II compactification on $\Zb$, $W_G$ is the 
superpotential å induced by NS and RR 3-form fluxes \TVsp, and $S$ the complex 
dilaton.
In heterotic compactifications, $W_G$ will be related below to the superpotential
of a compactification on non-K\"ahler manifolds with
$H$-flux \refs{\StromingerUH}.
Depending on the type of string theory and its 
compactification, the combined superpotential 
\eqn\ecc{W=W_{CS}+W_G\ ,} may
be exact or subject to various quantum corrections. 

The purpose of this note is to show how the methods of mirror symmetry of
refs.~\refs{\HV,\VafOS,\AV} when combined with Hodge theory can be used to 
compute effective couplings of these heterotic/type II compactifications, 
including the superpotential and the K\"ahler potential.
Hodge theory enters in two steps: A 'classical' theory
on the CY 3-fold, which computes the integrals on the 3-fold in \ecs,\ecsp, 
and a 'quantum' deformation of these 3-fold data defined by the 
(classical) Hodge variation on a 'dual' CY 4-fold. Physicswise, the 4-fold geometry represents 
the compactification manifold
of a dual F-theory or type IIA compactification. We will argue that 
the 4-fold result agrees with the 3-fold result when it should, but
gives more general results, including the case when the heterotic 3-fold
is not CY.

The first step on the three-fold 
can be realized by computing the Hodge variation on a {\it relative} cohomology 
group $H^3(\Zb,\cxH)$, which captures the brane/bundle data in addition to 
the geometry of $\Zb$. This was shown previously in the context of $B$-type branes
in \refs{\LMW,\JockersPE,\AlimRF,\AlimBX} and we generalize this relation here to 
heterotic 5-branes and general bundles, including the bundles 
on elliptically fibered 3-fold $\Zb$ constructed by 
Friedman, Morgan and Witten in \FMW\ (see also ref.~\BJPS). The 'classical' 
Hodge theory on the 3-fold gives an 
explicit evaluation of the 3-fold integrals in \ecs,\ecsp\
and a preferred choice of physical coordinates, which leads to the prediction of 
world-sheet corrections from sphere and disc instantons 
of the appropriately defined 
mirror theories.

The second step involves Hodge theory and mirror symmetry on a mirror pair of 
dual CY 4-folds. 4-folds enter the stage in two seemingly different
ways, in remarkable parallel with the two appearances of \ecs\ in heterotic and type II
compactifications on $\Zb$. Firstly, through the duality of heterotic strings on elliptically 
fibered CY 3-fold $\Zb$ to F-theory on a CY 4-fold $\Xb$ 
\refs{\vafaf,\MV}. This duality motivated the systematic construction 
of ``heterotic'' bundles on elliptically fibered $\Zb$ in refs.~\refs{\FMW,\BJPS}. 
Secondly, 4-folds appear in the computation of 
brane superpotentials of type II strings via an ``open-closed string duality'', which å 
associates a non-compact 4-fold geometry $\XXb$ 
to a $B$-type brane on a 3-fold $\Zb$ \refs{\MayrXK,\AlimRF,\AganagicJQ}. In this approach, the superpotential 
\ecs\ of the brane compactification on $(\Zb,E)$ is computed from the 
periods of the holomorphic $(4,0)$ form on the dual 4-fold $\XXb$.
Moreover, mirror symmetry of 4-folds relates the sphere instanton 
corrected periods
on the mirror 4-fold $\XXa$ of $\XXb$ to the disc instanton corrected superpotential of the compactification with $A$-type brane $L$ 
on the mirror manifold $\Za$ of $\Zb$. This
surprising relation between mirror symmetry of the 4-folds $\XXa$ and $\XXb$ 
and open string mirror symmetry of the brane geometries $(\Zb,E)$ and $(\Za,L)$ 
has been tested in various different contexts, see e.g. 
\refs{\LM,\AlimBX,\GrimmEF,\Yau}. 

As we will argue below, these two 4-fold strands are in fact connected 
by a certain physical and geometrical limit, that relates open-closed duality to 
heterotic/F-theory duality.\foot{A related explanation of type II open-closed duality based on 
T-duality of 5-branes \OoguriWJ\ has been recently given in ref.~\AganagicJQ.} 
In this limit part of the bundle degrees of freedoms decouple (in a physical sense) from the 
remaining compactification and the type II brane and the heterotic bundle are equalized.
Geometrically, this can be viewed as a local mirror limit in the open string sector
of type II strings or a local mirror limit for bundles considered in \refs{\KMV,\BMff}, respectively. 
In this limit, the F-theory/type IIA superpotential on the dual 4-fold $\Xb$ reduces to 
the 'classical' type II/heterotic superpotential \ecc\ å on the 3-fold $\Zb$, as has been 
observed previously in \AlimBX.

The result obtained from an F-theory/type IIA compactification on the dual 4-fold differs 
from the 3-fold result away from the decoupling limit. We assert that these deviations 
represent physical corrections to the dual type II/heterotic compactification 
from perturbative and instanton effects and describe how Hodge theory and mirror symmetry
on the 4-fold provides a powerful computational tool to determine these perturbative and non-perturbative
contributions. 
Depending on the point of view, the corrections computed by mirror symmetry of 4-folds 
describe world-sheet, D-brane or
space-time instanton effects in the dual type II and
heterotic compactifications. 

Finally we discuss the type II/heterotic duality in the context of non-compact 4-folds that
arise as two-dimensional ALE fibrations. For a particular choice of 
background fluxes these models admit a description in terms of certain Kazama-Suzuki
coset models \refs{\GVW,\EguchiFM}, whose deformation spaces coincide with the
deformation spaces of matrix factorizations of $\cx N=2$ minimal models \KnOm.
We give a physical interpretation of this relation via type II/heterotic duality and
we propose that this correspondence holds even more generally. 

The organization of this note is as follows. 
In sect.~2 we discuss the application of 
Hogde theory to the evaluation of the Chern-Simons functional \ecs\ with 
a focus on bundles on elliptic CY 3-fold constructed by 
Friedman, Morgan and Witten \FMW. For a perturbative bundle with structure group 
$SU(N)$ the superpotential captures obstructions to the deformation 
of the spectral cover $\Si$ imposed by a certain choice of line bundle.
We discuss also the case of a general structure group $G$ and heterotic 5-branes.
In sect.~3 we describe the decoupling limit in the type II and heterotic
compactifications and use it to relate open-closed string duality to 
F-theory/heterotic duality, giving an explicit map between type II and heterotic
compactifications. We discuss the relevant string dualities and the meaning of
the quantum corrections in the dual theories.
In sect.~4, we argue, that the F-theory superpotential on the 4-fold captures 
more generally the heterotic superpotential for a bundle compactification on a 
generalized Calabi--Yau manifold and describe the map from the F-theory superpotential
to the superpotential for heterotic bundles and heterotic 5-branes.
In sect.~5 we extend the previous discussion to the K\"ahler potential and the twisted
superpotential by studying the effective supergravity for the 
two-dimensional compactification of type IIA on the
4-fold and heterotic strings on $T^2\times \Zb$. 
In sect.~6 we start to demonstrate our techniques for an example of an $\cx N=1$
supersymmetric bundle compactification on the quintic. We discuss the perturbative 
heterotic theory, the general structure of the quantum corrections and give 
explicit results for the example.
In sect.~7 we consider other interesting examples, including heterotic 5-branes
wrapping a curve in the base of the heterotic CY manifold and bundles with non-trivial
Jacobians. In sect.~8 we connect via heterotic/type II duality the deformation spaces
of certain matrix factorizations to the deformation spaces
of type II on non-compact 4-folds that are ALE fibrations with fluxes.
Sect.~9 contains 
our conclusions. In the appendix we present further technical details
on the computations for the toric hypersurface examples analyzed in the main text.

\newsec{Hodge theoretic data and $\cx N=1$ superpotentials}
\subsec{Hodge variations in open-closed duality}
In the approach of refs.\refs{\LMW,\JockersPE,\AlimBX}, the superpotential
of $B$-type brane compactifications with 5-brane charge 
on a Calabi--Yau $\Zb$ is computed from the mixed Hodge variation
on a certain relative cohomology group $H^3(\Zb,\cxH)$. 
The superpotential is a linear combination of the period integrals of 
the relative (3,0) form $\ux \Om\in H^{3,0}(\Zb,\cxH)$ 
\eqn\spii{
W_{II}(\Zb,\cxH)=\sum_{\ga_\Si \in H_3(\Zb)} N_\Si \int_{\ga_\Si}\ux
\Om^{(3,0)} +\sum_{\ga_\Si \in H_3(\Zb,\cxH) \atop \cxH \supset\ \p\ga_\Si \neq 0}
\hx N_\Si \int_{\ga_\Si}\ux \Om^{(3,0)}\ .
}
The first term is the RR ``flux'' superpotential \refs{\TVsp,\GVW} 
on 3-cycles $\ga_\Si\in H_3(\Zb)$ and the second term an 
off-shell version of the brane superpotential \refs{\WitCh,\Katz,\AV} 
defined on 3-chains $\ga_\Si$ with non-empty boundary. Note that the 
superpotential $W_{II}(\Zb,\cxH)$ associated with the Hodge bundle 
does not include the NS part of the type II flux potential.

The boundary $\p\ga_\Si$ is required to lie 
in a hypersurface $\cxH\subset \Zb$, $\p\ga_\Si \in H_2(\cxH)$. 
The moduli of the 
hypersurface $\cxH$ parametrize certain deformations of the brane 
configuration $(\Zb,E)$. 
Infinitesimally, the accessible å deformations are described by 
elements in $H^{2,1}(\Zb,\cxH)$ and come in two varieties,
\eqn\defos{
\phi_a\in H^{2,1}(\Zb)\ ,\qquad 
\hx \phi_\al\in H^{2,0}(\cxH) \ .
}
Here $H^{2,1}(\Zb)$ captures the deformations of the 
complex structure of the 3-fold $\Zb$ and 
$H^{2,0}(\cxH)$ the deformations of the holomorphic 
hypersurface $i:\, D\hookrightarrow \Zb$. 

Mirror symmetry maps the $B$-type brane configuration $(\Zb,E)$ to an
$A$-type brane configuration $(\Za,L)$ on the mirror 3-fold $\Za$. å The flat
Gauss-Manin connection on $H^3(\Zb,\cxH)$ determines the mirror map $z(t)$ 
between the complex structure moduli $z$ of $(\Zb,E)$ and the K\"ahler moduli 
$t$ of $(\Za,L)$. Inserting the mirror map into \spii\ then gives the
disc instanton corrected superpotential of the $A$-type geometry near
a suitable large volume point of $(\Za,L)$ \AlimBX.

The relative cohomology problem and open string mirror symmetry is 
related to absolute cohomology and mirror symmetry of CY 4-folds 
by a certain open-closed string duality \refs{\MayrXK,\AlimRF,\AganagicJQ}. 
The constructions of these papers associate 
to a $B$-type brane compactification $(\Zb,E)$ and its mirror 
$(\Za,L)$ a pair of 
non-compact mirror 4-folds $(\XXa,\XXb)$, such that the ``flux'' superpotential of \GVW\ agrees with the combined ``flux'' 
and brane superpotential \spii\ of the three-fold compactification,
\eqn\Wnc{\eqalign{
W(\XXb)&=\sum_{\ga_\Si \in H_4(\XXb)} {\ux N}_\Si \int_{\ga_\Si}\Om^{(4,0)}\ = \ W_{II}(\Zb,\cxH)\ ,
}}
for appropriate choice of coefficients $N_\Si,\hx N_\Si,\ux N_\Si$. 
Open-closed string duality thus links the pure Hodge variation on 
$H^4_{hor}(\XXb)$ to the mixed Hodge variation on the relative 
cohomology space $H^3(\Zb,\cxH)\simeq H^3(\Zb)\oplus H^2_{var}(\cxH)$.
The relation between the pure Hodge spaces appearing in this relation is schematically

\eqn\ochr{
\xymatrix{
H^{3,0}(\Zb)\ar[r]^\dcl&H^{2,1}(\Zb)\ar[r]^\dcl&
H^{1,2}(\Zb)\ar[r]^\dcl&H^{0,3}(\Zb)&\cr
H^{4,0}(\XXb)\ar[r]^\dcl\ar[u]^\al&H^{3,1}(\XXb)\ar[r]^\dcl\ar[u]^\al\ar[d]^\be&
H^{2,2}_{hor}(\XXb)\ar[r]^\dcl\ar[u]^\al\ar[d]^\be&H^{1,3}(\XXb)\ar[r]^\dcl\ar[u]^\al\ar[d]^\be&H^{0,4}(\XXb)&\cr
&H^{2,0}(\cxH)\ar[r]^\dcl&
H^{1,1}_{var}(\cxH)\ar[r]^\dcl&H^{0,2}(\cxH)\cr
}
}
Here $\delta$ denotes universally a variation in the complex structure
of the respective geometries, represented by the Gauss-Manin derivative and
projecting onto pure pieces. 

The two maps $\al, \be:\ H^4_{hor}(\XXb)\to H^3(\Zb,D)$
identify an element of $H^4_{hor}(\XXb)$ either
with an element in $H^3(\Zb)$ of the closed string state space or 
an element in $H^2(\cxH)$ associated with the brane 
geometry $i:\, \cxH\hookrightarrow \Zb$. 
These maps can be explicitly 
realized on the level of 4-fold period integrals by integrating out 
certain directions of the 4-cycles $\Ga_\Si\in H_4(\XXb)$ \refs{\MayrXK,\AganagicJQ}. 
The map $\al:\ H^4_{hor}(\XXb)\to H^3(\Zb)$ 
can be represented as an integration over a particular $S^1$ in $\XXb$
and shifts the Hodge degree by $(-1,0)$. The other class of contours
produces a delta function on the hypersurface $D$ as in \HV, 
and leads to the map $\be:\ H^4_{hor}(\XXb)\to H^2(\cxH)$ that shifts by 
$(-1,-1)$.
Specifically, the infinitesimal deformations of the complex structure
of $\XXb$ split into the closed and open string deformations \defos\ as 
$$
H^{3,1}(\XXb)\simeq H^{2,1}(\Zb)\oplus H^{2,0}(\cxH)\ .
$$

The above deformation problem is a priori {\it unobstructed}, 
but becomes obstructed by the superpotential \Wnc\ upon adding the 
appropriate ``flux''. In the brane geometry $(\Zb,E)$ this
can be realized by a brane flux, adding a D5-charge $\tx \ga\in H^2(\cxH)$
\refs{\AlimBX,\GrimmEF,\AganagicJQ}. A
non-trivial obstruction in the open string direction arises for the choice
\eqn\deftxgamma{
\tx \ga\in H^2_{var}(\cxH)={\rm coker} \big( H^2(\Zb)\buildrel i^* \over \rightarrow H^2(\cxH)\big)\ .}
Restricting the open string moduli to the subspace where
the class $\tx\ga$ remains of type (1,1) leads to a superpotential for
the closed string moduli as in refs.~\refs{\Wa,\MW}. Note also that
a class $\tx \ga$ in the image of $i^*$ is always 
of type (1,1) and thus does not impose a restriction on the moduli
of $\cxH$, as the variation $\delta W_{II}$ of eq.\spii\ is automatically zero for
a holomorphic boundary $\p\Ga_\Si$.

\subsec{Hodge variations for heterotic superpotentials}
In the following we consider a similar Hodge theoretic approach to superpotentials of
``heterotic'' bundles on elliptically fibered Calabi--Yau manifolds 
constructed in \refs{\FMW,\BJPS}. 

In the framework Friedmann, Morgan and Witten, an $SU(n)$ bundle $E$ on
an elliptically fibered CY 3-fold $\pi_{\Zb}:\ \Zb\to B$ with section
$\sigma:\ B\to \Zb$ is described in terms of a spectral cover $\Si$, which is 
an $n$-fold cover $\pi_\Si:\, \Si\to B$, 
and certain twisting data specifying a line bundle on $\Si$.
Fixing the projection of the second Chern class of $E$ to the base $B$, 
the latter comprise a continuous part related to the 
Jacobian of $\Si$ and a discrete part from elements 
\eqn\defgamma{
\ga \in å 
\ker \big( H^{1,1}(\Si) \buildrel {\pi_{\Si_*}} \over \longrightarrow 
H^{1,1}(B)\big)
}
In the duality to F-theory on a 4-fold $\Xb$, the elements 
of the Hodge spaces of the spectral cover are related to those on $\Xb$ 
schematically å as \refs{\FMW,\BJPS,\CurioBVA}:
$$
{\vbox{\offinterlineskip\halign{
\strut \hfil~$#$~\hfil\vrule&\hfil~$#$~\hfil &~$#$\hfil \cr
\Si&\Xb\cr
\noalign{\hrule}
H^{2,0}&H^{3,1}\cr
H^{1,1}&H^{2,2}\cr
H^{1,0}&H^{2,1}\cr
}}}
$$

The first line identifies the infinitesimal deformations of $\Si$
with å infinitesimal deformations of the 4-fold.
The second relation relates the discrete data described by the class $\ga$
with 4-form flux in the F-theory compactification on $\Xb$.
The last relation reflects the isomorphism of the Jacobian of
$\Si$ and the corresponding Jacobian in $\Xb$ related to it by duality (see also \AspinwallBW).
Note that the heterotic/F-theory relation between $H^4(\Xb)$ and $H^2(\Si)$ 
is formally given by the same $(-1,-1)$ shift in Hodge degree as in the map $\be$ in
the open-closed duality relation \ochr. As argued below, this similarity is
not accidental, but a reflection of the fact, that the 
heterotic and type II data can be related by the afore mentioned decoupling limit.

Again the deformations of the spectral cover $\Si$ in $H^{2,0}(\Si)$ 
are unobstructed if $\ga$ is the ``generic'' $(1,1)$ class discussed 
in \refs{\FMW}.\foot{However, the existence of this class is a consequence of 
insisting on a section for $\pi_{\Si}:\, \Si\to B$.} 
Consider instead a class $\ga$ that is of type 
$(1,1)$ only on a subspace $\zh=0$ of the deformation space.
Twisting by $\ga$ then should obstruct the deformations of $\Si$ in the 
direction $\zh\neq 0$, which destroy the property 
$\ga \in H^{1,1}(\Si)$.

We propose that the heterotic superpotential describing 
this obstruction is captured by the chain integral

\eqn\wwiii{\eqalign{
W_{het}(\Zb,\Si,\ga)&=\int_{\Ga} \ux \Om^{3,0}\ ,
%=\int_{\Zb} \Om \wedge \tr (\fc{1}{2}A\wedge \bb \p A+\fc{1}{3}A\wedge A \wedge A)\ .
}}

\noi for $\Ga \in H_3(\Zb,\Si)$ a 3-chain with non-zero boundary on $\Si$. The dual space 
$H^3(\Zb,\cxS)\simeq H^3(\Zb)\oplus H^2_{var}(\cxS)$ is the
relative cohomology group defined by the spectral cover $\cxS$
with $H^2_{var}(\cxS)$ the mid-dimensional horizontal Hodge cohomology of $\cxS$.
Moreover the boundary 2-cycle $C=\p\Ga\subset\Si$ is the cycle
Poincar\'e dual to $\ga$.
The chain integral can then be computed å 
from the Hodge variation on the relative cohomology group, as has 
been used in refs.~\refs{\LMW,\JockersPE,\AlimBX} to
compute brane superpotentials in type II strings. As a first check on 
the relevance of the mixed Hodge variation on $H^3(\Zb,\cxS)$ for
the heterotic theory,
note that the deformation space $H^{2,0}(\cxS)$ is indeed captured by the Hodge space
$H^{2,1}(\Zb,\cxS)$, as in the type II case. 

In the type II context, the mixed Hodge variation gives more physical 
information than just the superpotential, specifically appropriate 
coordinates on the deformation space, which
lead to the interpretation of the superpotential as a disc 
instanton sum in the mirror $A$ model. The physical interpretation 
of the corrections in the heterotic theory will be discussed below.

The expression~\wwiii\ of the heterotic string can be argued for by relating
it to the holomorphic Chern-Simons functional~\ecs, which is the
holomorphic superpotential for the bundle moduli in the heterotic string \WittenBZ.
Before turning to the derivation for a genuine CY 3-fold of holonomy $SU(3)$,
it is instructive to reflect on the argument at
the hand of the simpler $\cx N=2$ supersymmetric case of dual compactifications 
of F-theory on $K3\times K3$ and heterotic string on $T^2\times K3$. The 
perturbative $F$-term superpotential associated with a heterotic flux on K3 in the $i$-th $U(1)$
factor å is \refs{\Luest,\JockersZY}
\eqn\whetnet{
W^{\cx N=2}_{het}=A_i\ \int_C \om^{2,0}\ ,
}
where $A_i$ is the Wilson line on $T^2$, $C$ the cycle Poincar\'e dual to 
the flux and $\om^{2,0}$ the holomorphic $(2,0)$ form on the heterotic K3. 
In this simple case, the spectral cover
is just points on the dual $T^2$ times K3, and the chain integral \wwiii\ over 
the holomorphic $(3,0)$ form $dz\wedge \om^{2,0}$ becomes
\eqn\hcskt{
W_{het}=\int_\Ga \Om = \int_0^{p_i} dz\ \int_C \om^{2,0}
= A_i\int_C\om^{2,0} \ ,
}
reproducing \whetnet. Here we used that the holomorphic Wilson lines
with periods $A_i\sim A_i+1\sim A_i+\tau$ appearing in \whetnet\ are
defined by the Abel-Jacobi map on $T^2$. Furthermore, $p_i$ denotes
the associated point in the Jacobian.
In the $\cx N=1$ case, the points $p_i$ 
vary over the base and the bounding 
2-cycles are not of the simple form $(0,p_i)\times C$. An important
consequence is that holomorphy of $C$ gets linked to the
deformations $A_i$.\foot{See ref.~\DonagiCA\ for a similar discussion.}

There is also a simple generalization of this $\cx N=2$ superpotential 
to the case, where the heterotic vacuum contains heterotic 5-branes \BMof, and 
this is also true for the $\cx N=1$ supersymmetric case studied below.
The 5-brane superpotential is in fact the most straightforward part starting 
from the results on type II brane superpotentials of 
refs.~\refs{\LMW,\JockersPE,\AlimBX}, as the
brane deformations of the type II brane map to the 
brane deformations of the heterotic 5-brane in a simple way. The 
type II/heterotic map providing this identification and explicit examples will
be discussed later on.

\subsec{Holomorphic-Chern Simons functional for heterotic bundles} 
The holomorphic Chern-Simons functional is (a projection of) the transgression 
of the Chern-Weil representation of the algebraic second Chern class for a supersymmetric
vector bundle configuration. Thus, in order to establish for a supersymmetric
heterotic bundle configuration that \ecs\ agrees with eq.~\wwiii\ on-shell, we need to show that
the boundary 2-cycle $C=\partial\Ga$ of the 3-chain $\Ga$ in eq.~\wwiii\ is given by
a curve representing the algebraic second Chern class of the
holomorphic heterotic vector bundle. The latter is encoded in the zero and
pole structure of a global meromorphic section $s_E : Z \rightarrow E$
of the supersymmetric holomorphic heterotic bundle $E$ \grothen. This is described in 
ref.~\ThomasAA\ for a general $SU(2)$ bundle and in ref.~\MW\ for a bundle associated with 
a matrix factorization.

\def\Zb{Z}
To apply this reasoning to the $SU(N)$ bundles of \FMW, we need to
construct an explicit representative for the algebraic Chern class.\foot{To avoid cluttering of 
notation, the heterotic manifold $Z_B$ is denoted simply by $Z$ in the following argument.}
As explained in ref.~\FMW, the spectral cover $\Si$ together with the class $\ga$ of
eq.~\defgamma\ defines the $SU(n)$ bundle $E$ over the elliptically fibered
3-fold $\pi_{\Zb} : \Zb \rightarrow B$ by
$$
E \,=\, \pi_{2*}\cx R \ , \qquad \cx R \,=\, \cx P_B\otimes \cx S \ ,
\qquad \cx R \rightarrow \Si \times_B \Zb \ .
$$
Here $\pi_2$ is the projection to the second factor of the fiberwise
product $\Si \times_B \Zb$ of the 3-fold $\Zb$ and of the spectral cover $\Si$
over the common base $B$. $\cx P_B$ is the restriction of the Poincar\'e
bundle of the product $Z\times_B Z$ to $\Si\times_B Z$, while $\cx S\to \Si$ denotes
the line bundle over the spectral cover $\Si$, which is given by\foot{For ease of
notation its pull-back to $\Si\times_B Z$ is also denoted by the same symbol $\cx S$.}
$$
\cx S \,=\, \cx N \otimes \cx L_\ga \ .
$$
The bundle $\cx N$ ensures that the first Chern class $c_1(E)$ of the $SU(n)$ bundle
vanishes and its explicit form is thoroughly analyzed in ref.~\FMW. The holomorphic
line bundle $\cx L_\ga$ with $c_1(\cx L_\ga)=\ga$ governs the twisting associated to the
class~$\ga$ in \defgamma, and it is responsible for the discussed obstructions to the deformations
of the spectral cover $\Si$. Note that, due to the property~\defgamma, the line bundle $\cx L_\ga$
does not further modify the first Chern class $c_1(E)$ \FMW.

In order to construct a section $s_E$ of the $SU(n)$-bundle, we need to push-forward a global
(meromorphic) section $s_{\cx R}=s_{\cx P}\cdot s_{\cx S}$ of the line bundle $\cx R$,
which in turn is the product of a section of the Poincar\'e bundle $\cx P_B$ and the
line bundle $\cx S$. The Poincar\'e bundle is given by $\cx P_B=\cx O(\Delta - \Si\times\sigma) \otimes K_B$,
where $\Delta$ is (the restriction of) the diagonal divisor in $Z \times_B Z$, $K_B$ is
the canonical bundle of the base (pulled back to $\Si\times_B Z$) and 
$\sigma: B \rightarrow \Zb$ the section of the elliptic fibration $\Zb$. Therefore
the section $s_{\cx P}=s_K\cdot s_F$ can be chosen to be the product of
the section $s_K$ of the canonical bundle of the base $B$ and the section $s_F$,
which has a (simple) zero set along the diagonal divisor $\Delta$ and a (simple) pole set along
the divisor $\Si \times_B\sigma$.
Finally, the zero set/pole set of the section $s_{\cx S}$ is induced from the 
(algebraic) first Chern class $c_1(\cx S)$ of the line bundle $\cx S$ over the
spectral cover $\Si$. Here we are in particular interested in the contribution
from the line bundle $\cx L_\ga$, whose global (meromorphic) section extended
to the fiber-product space $\Si\times_B\Zb$ is denoted by $s_\ga$.

For an $SU(n)$-bundle the projection map $\pi_2$ is an $n$-fold branched cover
of the 3-fold $\Zb$, and therefore in a open neighborhood $U\subset B$ of the
base the push-forward of the section $s_{\cx R}$ yields
\eqn\Esec{
s_E \,=\,\pi_{2*}s_{\cx R} \,=\, s_K \cdot ( s_F^1\cdot s_{\cx S}^1 \,, \, s_F^2\cdot s_{\cx S}^2\,, \, 
\ldots\,, \, s_F^n\cdot s_{\cx S}^n ) \ . }
As the section $s_K$ originates from the canonical bundle over the base, it appears
as an overall pre-factor of the bundle section $s_E$, while the entries $s_F^i$
and $s_{\cx S}^i$ arise from the $n$ sheets of the $n$-fold branched cover.
The entries $s_F^i$ restrict on the elliptic fiber to a section of 
$\oplus_{i=1}^n (\cx O(p_i)\otimes \cx O(0)^{-1})$ that have a simple zero
at $p_i$ and a simple pole at $0$. Here $0$ denotes the 
distinguished point corresponding to the section $\sigma:\, B\to \Zb$ and 
$\sum_ip_i=0$ for $SU(n)$.\foot{At branch points of the spectral cover
(at least) two points $p_i$ and $p_j$, $i\ne j$, coincide, and the restriction
of the bundle $E$ to the elliptic fiber becomes a sum of $n-2$ line bundles
plus a rank two bundle, which is a non-trivial extension of two line bundles \FMW.
However, due to the splitting principle the second algebraic Chern class is
insensitive to these non-trivial extension, and we can simply work with the
direct sum of $n$ line bundles.} 
The $n$ entries $s_{\cx S}^i$ arise again from the section $s_{\cx S}$ on the $n$ different 
sheets. Since the section $s_{\cx S}$ is induced from a line bundle over the spectral 
cover, the zeros/poles of the sections $s_{\cx S}^i$ correspond to co-dimension one 
sub-spaces on the base.

Now we are ready to determine the algebraic Chern classes of the $SU(n)$-bundle
$E$ from the global section~\Esec. By construction the first topological Chern
class is trivial, which implies
that also the first algebraic Chern class vanishes since the Abel-Jacobi map is trivial
for the simply-connected Calabi-Yau 3-folds discussed here. 
The second algebraic Chern class is determined by the ``transverse zero/pole sets''
of the section $s_E$, which correspond to the co-dimension two cycles of
the mutual zero/pole sets of distinct entries $s_E^i$ and $s_E^j$, $i\ne j$.

Since $s_E^i=s_F^i \cdot s_{\cx S}^i$, this computation exhibits $c_2(E)$ as a sum of 
three contributions: The joint vanishing of $s_F^i$ and $s_F^j$ is empty since
$p_i\neq p_j$ generically. The joint vanishing of $s_{\cx S}^i$ and $s_{\cx S}^j$
is a sum of fibers, which we may neglect since, moving in a rational family, they do 
not contribute to the superpotential.\foot{An equivalent way to see this is to note that
five-branes wrapped on the fiber on the elliptic threefold map under heterotic/F-theory
duality to mobile D3-branes which clearly have no superpotential.}

Equivalently, we may use the relation $ch_2(E)={1\over 2} c_1(E)^2- c_2(E)$
between the second Chern class and the second Chern character $ch_2(E)$, which
thanks to the vanishing of $c_1$ reduces to $ch_2(E)=-c_2(E)$, to compute
$c_2(E)$ from the transverse zero/pole sets of the local sections $s_F^k$
and $s_{\cx S}^k$ of the {\it same} entry $k$. This will more directly lead
to the desired boundary 2-cycle $C=\partial\Gamma$. (Again, we may neglect the self-intersections
of $s_F^k$ and $s_{\cx S}^k$.)

We focus now on the contribution $c_2(E_\ga)$ to the second algebraic Chern class,
which is associated to the intersection of the zero/pole sets of the local sections $s_{\ga}^k$
and the local sections $s_F^k$ for $k=1,\ldots, n$. As argued the obtained divisor is
rational equivalent to the (negative) boundary 2-cycle $C$ arising form the Poincar\'e dual
of the 2-form $\ga$ on the spectral cover $\Si$, and we obtain for the second algebraic Chern
class
\eqn\Ectwo{ c_2(E) = c_2(E_\ga) + c_2(V) \ , \qquad 
c_2(E_\ga)\,=\,- [ C ] å \ , }
where we denote by $[C]$ the cycle class, which arises from embedding the two-cycle $C$
of the spectral spectral cover $\Si$ into the Calabi-Yau 3-fold $Z$. Due to the property~\defgamma\
the curve associated to $c_2(E_\ga)$ is (up to a minus sign) rational equivalent to the boundary
of the same 3-chain $\Ga$ appearing in eq.~\wwiii.
The other piece $c_2(V)$, which is (locally) independent of the analyzed deformations of the spectral
cover, is discussed in detail in ref.~\FMW. In general it gives rise to 
a non-trivial second topological Chern class. In a globally consistent heterotic string
compactification this contribution is compensated by the second topological Chern
class of the tangent bundle as dictated by the anomaly equations of the
heterotic string.\foot{In
generalized Calabi-Yau compactifications of the heterotic string additional contributions
enter into the anomaly equation due to non-trivial background fluxes and the modified
generalized geometry \refs{\StromingerUH}.}

Thus, by reproducing the 3-chain $\Gamma$ from the second algebraic Chern class of
the holomorphic $SU(n)$ bundles, the holomorphic Chern-Simons functional is demonstrated
to be agreement with the holomorphic superpotential~\wwiii. Analogously to the non-supersymmetric
off-shell deformations of branes in type~II compactifications \refs{\AlimBX,\AganagicJQ}, 
we propose that the correspondence between the superpotential~\wwiii\ and the
Chern-Simons functional even persists along deformations of the spectral cover, which
yield non-supersymmetric $SU(n)$ bundle configurations. 

To illustrate the presented construction, we briefly return to the $\cx N=2$ compactification
of the heterotic string on $T^2\times K3$. For this example the spectral cover of an $SU(n)$
bundle is a disjoint union of $n$ K3 surfaces $\coprod_{i=1}^n \{p_i\} \times K3$ embedded
into $T^2\times K3$.
A class $\ga$ fulfilling the property~\defgamma\
can be thought of as a non-trivial (1,1)-form $\om_\ga$, which appears in the component
$p_i \times K3$ and $p_j \times K3$, $i\ne j$, with opposite signs. Then the å 
Poincar\'e dual curve $C$ of $\ga$ embedded into $T^2\times K3$ is the boundary of
the 3-chain $\Gamma = (p_i, p_j) \times C$,
where $(p_i,p_j)$ denotes the 1-chain on the torus bounded by the points $p_i$ and $p_j$.
The resulting chain integral over $dz\wedge \om^{2,0}$ exhibits the
same structure as the naive equation \whetnet.

%%%%%%%%%%%%%%%%%%%
\lref\TBA{to appear.}

\subsec{Chern Simons vs. F-theory/heterotic duality}
\def\bA{{\bf A}}\def\bE{{\bf E}}
In the next section we will consider a dual 
F-theory compactification on a 4-fold and argue that mirror symmetry of the 4-fold 
computes interesting quantum corrections to the Chern-Simons functional.
Here we want to motivate the following 'classical' relation between the 4-fold periods and the Chern-Simons
functional \ecs\ 
\eqn\CSrel{
\int_{\Xb} \Om^{4,0}\wedge G_\bA=\int_{\Zb} \Om^{3,0} \wedge \tr (\fc{1}{2}A\wedge \bb \p A+\fc{1}{3}A\wedge A \wedge A)
\ +\ \cx O(S^{-1},e^{2\pi i S}) \ .
}
In the above, $\Xb$ is a CY 4-fold which will support the F-theory compactification 
dual to the heterotic compactification on the 3-fold $\Zb$
and  $G_\bA$ is a 4-form 'flux' related to the connection $A$ of a bundle $E\to\Zb$ as described below.
Moreover $S$ is a distinguished complex structure modulus of the 4-fold $\Xb$ such that $\Im S\to \infty$
imposes a so-called stable degeneration (s.d.) limit in the complex structure of $\Xb$. 
In this limit the 4-fold $X$ degenerates into two components
$$
 X \ {\buildrel\Im S\to\infty\over\longrightarrow} \ X^\sharp = X_1 \cup_Z X_2  \ ,
$$
intersecting over the elliptically fibered heterotic 3-fold $\Zb\to B_2$ \refs{\MV,\FMW,\BJPS,\AMsd}. 
The two 4-fold components $X_i$ are also fibered over the same base $B_2$ and 
capture (part of) the bundle data of the
two $E_8$ factors of the heterotic string, respectively.

The idea is now to view $\Zb$ as a complex boundary within one of the components $X_i$ and 
to apply a theorem of \ThomasAA, which relates the holomorphic Chern-Simons functional on a 
3-fold $\Zb$ to an integral of the Pontryagin class of a connection $\bA$ on an extension 
$\bx E\to X'$ of the bundle $E\to \Zb$ defined over a Fano 4-fold $X'$:
\eqn\intComp{
\int_{X'} \tr\left( F^{0,2}_\bA \wedge F^{0,2}_\bA\right) \wedge s_1^{-1} \ = \
CS(\Zb,A) \ . }
Here $CS(\Zb,A)$ is short for the Chern-Simons functional on the r.h.s. of \CSrel\ without the finite $S$ corrections.
Moreover $s\in H^0(K_{X'}^{-1})$ is a section of the anti-canonical bundle of $X'$ whose zero set defines 
the 3-fold $\Zb$ as a 'boundary' of $X'$.

Now it is straightforward to show, that the components $X_i$ of the degenerate F-theory 4-fold $X^\sharp$ are Fano 
in the sense required by the theorem and moreover that the heterotic Calabi-Yau 3-fold $\Zb$ can be defined as
the zero set of appropriate sections $s_i$ of the anti-canonical
bundles $K^{-1}_{X_i}$, as required by the theorem. This will be discussed in more
detail in sect.~4.2\yyy, where we explicitly discuss hypersurface representations for $X^\sharp$ 
to match the F-theory/heterotic deformation spaces.

The above line of argument then leads to a relation of the form \CSrel, provided one 
identifies the 4-form flux $G_\bA$ with the Pontryagin class of a gauge connection 
$\bA$ on an extension $\bE$ of the bundle over the component $X_1$. 
Up to terms of lower Hodge type, we shall have
\eqn\Gid{
G_\bA|_{X_1} \sim \tr\left( F^{0,2}_\bA \wedge F^{0,2}_\bA\right) \ .
}
Note that this identification of the 4-form flux is a non-trivial prediction
of the outlined duality.

The real challenge posed by the relations \Gid,\CSrel\ is not the on-shell relation, which has been
argued for in a special case in the previous section, but a proper off-shell extension of both sides.
On the 4-fold side, the standard lore of string compactifications is 
to not fix the Hodge type of $G$, but rather to 
view the flux superpotential as a  potential on the moduli space of the 4-fold $X$, which 
fixes the moduli to the critical locus. The idea is, that the periods $\int_X \Om^{4,0}$ on the
l.h.s. of \CSrel\ have a well-defined meaning as the section of a bundle 
over the unobstructed complex structure moduli space 
$\cx M_{CS}(X)$ of the 4-fold {\it before} turning on a 
the flux; in particular they define the K\"ahler metric on $\cx M_{CS}(\Xb)$.
In this way, viewing non-zero $G$ as a 'perturbation' on top of an unobstructed moduli space, 
the section $W(\Xb)$ is considered as an off-shell 
potential for fields parametrizing $\cx M_{CS}(X)$. Although it is not clear
in general under which conditions it is valid to restrict the effective 
field theory to the fields parametrizing $\cx M_{CS}(X)$ and to interprete $W(\Xb)$ as 
the relevant low energy potential for the light fields, this working definition for an 
off-shell deformation space seems to make sense in many situations.\foot{There is a
considerable literature on this subject. We suggest ref.~\refs{\VafaWI} 
for a justification in the context of type IIA flux compactifications on 3-folds,
ref.~\GiddingsYU\ in the type IIB context, ref.~\BeckerKS\ in non-geometric phases,
and ref.~\DouglasZN\ for a recent general discussion.}

The relation \CSrel\ suggests that it should be possible to give a sensible notion of 
a distinguished, finite-dimensional 'off-shell moduli space' for non-holomorphic bundles and to treat 
the obstruction induced by the Chern-Simons superpotential as some sort of 'perturbation' to
an unobstructed problem. This is also suggested by the recent success to compute off-shell superpotentials
for brane compactifications from open string mirror symmetry. We plan to circle around these
questions in the future.

%%%%%%%%%%%%%%%%%%%%

\def\Zb{Z_B}\def\Xb{X_B}
\newsec{Quantum corrected superpotentials in F-theory from mirror symmetry of 4-folds}
In this section we show, that the various Hodge theoretic computations of superpotentials in CY 
3-fold and 4-fold compactifications discussed above are in some cases linked together by a chain 
of dualities. The unifying framework is the type IIA compactification on a pair $(\Xa,\Xb)$ 
of compact mirror CY 4-folds and its F-theory limits. 
As will be argued below, mirror symmetry of the 4-folds computes interesting quantum corrections,
most notably D-instanton corrections to type II orientifolds and world-sheet corrections to 
heterotic (0,2) compactifications, which are hard to compute 
by other means at present. Another interesting 
connection is that to the heterotic superpotential for generalized Calabi--Yau manifold.
The purpose of this section is to study the general framework, which 
involves a somewhat involved chain of dualities, while 
explicit examples are given in sects.~6, 7\yyy.

\subsec{Four-fold superpotentials: a first look at the quantum corrections}
For orientation it is useful to keep in mind the 
concrete structure of the superpotential on compact 4-folds that we want to study, as 
it links the different dual theories discussed below at the level of effective 
supergravity. The compact 4-fold $\Xb$ for F-theory compactification is
obtained from the non-compact 4-fold $\XXb$ of open-closed in eq.\Wnc\ by a 
simple compactification \refs{\AlimRF,\AlimBX,\GrimmEF}, discussed in more detail later on.
In a certain decoupling limit defined in \AlimBX, 
the F-theory superpotential on $\Xb$ reproduces the type II superpotential \spii\ 
plus further terms: 
\eqn\wwii{\eqalign{
W_{F}(\Xb)&\hskip12pt =\ \ \sum_{\ga_\Si \in H_4(\Xb)} {\ux N}_\Si \int_{\ga_\Si}\Om^{(4,0)}\cr
&\buildrel \Im S \to \infty \over =\hskip-10pt\sum_{\ga_\Si \in H_3(\Zb)} (N_\Si+S\ M_\Si) \int_{\ga_\Si}\ux\Om^{(3,0)} +\sum_{\ga_\Si \in H_3(\Zb,\cxH) \atop \p\ga_\Si \neq 0}
\hx N_\Si \int_{\ga_\Si}\ux \Om^{(3,0)}+\ldots .\cr
}
}
The essential novelty in the superpotential of the compact 4-fold, 
as compared to the previous result \spii, is the additional dependence
on the new, distinguished å complex structure modulus $S$ of the compactification $\Xb$ of $\XXb$. This modulus is
identified in \AlimBX\ with the decoupling limit 
\eqn\delim{\Im S \sim 1/g_s \to \infty\ .} 
A similar weak coupling expansion of the 4-fold K\"ahler potential leads to a 
conjectural K\"ahler potential for the open-closed deformation space, as will be discussed in more detail in sect.~5.\yyy

Note that the flux terms $\sim S\, å M_\La$ in the 4-fold superpotential $W_{F}(\Xb)$ 
correspond to NS fluxes in the type II string on $\Zb$, which were missing in 
\spii.\foot{This has been observed already earlier in a related context in ref.~\HaackDI, 
see also the discussion in sect.~5\yyy\ below.} 
In addition there are subleading corrections for finite $S$, 
denoted by the dots in \wwii, which 
include an infinite sum of exponentials with the characteristic weight $e^{-1/g_s}$ of
D-instantons.
Before studying these corrections in detail, it is instructive to consider the dualities
involved in the picture, which leads to a somewhat surprising reinterpretation of the open-closed duality
of \refs{\MayrXK,\AlimRF}.

\subsec{$\cx N=1$ Duality chain}
The relevant duality chain for understanding the quantum corrections in \wwii,
and the relation to open-closed duality, relates the following $\cx N=1$ supersymmetric 
compactifications:\foot{In this note, for ease of notation and to emphasize the relation
to four-dimensional theories, $\cx N=1$ compactifications
to two space-time dimensions also refer to low energy effective theories with four supercharges.}
\vskip-4pt å 
\eqn\netdualiia{\eqalign{
{\hbox{\ninerm type\ II \ OF} \over T^2 \times \Zb} 
\ \sim \
{\hbox{\ninerm F-theory} \over K3\times \Zb} 
\ \sim \ 
{\hbox{\ninerm heterotic} \over T^2\times \Zb} 
\ \sim \
{\hbox{\ninerm type IIA} \over \Xb/\Xa} 
\ \sim \
{\hbox{\ninerm F-theory} \over \Xb\times T^2} å 
}}
\vskip8pt
\noi
where $\Zb$ is a CY 3-fold and $(\Xa,\Xb)$ a mirror pair of 4-folds
which is related to the heterotic compactification on $\Zb$ by 
type IIA/heterotic duality.
Here and in the following it is assumed that the 3-fold $\Zb$ and the 4-fold $\Xb$ have suitable elliptic fibrations,
in addition to the K3 fibration of $\Xb$ required by heterotic/type IIA duality \refs{\VafaGM}. This guarantees
the existence of the F-theory dual in the last step. 
For an appropriate choice of bundle one can take the large volume of the 
$T^2$ factor to obtain the four-dimensional duality between heterotic 
on $\Zb$ and F-theory on $\Xb$ \MV. 

The remaining section will center around the identification of the limit \delim\ in the various
dual theories. 
Note that there are two different F-theory compactifications involved in the duality chain \netdualiia, namely on 
the manifolds $K3\times \Zb$ and $\Xb\times T^2$, respectively. 
The identification \delim\ is associated with the F-theory compactification on $K3\times \Zb$,
or the type II orientifold on $T^2\times \Zb$, in the orientifold limit \SenBP.
The decoupling limit describes also a certain limit of the heterotic compactification {\it on the same 3-fold $\Zb$}, 
which will be identified as a large fiber limit of the elliptic fibration $\Zb$ below.

In order to make contact with the brane configuration $(\Zb,E)$
discussed in sect.~2.1\yyy, we combine the orientifold limit of F-theory
with a particular Fourier-Mukai transformation \refs{\SenPM,\AndreasVE}
$$
{\hbox{\ninerm type\ II \ OF} \over \check T^2 \times \check\Zb} 
\ \sim \
{\hbox{\ninerm type\ II \ OF} \over T^2\times \Zb} 
\ \sim \ 
{\hbox{\ninerm F-theory} \over K3\times \Zb} \ .
$$
The relevant Fourier-Mukai transformation is discussed in detail in
ref.~\AndreasVE. Heuristically,
it implements $T$~duality in both directions of the torus $T^2$ to
the dual torus $\check T^2$ together with a fiberwise $T$~duality in both
directions of the elliptic fibers of the 3-fold $\Zb$ to the 3-fold $\check\Zb$
with dual elliptic fibers. This operation does not change the complex structure
of the bulk geometry, but instead it transforms the brane configuration
to the open-closed geometry $(\Zb,E)$.
These orientifold limits of F-theory, the type II and heterotic compactifications on $\Zb$
can be also connected as:
\eqn\netdualiic{\eqalign{
{\hbox{\ninerm type\ II \ OF} \over \check T^2 \times \check \Zb} 
\ & \sim \
{\hbox{\ninerm type\ I} \over \check T^2\times \Zb} 
\ \sim \ 
{\hbox{\ninerm heterotic} \over T^2\times \Zb} \cr
&\qquad\ \ \ \ \wr \cr
&\quad {\hbox{\ninerm type\ II \ OF} \over T^2 \times\Zb} 
}}
Here $S$ duality associates the type I to the heterotic string,
$T$ duality on $\check T^2$ relates the type I compactification to the type II
orientifold on $T^2\times \Zb$, while the afore mentioned Fourier Mukai
transformation, which realizes fiberwise $T$ duality, applied to the 3-fold $\Zb$
of the type I theory maps to the type II orientifold
on $\check T^2\times\check\Zb$ \refs{\SenBP,\SenPM,\AndreasVE}.

\subsec{The decoupling limit as a stable degeneration}
The meaning of the decoupling limit in the mirror pair $(\Xa,\Xb)$ of 4-folds and the dual heterotic 
string on $\Zb(\times T^2)$ can be understood with the help of the following two 
propositions obtained in the study of F-theory/heterotic duality and mirror symmetry on toric 4-folds 
in ref.~\BMff. It is shown there that \foot{For concreteness, we quote the result for F-theory 
on a 4-fold, although it applies more generally to $n$-folds, as will be also used below.}

\item{$(C1)$} If F-theory on the 4-fold $\Xb$ is dual 
to a heterotic compactification on a 3-fold $\Zb$ then the 
mirror $4$-fold $\Xa$ is a fibration $\Za\to \Xa\to \IP^1$, where
the generic fiber $\Za$ is the 3-fold mirror of $\Zb$. 

\item{$(C2)$} In the above situation, the large base limit in the K\"ahler moduli 
of the fibration $\Xa\to \IP^1$ maps under mirror symmetry to 
a ``stable degeneration'' limit in the complex structure moduli of the mirror $\Xb$.

\noi 
The first part applies, since the 4-fold duals constructed in the context of open-closed string duality 
have precisely the fibration structure required by $(C1)$; indeed the mirror pair $(\XXa,\XXb)$ of 
open-closed dual 4-folds, dual to an $A$-brane geometry $(\Za,L)$ and its mirror $B$-brane geometry 
$(\Zb,E)$, is constructed in refs.~\refs{\MayrXK,\AlimRF}\ as a fibration over the complex plane, 
where the generic fiber is the CY 3-fold $\Za$:
\eqn\epiL{
\xymatrix{
\Za \ar[r] 
&\XXa \ar[d]^{\pi(L)} \ar[rrr]^{4-fold}_{mirror\ symmetry}&&& \XXb \ar[lll] \cr
&\IC
}
}
The notation $\pi(L)$ for the fiber projection is a reminder of the fact that the data of the bundle 
$L$ are encoded in the singularity of the central fiber as described in detail in 
refs.~\refs{\MayrXK,\LM,\AlimRF,\AlimBX}. The manifold 
$\XXb$ may be defined as the 4-fold mirror of the fibration $\XXa$. Since the pair of compact 
4-folds $(\Xa,\Xb)$ is obtained by a simple compactification of the base to a $\IP^1$ 
\refs{\AlimBX,\GrimmEF}, it follows that the
F-theory 4-fold $\Xb$ has a mirror $\Xa$, which is a 3-fold fibration $\pi:\, \Xa\to \IP^1$ with generic fiber $\Za$. The multiple fibration structures are summarized below:
\eqn\dfib{\vbox{\offinterlineskip
\halign{\vrule\strut~#~\hfil\vrule&\hfil~#~\hfil\vrule&\hfil~#~\hfil\vrule\cr
\noalign{\hrule}
\phantom{$\pmatrix{1\cr1}$}& ${\hbox{\ninerm F-theory} \over \Xb}\ \sim \ {\hbox{\ninerm heterotic} \over \Zb}$ &
\quad${\hbox{\ninerm closed} \over \Xa}\ \sim \ {\hbox{\ninerm open} \over (\Za,L)}$\quad \cr
\noalign{\hrule}
\vbox{\halign{#\cr\phantom{$X\over X$}\cr Elliptic Fib.\cr\phantom{$X\over X$}\cr}}& 
\vbox{\halign{\hfil#\hfil&\hfil#\hfil&\hfil#\hfil\cr$T^2$&$\rightarrow$&$\Xb$\cr&&$\downarrow$\cr&&$B_3$\cr}} \quad
\vbox{\halign{\hfil#\hfil&\hfil#\hfil&\hfil#\hfil\cr$T^2$&$\rightarrow$&$\Zb$\cr&&$\downarrow$\cr&&$B_2$\cr}}&
\vbox{\halign{#\cr\phantom{X}--\cr\phantom{X}\cr}} å 
\cr
\noalign{\hrule}
\vbox{\halign{#\cr\phantom{$X\over X$}\cr K3 Fib.\cr\phantom{$X\over X$}\cr}}&
\vbox{\halign{\hfil#\hfil&\hfil#\hfil&\hfil#\hfil\cr$K3$&$\rightarrow$&$\Xb$\cr&&$\downarrow$\cr&&$B_2$\cr}}&
\vbox{\halign{#\cr\phantom{X}--\cr\phantom{X}\cr}}
\cr
\noalign{\hrule}
\vbox{\halign{#\cr\phantom{$X\over X$}\cr 3-fold Fib.\cr\phantom{$X\over X$}\cr}}&
\vbox{\halign{\hfil#\hfil&\hfil#\hfil&\hfil#\hfil\cr$\Za$&$\rightarrow$&$\Xa$\cr&&$\downarrow$\cr&&$\IP^1$\cr}}&
\vbox{\halign{\hfil#\hfil&\hfil#\hfil&\hfil#\hfil\cr$\Za$&$\rightarrow$&$\Xa$\cr&&$\downarrow$\cr&&$\IP^1$\cr}} \cr
\noalign{\hrule}
}}}
Here $B_3$ and $B_2$ denote the corresponding three- and two-dimensional base spaces,
where $B_2$ is common to the F-theory manifold and the heterotic dual.
The crucial link is the 3-fold fibration of $\Xa$, which is required by both, 
F-theory/heterotic {\it and} open-closed duality. $(C1)$ then implies 
that F-theory on $\Xb$ has an open-closed dual interpretation
as a $B$-type brane on a 3-fold $\Zb$ and an $A$-type brane on the mirror $\Za$.
The reverse conclusion, namely that an open-closed dual pair $(\Xa,\Xb)$ also has an
F-theory/heterotic interpretation, requires the additional condition, that 
$\Xb$ is elliptic and K3 fibered. This leaves the possibility, that open-closed
duality holds for more general 4-fold geometries than F-theory/heterotic duality.
For simplicity we impose in the following, that $\Xb$ is elliptically and K3 fibered,
which implies that $(C1)$ holds also in the reverse direction.

Part two of the proposition 
applies, since the decoupling limit $\Im S\to \infty$ in the complex structure
of $\Xb$ was defined in ref.~\AlimBX\ as the mirror of the large base volume in the K\"ahler 
moduli of the fibration $\pi:\, \Xa\to \IP^1$. The image 
of this limit under the mirror map in the complex structure of $\Xb$ is a local mirror limit in the sense of \KMV\
and effectively imposes the stable degeneration (s.d.) limit of $\Xb$ studied 
in refs.~\refs{\MV,\FMW,\AMsd}. Under F-theory/heterotic duality, the s.d. limit maps 
to a large fiber limit of the heterotic string compactification
on the elliptic fibration $\Zb$ and this is the sought for identification of 
limit \delim\ in the heterotic string. The meaning as a physical decoupling limit of 
a sector of the heterotic string can be understood from both, the world-sheet and the effective supergravity
point of view, as will be discussed in sect.~5. Explicit examples for the relation between 
the hypersurface geometries $\Xb$ and $\Zb$ 
in the s.d. limit will be considered in sects.~6,7\yyy.

\subsec{Open-closed duality as a limit of F-theory/heterotic duality}
The relation in \netdualiia\ between the type II orientifold on $\Zb$ and type IIA on the 4-folds $(\Xb,\Xa)$ is
similar as in the open-closed duality of refs.~\refs{\MayrXK,\AlimRF,\AganagicJQ}. These papers
claim to compute the type II superpotential for a $B$-type brane compactification on $\Zb$ with a
given 5-brane charge from the periods of a dual (non-compact) 4-fold $\XXb$. As explained
in refs.~\refs{\AlimBX,\GrimmEF,\AganagicJQ},
this 5-brane charge can be generated by non-trivial fluxes on higher dimensional branes. The only difference 
to the type II orientifold on $T^2\times \Zb$ appearing in \netdualiia\ is the extra $T^2$ compactification and the presence of 
7-branes wrapping $\Zb$, which does not change the superpotential associated with the 5-brane charge.

In the decoupling limit $\Im S \to \infty$, which sends $\Xb$ to the non-compact manifold $\XXb$, 
the ``local'' $B$-type brane with 5-brane charge decouples from the global orientifold
compactification and we recover the type II result $W_{II}(\Zb)$ in eq.~\ecsp.\foot{In the type II string without branes/orientifold, $\hx N_\Si=0$ and the subleading corrections to the 
superpotential would be absent \TVsp.}
Note that in this limit there are two different paths connecting the $B$-type orientifold
to the non-compact open-closed string dual $\XXb$. The first
one goes via the open-closed string duality of refs.~\refs{\MayrXK,\AlimRF,\AganagicJQ}, while the
second goes via F-theory/heterotic/type IIA duality of eq.~\netdualiia.
\eqn\diahetop{
\xymatrix{
\hbox{${\ninerm type\ II \ OF} \atop {T^2\times \Zb}$}
\ar[rr]^{\sevenrm F/het/IIA}_{\sevenrm duality}
\ar[d]^{g_s\to 0}
&&
\hbox{${\ninerm type\ IIA } \atop {\Xb}$}
\ar[d]^{\Im S \to \infty} 
\cr
\hbox{${\ninerm local\ B-brane} \atop {(\Zb,E)}$}
\ar[rr]^{\sevenrm open-closed}_{\sevenrm duality}
&&
\hbox{${\ninerm type\ IIA } \atop {\XXb}$ }
\cr
}
}
Commutativity of the diagram 
implies that for this special case, open-closed duality of refs.~\refs{\MayrXK,\AlimRF,\AganagicJQ} 
coincides with heterotic/F-theory duality in the decoupling limit.

Note that the duality \netdualiic\ maps a D3 brane wrapping a curve $C$ in $\Zb$ 
in the orientifold to a heterotic 5-brane wrapping the same curve $C$ in 
the heterotic dual $\Zb$. The heterotic 5-brane can be locally viewed as an 
M-theory 5-brane \Witsi, which is in turn related to the type IIA 5-brane 
used in \AganagicJQ\ to derive open-closed string duality from T-duality. 

The original observation of open-closed string duality of ref.~\MayrXK\ is that it maps the disc instanton
generated superpotential of the brane geometry $(\Za,L)$ (mirror to $(\Zb,E)$) to the sphere instanton generated 
superpotential for the dual 4-fold $\XXa$ (mirror to $\XXb)$. At tree-level, 
this map is {\it term by term}, that is it maps an individual 
Ooguri--Vafa invariant for a given class $\be\in H^2(\Za,L)$ to a Gromov-Witten 
invariant for a related class $\be'\in H^2(\XXa)$. This genus zero correspondence left the important question, whether there is 
a full string duality, that extends this relation between the 3-fold and the 4-fold data 
beyond the superpotential. From the above diagram
we see, that there is at least one true string duality which reduces to open-closed string
duality of refs.~\refs{\MayrXK,\AlimRF,\AganagicJQ} at $g_s=0$ and extends it to a true string duality: F-theory/heterotic duality!

\subsec{Instanton corrections and mirror symmetry in F-theory}
The above discussion has lead to the qualitative identification of the dual interpretations of the 
expansion in \wwii\ in terms of a weak coupling limit of the type II orientifold, 
a large fiber volume of the heterotic string on the elliptic fibration
$\Zb$, a stable degeneration limit of the F-theory 4-fold $\Xb$ and a large base limit of the 
3-fold fibration $\Xa\to\IP^1$. We will now argue that the quantum corrections 
computed by 4-fold mirror symmetry can be tentatively assigned to the two 4-fold superpotentials 
in refs.~\refs{\GVW,\GukovIQ} as 
\eqn\ffsps{\vbox{\offinterlineskip\halign{
\strut # 
&~#~\hfil
&\hfil~#~\hfil
&\hfil~#~\hfil
\cr
$W(\Xb)=\int_{\Xb} \Omega \wedge F_{hor}$ &$\leftrightarrow$ &
\hbox{\ninerm D-1,D1/finite-fiber corrections in type II~OF/Het}\cr
\cr\cr
$\widetilde W(\Xb) \,=\, \int_{\Xb} e^{B+i J} \wedge F_{ver}$&$\leftrightarrow$&
\hbox{\ninerm D3/space-time instantons in type II~OF/Het}\cr
}}
}
Here $W(\Xb)$ is the 4-fold superpotential of eq.~\wwii, while 
$\widetilde W(\Xb)$ is the twisted superpotential associated with the type IIA compactification on $\Xb$.\foot{See 
the discussion in sect.~5 below.} The latter 
computes also world-sheet instanton corrections to the large volume limit of the type II/heterotic compactification.

The details of the argument are somewhat involved and may be skipped on a first reading.
It is again instructive to first 
consider the simpler case of a closely related duality chain with $\cx N=2$ supersymmetry:
\eqn\netduali{\eqalign{
&{\hbox{\ninerm type\ II \ OF} \over T^2\times \Kh} 
\ \sim \
{\hbox{\ninerm F-theory} \over \Kv\times \Kh} 
\ \sim \ 
{\hbox{\ninerm heterotic} \over T^2\times \Kh} 
\ \sim \
{\hbox{\ninerm type IIA/IIB} \over \Xb/\Xa} 
\ \sim \
{\hbox{\ninerm F-theory} \over \Xb\times T^2} 
}}
where $\Kv$, $\Kh$ are two K3 manifolds and $(\Xa,\Xb)$ denotes a mirror pair of CY 3-folds; differently 
then in \netdualiia, mirror symmetry of the 3-folds exchanges the IIA compactification on $\Xb$
with a type IIB compactification on $\Xa$.
As before, we assume that the 3-fold $\Xb$ 
is elliptically fibered, such that one
can decompactify the $T^2$ of the heterotic string to obtain F-theory in six dimensions. 
Note that the $\cx N=1$ duality chain \netdualiia\ can be heuristically thought of as a 
chain of dualities obtained by ``fibering'' \netduali\ 
over $\IP^1$, so that some observations from the $\cx N=2$ supersymmetric
case will carry over to $\cx N=1$.

The two basic questions that we want to study in this simpler setup are 
the meaning of mirror symmetry in F-theory and the identification of quantum corrections
computed by it. It will turn out that, under favorable conditions, the distinguished 
modulus $S$ has a mirror partner $\rho$ and mirror symmetry of the CY manifolds
$\Xa$ and $\Xb$ exchanges the two weak coupling expansions in $\Im S$ and $\Im \rho$.

The quantum corrections to the $\cx N=2$ supersymmetric duality chain \netduali\
have a rich structure studied previously in \refs{\BMof,\HalmagyiWI}. 
The F-theory superpotential
for the $K3\times K3$ compactification, which arises in the effective
$\cx N=2$ supergravity theory from certain gaugings in the hypermultiplet
sector, can be written as a bilinear
in the period integrals on the two K3 factors \refs{\ferrara,\Luest}
\eqn\wf{
W_{F,pert}=\sum_{I,\La}\ \big( \int_{\Kh} \om^{2,0}\wedge \mu^I\big) \, \g_{I\La}\, \big( \int_{\Kv} \om^{2,0}\wedge \tx \mu^\La\big) \ .
}
Here $\g_{I\La}$ labels the 4-form flux in F-theory, decomposed on a basis $\{\tx \mu^\La\}$ 
for $H^2_{prim}(\Kv)$ and $\{\mu^I\}$ for $H^2_{prim}(\Kh)$
as 
$G=\sum_{I,\La}\g_{I\La}\mu^I\wedge \tx \mu^\La$. 

The periods on $\Kh$ depend on $\cx N=2$ hyper multiplets and are mapped under duality to
the type IIA/F-theory compactification on $\Xb$ to the 3-fold periods, by a similar relation as \wwii:
\eqn\tfcor{
\eqalign{
\int_{\Xb} \om^{3,0}\wedge \ga^I&=\int_{\Kh} \om^{2,0}\wedge \mu^I + \cx O(e^{2\pi i S},S^{-1})\, .\cr
%\int_{\Xa} \om^{3,0}\wedge \tx \ga^\La&=\int_{\Kv} \om^{2,0}\wedge \tx \mu^\La + \cx O(e^{2\pi i \rho},\rho^{-1})\ .\cr
}}
This equation describes,
how the periods on the F-theory 3-fold $\Xb$ defined on the basis $\ga^I\in H^3(\Xb,\IZ)$
compute finite $S$ corrections to the periods on the 2-fold $\Kh$ of the dual 
type II compactification. As explained in the 4-fold å case, $(C2)$ says that 
these are corrections to the s.d. limit in the complex structure of $\Xb$. 

Note that $\wf$ is apparently symmetric in the periods of the two K3 factors. This is somewhat
misleading, as the periods on $\Kv$ depend on $\cx N=2$ vector multiplets.\foot{See refs.~\refs{\ferrara} 
for a discussion of the 
effective supergravity theory for the orientifold limit of $K3\times K3$.} 
It was argued in \BMof, that there is also a similar relation as \tfcor\ for the second period vector on $\Kv$ \tfcor,
\eqn\tfcorii{
\eqalign{
\int_{\Xa} \om^{3,0}\wedge \tx \ga^\La&=\int_{\Kv} \om^{2,0}\wedge \tx \mu^\La + \cx O(e^{2\pi i \rho},\rho^{-1})\ ,\cr
}}
where $\rho$ is a distinguished vector multiplet related to the heterotic
string coupling as discussed below.
This relation describes corrections to the result \wf\ computed by the periods {\it of the mirror manifold $\Xa$}.
Here it is understood, that one uses mirror symmetry to map the periods of the holomorphic (3,0) form 
on $H^3(\Xa,\IZ)$ defined on the basis $\tx \ga^\La\in H^3(\Xa,\IZ)$ to the periods of the K\"ahler form on a dual basis 
$\ga^\La\in \oplus_k H^{2k}(\Xb,\IZ)$,
\eqn\msrel{
\int_{\Xa} \om^{3,0}\wedge \tx \ga^\La \ \longrightarrow\ å \int_{\Xb} \fc{1}{k!}J^k \wedge \tx \ga^\La\ .
}
Note that these 'K\"ahler periods' of $\Xb$ are the 3-fold equivalent of the integrals appearing in 
the twisted superpotential $\widetilde W(\Xb)$ in \ffsps. However, replacing the K3 periods in $\wf$ by 
the quantum corrected expressions \tfcor,\tfcorii, we get a superpotential that is proportional to both, the
periods of the manifold $\Xb$ {\it and} of its mirror $\Xa$. It was argued in \BMof, that this 'quadratic' superpotential
in the 3-fold periods is in agreement with the $S$-duality of topological strings predicted in ref.~\NekrasovJS. 
Similar expressions have been obtained in refs.~\refs{\DAuriaTR,\GranaHR} from the study of å type II compactification on generalized CY manifolds.

The similarity of the two expansions \tfcor,\tfcorii\ is no accident. By $(C2)$, the s.d. limit $\Im S \to \infty$ is 
mirror to the large base limit of the fibration $\Xa\to\IP^1$, which is a K3 fibration by $(C1)$ in the 3-fold case.
By type IIA/heterotic duality, $\Xb$ is also a K3 fibration $\Xb\to\IP^1$ and eq.~\tfcorii\ represents the large base limit $\Im \rho \to \infty$ of $\Xb$, where $\rho$ is the K\"ahler volume of the base $\IP^1$. 
By heterotic/type IIA duality, the K\"ahler volume of the base of $\Xb$ is identified with the four-dimensional 
heterotic string coupling \refs{\AspinwallVK}. Adding the identification of $S$ provided by $(C2)$, 
we get the following heterotic interpretation of the volumes $V_{A/B}$ of the base $\IP^1$'s of 
the fibrations $X_{A/B}\to \IP^1$:
\eqn\hetlam{V_B=\la_{4,het}^{-2}=\Im \rho\ ,\qquad V_{A}=V_{\Ehet}=\Im S\ .}
Here $V_{\Ehet}$ denotes the volume of the elliptic fiber of $\Kh$ in the heterotic compactification in \netduali.
Clearly, mirror symmetry exchanges the two expansions \tfcor\ and \tfcorii\ associated with a compactification on 
$\Xa$ or on $\Xb$, respectively
\eqn\exchange{
{S\atop \tfcor} \quad \matrix{\hbox{\sevenrm mirror}\cr \longleftrightarrow\cr \hbox{\sevenrm symmetry}}\quad å {\rho \atop \tfcorii} .}

In the dual F-theory compactification on $K3\times K3$, mirror symmetry represents the exchange of the
two K3 factors \refs{\AspinwallQW,\HalmagyiWI}, which gives rise to two
dual heterotic $T^2\times K3$ compactifications. å å 
Starting from the duality relation between M-theory on $K3\times K3$
and heterotic string on $T^2\times S^1\times K3$ \refs{\WittenEX},
it is shown in ref.~\HalmagyiWI, that the 
exchange of the two K3 factors in M-theory generates the following $\IZ_2$ transformation on the moduli 
of the two heterotic duals:
$$
V_{\Ehet'}=\la_4^{-2},\qquad \la_4^{'-2}=V_{\Ehet}\ .
$$
Comparing with the relation \hetlam\ between the four-dimensional 
heterotic coupling and the volumes of the bases of the fibrations $(\Xa,\Xb)$,
one concludes that the result of \HalmagyiWI\ is in accord with the claim $(C2)$ of \BMff\ and its consequence 
\exchange\ in this case.
It is reassuring to observe that these conclusions, reached by rather different arguments in 
refs.~\refs{\BMff,\BMof}\ and \HalmagyiWI, agree so nicely. 

As further argued in \BMof, the expansion \tfcorii\ computed from mirror symmetry of the 
3-folds $\Xb$ and $\Xa$ computes D3 instanton corrections to the orientifold on $K3\times T^2$ (or F-theory on 
$K3\times K3$). The basic instanton is a D3 brane wrapping $K3$, which is mapped under the duality 
\netdualiic\ to a 5-brane instanton of the heterotic brane wrapping $T^2\times K3$.
In the type II orientifold, $\rho$ is the K3 volume.

Compactifying the $\cx N=2$ chain on a further $\IP^1$, the previous arguments leads to 
the assignements \ffsps. In particular the identification of D3 instantons in \BMof\ continues to hold
with the appropriate replacement of K3 with 4-cycles in $\Zb$. 
The above argument based on $(C2)$ is in fact independent of the dimension 
and can be phrased more generally as the following statement on 
mirror symmetry in F-theory. 
Let $\Xb$ be an F-theory $n$-fold with heterotic dual $(\Zb,V_B)$,
where $V_B$ denotes the gauge bundle. å 
If the mirror $\Xa$ of $\Xb$ is also elliptically and K3 fibered, we have the following relations
between the F-theory compactifications on $(\Xa,\Xb)$ and heterotic compactifications on $(\Za,\Zb)$:
\eqn\hethet{
\xymatrix{
\hbox{${\ninerm F-theory} \atop {\Za\to \Xb\to \IP^1}$}
\ar[rr]^{\sevenrm mirror}_{\sevenrm symmetry}
\ar[d]
&&
\ar[ll]\hbox{${\ninerm F-theory} \atop {\Zb\to \Xa\to \IP^1}$}
\ar[d]
\cr
\hbox{${\ninerm heterotic} \atop {(\Zb,V_B)}$}
\ar[rr]^{\sevenrm het/het}_{\sevenrm map}\ar[rru]^{\fiverm (C1)}
&&
\ar[ll]\ar[llu]\hbox{${\ninerm heterotic} \atop å {(\Za,V_A)}$}
\cr
}
}
Under mirror symmetry, the s.d. limit and the 
large base limit are exchanged:
\eqn\hethetii{
\xymatrix{
{\Za\to \Xa\to \IP^1}
&&
{\Zb\to \Xb\to \IP^1}
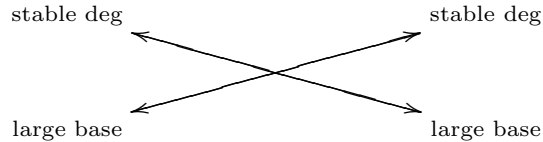
\cr 
\hbox{\sevenrm stable\ deg}
\ar[rrd]
&&
\hbox{\sevenrm stable\ deg}
\ar[lld]
\cr
\hbox{{\sevenrm large base}}
\ar[rru]
&&
\hbox{{\sevenrm large base}}
\ar[llu]
\cr
}
}
Note that the two theories on the left and on the right are in general {\it not} dual 
but become dual after further circle compactifications. 

The simplest example is
F-theory on a K3 $\Xb$ dual to heterotic on $(\Zb=T^2,V_G)$, where $V_G$ denotes a flat gauge bundle
on $T^2$ with structure group $G$. The eight-dimensional heterotic compactification has an unbroken gauge group
$H$, where $H$ is the centralizer of $G$ in the ten-dimensional heterotic gauge group. In a further compactification on $T^2$ 
one has to choose a flat $H$ bundle on the second $T^2$. Assuming that the bundles factorize, 
one can exchange the two $T^2$ factors and thus $H$ and $G$. In F-theory this exchange corresponds to mirror symmetry
of K3 and this was used in \refs{\KMV,\BMff} to construct local mirrors of bundles on $T^2$ from local ADE singularities.

The next simple example is the above $\cx N=2$ supersymmetric case, where $\Xb$ is the 3-fold 
in \netduali, with a heterotic dual compactified on $K3\times T^2$.
Assuming a suitable factorization of the heterotic bundle, the 
action of 3-fold mirror symmetry maps to the exchange of the two K3 factors $(\Kv,\Kh)$ in the dual F-theory 
compactification in \hethet. In the heterotic string this symmetry
relates two {\it different} K3 compactifications $(\Kh,V)$ and $(\Kv,V')$
which become dual after compactification on $T^2\times S^1$ \refs{\PerevalovHT,\BMff}.\foot{One needs
the $T^2$ compactification to get two type IIA compactifications on the mirror pair $(\Xa,\Xb)$, 
which become T-dual after a further circle compactification.}

In the 4-fold case, the fibrations required by the above arguments are not granted, since $(C1)$ now implies 
that the 4-fold $\Xa$ is a 3-fold fibration $\Xa\to\IP^1$ (as opposed to the K3 fibration in the 3-fold case).
{\it If} $\Xa$ is K3 fibered, the $\cx N=1$ chain can be viewed as a $\cx N=2$ chain fibered over $\IP^1$
and the above arguments apply, leading to the assignment \ffsps. 
In the other case, the large $\Im S$
expansion of $W(\Xb)$ always exists, but there is no corresponding large $\rho$ expansion of the 
twisted superpotential $\widetilde W(\Xb)$.

\newsec{Heterotic superpotential from F-theory/heterotic duality}
Having identified the limit $S\to i\infty$ as a large fiber limit in the heterotic
interpretation, the next elementary question is to identify the ``flux quanta''
of the 4-fold superpotential \wwii\ in the context of the heterotic string. 
This task
can be divided into identifying the origin of the 
terms $\sim N_\Si,\ M_\Si$ captured by the bulk periods 
and the terms $\sim \hx N_\Si$ proportional to chain integrals.

\subsec{Generalized Calabi--Yau contribution to $W_F(\Xb)$}
The back-reaction of the bulk background fluxes in the heterotic string requires
the compactification space to be a generalized Calabi-Yau
space \refs{\StromingerUH,\HullKZ,\BarsQQ,\BBDG,\LopesCardosoSP,\GurrieriDTJG,\BenmachicheMA}.
Using dimensional reduction techniques of the
heterotic string on such generalized Calabi-Yau geometries $\ZG$ reveals that the
flux-induced superpotential
reads \refs{\BeckerGQ,\LopesCardosoAF,\GurrieriDTJG,\BenmachicheMA,\AndriotFP}
\eqn\hetflux{ W_{het}\,=\,\int_\ZG\tilde\Om \wedge \left(H -i\,d\tilde J \right) \ , }
where $H$ is the non-trivial NS 3-form flux and $d\tilde J$ is often called the
geometric flux of the generalized 3-fold $\ZG$. The 3-forms $\tilde\Om$
and the 2-form $\tilde J$ are the generalized counterparts of the
holomorphic 3-form $\Om$ and the (complexified) K\"ahler form $J$ of the associated
Calabi-Yau 3-fold $\Zb$.\foot{In the context of generalized Calabi-Yau spaces
$\tilde J$ and $\tilde\Om$ are in general not closed with respect to the de~Rahm
differential $d$.} In general the direct evaluation of the heterotic
superpotential~\hetflux\ of the 3-fold $\ZG$ is rather complicated, therefore we
argue here that under certain circumstances the heterotic fluxes can be
computed from the periods of the original 3-fold $\Zb$.

It is instructive to examine first the fluxes of the heterotic string
compactified on the $\cx N=2$ background $T^2\times K3$. For this particular
geometry the analyzed fluxes induce a deformation to the non-K\"ahler geometry $\tilde K$,
which is a non-trivial toroidal bundle $\pi: T^2\to \tilde K \to K3$ over the
$K3$ base \refs{\DasguptaSS,\FuSM,\BeckerET}.

In order to show the relation to the superpotential~\hetflux\ we first construct
the cohomology classes, which capture the twisting to the toroidal bundle $\tilde K$.
Choosing a good open covering $\cx U=\{ U_\alpha \}$ of the $K3$ base together with a trivialization of
the toroidal bundle, the non-trivial bundle structure is captured by 
transition funcions $\varphi^{(k)}_{\alpha\beta} : U_{\alpha\beta} \rightarrow \IR, k=1,2$,
in the open sets $U_{\alpha\beta}\,=\, U_\alpha\cap U_\beta$. These transition functions
patch together the angular coordinates of the two circles $S^1\times S^1$ in
the torodial fibers. Due to the periodicity of the angular variables the transition
functions fulfill on triple overlaps $U_{\alpha\beta\gamma}=U_\alpha\cap U_\beta\cap U_\gamma$
the condition
$$
\varepsilon^{(k)}_{\alpha\beta\gamma} \,=\,
{1\over 2\pi}\left( \varphi^{(k)}_{\alpha\beta} - \varphi^{(k)}_{\alpha\gamma} + \varphi^{(k)}_{\beta\gamma} \right) \in \IZ \ , \qquad k=1,2 \ .
$$
The constructed functions $\varepsilon^{(k)}: U_{\alpha\beta\ga}\to \IZ$ specify
2-cocycles in the \v{C}ech cohomology group $\check H^2(K3,\IZ)$. The classes
$\varepsilon^{(k)}$ correspond to the Euler classes $e^{(k)}$ of the two circular
bundles in the integral de~Rham cohomology $H^2(K3,\IZ)$.\foot{For details and
background material on \v{C}ech cohomology and on the construction of the Euler
classes we refer the interested reader, for instance, to ref.~\BottTu.} 

The non-K\"ahler manifold $\tilde K$ is equipped with the hermitian form $\tilde J$ and the
holomorphic 3-form $\tilde\Om$ \refs{\FuSM,\BeckerET}\foot{For simplicity, we ignore
a warp factor in front of the K\"ahler form $J_{K3}$, as it is not relevant for
the analysis of the superpotential. Also note
that in our conventions the imaginary part of $\tilde J$ corresponds to the hermitian
volume form.}
$$
\tilde J = \pi^*J_{K3} -S\,i\,\theta^{(1)}\wedge\theta^{(2)} \ , \qquad
\tilde \Om = \om^{2,0} \wedge (\theta^{(1)} + i \theta^{(2)}) \ .
$$
Here $\theta^{(k)}, k=1,2,$ are the two 1-forms of the toroidal fibers, while
$J_{K3}$ is the (complexified) K\"ahler form
and $\om^{2,0}$ is the holomorphic 2-form of the $K3$ base.
$S$ is the (complexified) Volume modulus of the toroidal fiber. On-shell the value
of the volume modulus $S$ becomes stabilized at $S=i$ \BeckerET, since
the equations of motions impose the torsional
constraint \refs{\StromingerUH,\HullKZ,\BarsQQ}\foot{The stabilization of volume
moduli in the context of heterotic
string compactifications with fluxes is also discussed in refs.~\refs{\BBDG,\LopesCardosoAF}.}
\eqn\Htor{ H = (\p-\bar\p) \tilde J \ . }

As the two-forms $d\theta^{(k)}$ restrict to the Euler
classes $e^{(k)}$ on the $K3$ base, the non-K\"ahler 3-fold $\tilde K$
encodes the background fluxes
$$
d\tilde J\,=\,-i\,S( \pi^*e^{(1)} \wedge \theta^{(2)} - \pi^*e^{(2)}\wedge \theta^{(1)} ) \ , \qquad
H \,=\, \pi^*e^{(1)} \wedge \theta^{(1)} + \pi^*e^{(2)}\wedge \theta^{(2)} \ , å 
$$ å 
where the $H$-flux is determined by imposing the torsional constraint~\Htor\ for the on-shell
value $S=i$ of the fiber volume. Then evaluating the superpotential \hetflux\ with
these fluxes yields
\eqn\WtK{
W_{het} \,=\, \int_{\tilde K}\tilde\Om \wedge (H -i\,d\tilde J)\,=\,
\int_{C_H} dz\wedge\om^{2,0} -i S \int_{C_J} dz\wedge \om^{2,0} \ . }
In the last equality the toroidal fibers of the twisted
manifold $\tilde K$ are integrated out, and in a second step the resulting
period integrals of the $K3$ base are transformed into periods of the
holomorphic 3-form $dz\wedge\om^{2,0}$ on the original 3-fold $T^2\times K3$
with respect to the 3-cycles $C_H$ and $C_J$, which are Poincar\'e dual to
the integral 3-forms $e^{(1)} \wedge dy - e^{(2)}\wedge dx$
and $e^{(1)} \wedge dx + e^{(2)}\wedge dy$.

Note that the structure of the derived superpotential is
in agreement with the superpotential periods obtained in ref.~\BMof.

The idea is now to generalize the construction by ``twisting'' the fibers of the
elliptically fibered 3-fold $\pi: \Zb \to B$ with a section $\sigma: B \to \Zb$, such
that we arrive at the generalized Calabi-Yau 3-fold $\ZG$. In order to eventually
relate the periods of the two manifolds $\Zb$ and $\ZG$,
we first translate the 3-form cohomology of the 3-fold $\Zb$ to 
appropriate cohomology groups on the common base $B$.
This is achieved with the Leray-Serre spectral sequence, which associates the
cohomology of a fiber bundle to cohomology groups on the base.

Let ${\cal U} = \{ U_\alpha \}$ be a good open covering of the base $B$. Then the cohomology
group $H^k(\Zb,\IZ)$ is iteratively approximated by the Leray-Serre spectral sequence. The
leading order $E_2$ terms of the spectral sequence read \BottTu
$$
E_2^{p,q} \,=\, \check H^p(B,{\cal H}^q ) \,\simeq\,H^p(B,{\cal H}^q) \ .
$$
Here the (pre-)sheaf ${\cal H}^q$ of the base $B$ is defined by
assigning to each open set $U$ the group ${\cal H}^q(U)\,=\,H^q(\pi^{-1}U,\IZ)$,
and the inclusion of open sets $\iota^V_U : V \hookrightarrow U$ induces the
homomorphism ${\iota^V_U}^* : {\cal H}^q(U) \rightarrow {\cal H}^q(V)$ via pullback of forms.
Then the spectral sequence abuts to $H^3(\Zb,\IZ)$, and we get\foot{Strictly speaking
the first relation is not an equality `$\simeq$' but an inclusion `$\subseteq$', because we ignore the
``higher order corrections'' from the spectral sequence. This implies that some of the
elements on the right hand side might actually be trivial in $H^3(\Zb,\IZ)$.}
$$
H^3(\Zb,\IZ) \,\simeq\, \bigoplus_{n=0}^3E_2^{n,3-n} \,=\, \bigoplus_{n=0}^3 H^n(B, {\cal H}^{3-n}) \ .
$$ å å 
Due to the simple connectedness of the examined Calabi-Yau 3-fold $\Zb$ we arrive at the simplified relation
\eqn\BCoh{ å H^3(\Zb,\IZ) \,\simeq\, E_2^{2,1} \,=\, \check H^2(B, {\cal H}^1)\simeq H^2(B,{\cal H}^1) \ . }
Note that the (pre)sheaf ${\cal H}^1$ is not locally constant, because the dimension of the sheaf ${\cal H}^1$
differs at a singular fiber from the dimension of the sheaf ${\cal H}^1$ at a generic regular fiber. 

In terms of the open covering $\cx U$ a \v{C}ech cohomology element $\varepsilon$
in $\check H^2(B, {\cal H}^1)$ is a map that assigns to each
triple intersection set $U_{\alpha\beta\gamma}$ an element
in ${\cal H}^1(U_{\alpha\beta\gamma})$ and fulfills the cocycle
condition on quartic intersections $U_{\alpha\beta\gamma\delta}$
$$
0\,=\, (\rho_\delta \circ \varepsilon)(U_{\alpha\beta\gamma}) - (\rho_\gamma \circ \varepsilon)(U_{\alpha\beta\delta})+
(\rho_\gamma \circ \varepsilon)(U_{\alpha\beta\delta}) - (\rho_\alpha \circ \varepsilon)(U_{\beta\gamma\delta}) \ .
$$
The map $\rho_\delta$, for instance, is the pull-back induced from the
inclusion $\iota_\delta : U_{\alpha\beta\gamma} \hookrightarrow U_{\alpha\beta\gamma\delta}$. 
Then the cohomology element $\varepsilon$ is called a 2-cocycle
with coefficients in the (pre-)sheaf ${\cal H}^1$, and it is non-trivial
if it does not arise form a 1-cochain on double intersections $U_{\alpha\beta}$. 

To proceed we assume that the generalized Calabi-Yau manifold $\ZG$ is
also fibered $\tilde\pi: \ZG \rightarrow B$ over the same base $B$ and
that it arises from ``twisting'' the elliptic fibers of the 3-fold $\Zb$. 
This ``twist'' is measured by the 1-cochain $\varphi$, which assigns to each double
intersection $U_{\alpha\beta}$ an element in ${\cal H}^1(U_{\alpha\beta})\otimes_\IZ \IR$
and which captures the distortion of the angular variable of the 1-cycles in
the elliptic fibers of the original 3-fold $\Zb$.

In general the 1-chain $\varphi$ does not fulfill the cocycle condition due
to the periodicity of the angular variables of the 1-cycles. Instead we find
on triple intersections $U_{\alpha\beta\gamma}$
$$
\varepsilon : U_{\alpha\beta\gamma} \mapsto {1\over 2\pi}\left[(\rho_\gamma \circ \varphi)( U_{\alpha\beta} )-(\rho_\beta \circ \varphi)( U_{\alpha\gamma} )+
(\rho_\alpha \circ \varphi)( U_{\beta\gamma} ) \right] \in {\cal H}^1(U_{\alpha\beta\gamma}) ,
$$
which defines a 2-cocycle in $\check H^2(B,{\cal H}^1)$ characterizing the ``twist'' of the 3-fold $\ZG$. 

Analogously the element $e$ in $H^3(\Zb,\IZ)$, which corresponds to the \v{C}ech cohomology element
$\varepsilon$ in $\check H^2(B,\cx H^1)$, is explicitly constructed. Namely,
there are 1-forms $\xi_\alpha$ defined on the open sets $U_\alpha$, which are exact
on double overlaps $U_{\alpha\beta}$ \BottTu
\eqn\Eul{ {1\over 2\pi} d \varphi( U_{\alpha\beta} ) \,=\, \rho_\beta( \xi_\alpha ) - \rho_\alpha( \xi_\beta) \ . }
Therefore the 2-forms $d \xi_\alpha$ patch together to a global 2-form $s_e$ in $H^2(B,\cx H^1)$, which
in turn can be identified with the 3-form $e$ in $H^3(\Zb,\IZ)$ according to $\BCoh$.

In order to extract the geometric flux from the 3-fold $\ZG$, we need to get a handle on the
2-form $\tilde J$ in the superpotential~\hetflux. Due to the fibered structure of the 3-fold $\Zb$ 
the K\"ahler form $J$ splits into two pieces
$$
J = \pi^*J_B + J_F \ ,
$$
where $J_B= \sigma^* J$, $J_F = J - \pi^*J_B$, and $J_F\,=\, S\,\omega_F$ in terms of the
integral generator $\om_F$ and the (complexified) K\"ahler volume of the generic elliptic fiber.
Then upon the ``twist'' to the 3-fold $\ZG$ the K\"ahler form $J$ is transformed into the 2-form
$$
\tilde J \,=\, \tilde\pi^*J_B + \tilde J_F \,=\,\tilde\pi^*J_B + S\, \tilde\omega_F \ .
$$
The 2-form $\tilde\omega_F$ is defined on each open-set $\tilde\pi^{-1}U_\alpha$ by
$$
\tilde\omega_F|_{\tilde\pi^{-1}U_{\alpha} }= \omega_F|_{\tilde\pi^{-1}U_\alpha} + \xi_\alpha \ ,
$$
where we now view $\xi_\alpha$ as a two form in the open set $\tilde\pi^{-1}U_\alpha$. Due to the ``twist'' 
the 2-forms $\tilde\omega_F$, which are defined on open sets, patch together to a global 2-form on the 3-fold $\ZG$.
Furthermore, as a consequence of eq.~\Eul\ we observe that
\eqn\dJ{ å d \tilde J = S\,d\tilde\omega_F = S \, s_e \ , }
in terms of the element $s_e$ in $H^2(B,{\cal H}^1)$. 

In order to evaluate the heterotic superpotential~\hetflux\ we express
the 3-forms of $\tilde\Omega$, $H$ and $d\tilde J$ of the ``twisted'' 3-fold $\ZG$
as elements $s_\Om$, $s_H$ and $s_e$ of the sheaf cohomology $H^2(B,{\cal H}^1\otimes\IC)$.
Using eq.~\BCoh\ we induce $s_\Om$ from the holomorphic 3-form $\Om$ in $H^{3,0}(\Zb)$ and
the NS flux $s_H$ from an integral 3-form in $H^3(\Zb,\IZ)$. 
Furthermore, we also inherit the the pairing $\langle \cdot , \cdot \rangle$ on $H^2(B,{\cal H}^1\otimes\IC)$
from the 3-form pairing $\int_{\Zb}\cdot\, \wedge\, \cdot\,$ on the Calabi-Yau 3-fold $\Zb$.
Then the superpotential~\hetflux\ for the ``twisted'' manifold $\ZG$ becomes
\eqn\WGZ{
W_{het}\,=\,\langle s_\Omega, s_H \rangle - i\,S \, \langle s_\Omega, s_e \rangle
\,=\, \int_{C_H} \Omega - i S \int_{C_J} \Omega \ . }
In the last step we have again related the integral elements $s_H$ and $s_e$ to their
Poincar\'e dual 3-cycles $C_H$ and $C_J$ in the original Calabi-Yau manifold $\Zb$.

In the context of heterotic string 
compactifications on the 3-fold $\Zb$ the presented arguments provide
further evidence for the encountered structure of the closed-string periods
in eq.~\wwii. In particular we find that the complex modulus $S$ should be
identified with the complexified volume of the generic elliptic fiber.

There is, however, a cautious remark overdue. We tacitly assumed that
the manifold $\ZG$ can be constructed by simply ``twisting'' the elliptic fibers of $\Zb$. In
general, however, we expect that such a construction is obstructed and additional modifications are
necessary to arrive at a ``true'' generalized Calabi-Yau manifold. A detailed analysis of such
obstructions is beyond the scope of this note.
However, we believe that the outlined construction is
still suitable to anticipate the (geometric) flux quanta, which are responsible for the transition to the
generalized Calabi-Yau manifold $\ZG$ to leading order. From the duality perspective of the previous section
we actually expect further corrections to the superpotential~\WGZ. These corrections should be 
suppressed in the large fiber limit $\Im S\to \infty$. It is in this limit, in which we
expect the ``twisting'' construction to become accurate.

\def\bA{{\bf A}}\def\bE{{\bf E}}\def\Xsd{{X^\sharp}}\def\Xbsd{{\Xb^\sharp}}\def\st{\tx s}
\def\Xabms{\hx{X}}

\subsec{Chern-Simons contribution to $W_F(\Xb)$}
The F-theory prediction from the last term in \wwii\ is the equality, up to finite $S$ corrections, of
certain 4-fold period integrals on $\Xb$ and the Chern-Simons superpotential on $\Zb$, 
for appropriate choice of $G\in H^4(\Xb)$ and a connection on $E\to \Zb$. The general relation
of this type has already been described in sect.~2.4 where we used that the 3-fold $\Zb$ may be 
viewed as a 'boundary' within å the F-theory 4-fold $\Xb$ in the s.d. limit.
Here we complete the argument and discuss the map of the deformation spaces
by using hypersurface representations for $\Xsd$ and $\Zb$. This will also lead to a direct identification of the 
open-closed dual 4-fold geometries for type II branes and the local mirror geometries for (heterotic) bundles
of \refs{\KMV, \BMff}.

To this end, we represent the s.d. limit $\Xbsd$ of the F-theory 4-fold $\Xb$ 
as a reducible fiber of a CY 5-fold $W$ obtained by fibering $\Xb$ over
$\IC$ as in \refs{\AMsd,\BMff}. Let $\mu$ be the local coordinate on the base $\IC$ which 
serves as a deformation space for the 4-fold fiber $\Xb$.
We start from the Weierstrass form 
$$
p_W=y^2+x^3+x\sum_{\al,\be} s^{4-\al}\st^{4+\al}\mu^{4-\be}a_{\al,\be}f_{\al}(x_k) 
+\sum_{\al,\be} s^{6-\al}\st^{6+\al}\mu^{6-\be}b_{\al,\be}g_{\al}(x_k) ,
$$
where $f_\al(x_k)$ and $g_\al(x_k)$ are functions of the coordinates on the two-dimensional base $B_2$
of the K3 fibration of the 4-fold $\Xb$. Moreover $(y,x)$ and $(s,\st)$ can be thought of as 
(homogeneous) coordinates on the elliptic fiber and the base $\IP^1$ of the K3 fiber, respectively.
Finally 
$a_{\al,\be},\ b_{\al,\be}$ are some complex constants entering the complex structure of $W$.
The fiber of $W\to\IC$ over a point $p\in\IC$ å represents a smooth F-theory 4-fold $\Xb$
with a complex structure determined by the values of the constants $a_{\al,\be},\ b_{\al,\be}$ and of the coordinate
$\mu$ at $p$.

Tuning the complex structure of $W$ by choosing $a_{\al,\be}=0$ for $\al+\be>4$
and $b_{\al,\be}=0$ for $\al+\be>6$, the central fiber of $W$ at $\mu=0$ acquires
a non-minimal singularity at $y=x=s=0$, which can be blown up by 
$$
y=\rho^3y,\ x=\rho^2x,\ s=\rho s,\ \mu = \rho \mu,
$$
to obtain the hypersurface\foot{The non-zero constants å $a_{\al,\be},\ b_{\al,\be}$ are set to one in the following.}
\eqn\fivefold{
p_{W^\sharp}=y^2+x^3+x\sum_{\al,\be} s^{4-\al}\st^{4+\al}\mu^{4-\be}\rho^{4-\al-\be}\, f_{\al}(x_k) 
+\sum_{\al,\be} s^{6-\al}\st^{6+\al}\mu^{6-\be}\rho^{6-\al-\be}g_{\al}(x_k) ,
}
The singular central fiber has been replaced by a fiber $\Xsd = X_1\cup X_2$ with two components
$X_i$ defined by $\rho=0$ and $\mu=0$, respectively. The component $\rho=0$ is described by 
\eqn\fivefoldii{\eqalign{
p_{X_1}&=p_0+p_+\, ,\cr
p_0&=y^2+x^3+xf_0(x_k)+g_0(x_k)\, ,\cr
p_+&=x\sum_{\al>0} s^{4-\al}\mu^{\al}f_{\al}(x_k)
+\sum_{\al>0} s^{6-\al}\mu^{\al}g_{\al}(x_k)\, ,
}}
where we have collected the terms with zero and positive powers in $\mu$ into 
the two polynomials $p_0$ and $p_+$ for later use.
The hypersurface $X_1$ is a fibration $X_1 \to B_2$ with fiber a rational elliptic surface $S_1$. The expressions 
in \fivefoldii\ are sections of line bundles, specifically the anti-canonical bundle $\cx L=K_{B_2}^{-1}$, 
a line bundle $\cx M$ over $B_2$ that enters the definition of the fibration $\Xb\to B_2$ and a bundle $\cx N$ 
associated with a $\IC^*$ symmetry acting on the homogeneous coordinates $(y,x,s,\mu)$. The powers
of the line bundles appearing in these sections are 
\eqn\cvex{\vbox{\offinterlineskip\halign{
\strut \hfil~$#$~\hfil\vrule 
&\hfil~$#$~\hfil\vrule &\hfil~$#$~\hfil&\hfil~$#$~\hfil
&\hfil~$#$~\hfil&\hfil~$#$~\hfil&\hfil~$#$~\hfil&\hfil~$#$~\hfil
\cr
&p_{X_1}&y&x&s&\mu&f_\al(x_k)&g_\al(x_k)\cr
\noalign{\hrule}
\cx L&6&3&2&0&0&4&6\cr
\cx M&6&3&2&1&0&\al&\al\cr
\cx N&6&3&2&1&1&0&0\cr
}}
}
e.g. å $p_{X_1}\in\Gamma(\cx L^6\otimes\cx M^6\otimes\cx N^6)$.

The hypersurface $X_1$ has a positive first Chern class $c_1(X_1)=c_1(\cx N)$ and the 
CY 3-fold $\Zb$ is embedded in $X_1$ as the divisor $\mu=0$,
$$
p_{\Zb}=p_{X_1}\cap \{\mu=0\}=p_0\, ,
$$
verifying a claim that was needed in the argument of sect.~2.4.
According to the picture of F-theory/heterotic duality developped in \refs{\MV,\FMW}, 
the polynomial $p_+$ containing the positive powers in $s$ describes part of the bundle data 
in a single $E_8$ factor of the heterotic string compactified on $\Zb$. Using a different argument, based
on the type IIA string compactified on fibrations of ADE singularities, more general $n$-fold 
geometries $\Xabms$ of the general form \fivefoldii\ have been 
obtained in \refs{\KMV,\BMff}\ as local mirror geometries of bundles å with arbitrary structure group on elliptic fibrations. 
Mirror symmetry gives an entirely explicit map between the moduli of a given toric $n$-fold
and the geometric data of a $G$ bundle on a toric $n-1$-fold $\Zb$, 
which applies to any geometry $\Xabms$\ of the form \fivefoldii\ \BMff. The application of 
these methods will be illustrated at the hand of selected examples in sects.~6 and 7\yyy.

A special case of the above discussion is the one, where the heterotic gauge sector is not 
a smooth bundle, but includes also non-perturbative small instantons \Witsi. The F-theory
interpretation of these heterotic 5-branes as a blow up of the
base of elliptic fibration $\Xb\to B_3$ has been studied in detail in
\refs{\MV,\AMsd,\BJPS}; see also refs.~\refs{\BMff,\RajeshIK} for details
in the case of toric hypersurfaces and ref.~\DonagiXE\ for an elegant
discussion of the moduli space in M-theory.

From the point of Hodge variations and brane superpotentials
this is in fact the most simple case,
starting from the approach of \refs{\LMW,\JockersPE,\AlimBX},
as the brane moduli of the type II side map to moduli of the heterotic
5-brane. An explicit example from \AlimRF\ will be discussed in sect.~7

\subsec{Type II / heterotic map}
The above argument also provides a means to describe an explicit map between a type II brane compactification 
on $\Zb$ and a heterotic bundle compactification on $\Zb$.
The key point is again the afore mentioned relation $(C2)$ between the large volume limit of the 
fibration $\pi:\, \Xa\to \IP^1$ and the s.d. limit of the F-theory 4-fold $\Xb$. 
The relation between the F-theory 4-fold geometry,
the heterotic bundle on $\Zb$ and the type II branes on $\Zb$ is concisely summarized
by the following diagram:
\eqn\diaCS{
\xymatrix{
\Za\to\Xa\to\IP^1
\ar[d]_{\sevenrm mirror\ symmetry}
&
\ar[r]^{\sevenrm large \ base}_{+\ local\ limit}
&&
\Za\to\XXa(L)\to\IC^1
\ar[d]_{\sevenrm local}^{\sevenrm mirror\ symmetry}
\cr
\Xb
&
\ar[r]^{\sevenrm stable \ deg}_{\sevenrm +\ local\ limit}
&&
\Xabms(E) 
\cr}
}
The upper line indicates how the open-closed string dual $\XXa(L)$ of 
an $A$-type bundle $L$ on the 3-fold $\Za$ sits in the 
compact 4-fold $\Xa$ mirror to $\Xb$.
The details of the bundle $L$ are encoded in the toric resolution of the central
fiber $\Za^0$ at the origin $0\in\IC^1$, as described in terms of toric polyhedra 
in refs.~\refs{\MayrXK,\LM,\AlimRF}.
The limit consists of concentrating
on a local neighbourhood of the point $0\in \IP^1$ and
taking the large volume limit of $\IP^1$ base.

The lower row describes how the heterotic bundle $E$ on the 
elliptic manifold $\Zb$ dual to F-theory on $\Xb$ is captured by a local mirror geometry of the form \fivefoldii.
Assuming that the large base/local limit commutes with mirror symmetry, 
the diagram is completed to the right by another vertical arrow, which represents 
local mirror symmetry of the non-compact manifolds. 
The mirror of the open closed dual $\XXa(L)$ has been previously 
called $\XXb(E)$, and we see that commutativity of the diagram requires
that the open-closed dual $\XXb(E)$ {\it is the same} as the heterotic dual $\Xabms(E)$.
Indeed, the hypersurface equations for $G=SU(N)$ given in ref.~\BMff\ for the heterotic 4-fold $\Xabms$
and in ref.~\refs{\AlimRF}\ for the open-closed 4-fold $\XXb$ can be both written in the 
form 
\eqn\ncffs{\eqalign{
p(\Xabms)&=p_0(\Zb)+v\, p_+(\cxS)\, \qquad{\hbox{\ninerm (heterotic/F-theory\ duality)}}
\cr
p(\XXb)&=P(\Zb)+v\, Q(\cxH)\, \qquad {\hbox{\ninerm\ (open-closed\ duality)}\phantom{\matrix{1\cr1}}}
}}
where $v$ is a local coordinate defined on the cylinder related to $s$ in \fivefoldii.
In both cases, the $v^0$ term specifies the 3-fold $\Zb$ on which the type II/heterotic string is compactified. 
In the type II context, $Q(\cxH)=0$ is the hypersurface $\cxH\subset \Zb$, which is part of the
definition of the $B$-type brane \refs{\MayrXK,\LM,\AlimRF}.
In the heterotic dual of \BMff, $p_+(\cxS)=0$ specifies the $SU(n)$ spectral cover
\FMW. 

The agreement of the local geometries dual to the type II/heterotic compactification on $\Zb$ predicted by 
the commutativity of \diaCS\ is now obvious with the identification
\eqn\idi{
\hskip-1cm \hbox{\ninerm type\ II/heterotic\ map:}\qquad P(\Zb)=p_0(\Zb),\qquad 
Q(\cxH)=p_+(\cxS)\, .
}
This map between the dual 4-folds in \ncffs\ can be interpreted as a geometric reflection of the physical fact 
that the decoupling limit conforms the heterotic and type II bundles.

Note that, with the identification \idi,
the proofs of refs.~\refs{\MayrXK,\LM,\AlimBX,\AganagicJQ},
which relate the relative periods $H^3(\Zb,\cxH)$ to the
periods of the 4-fold $\XXb$ in the context of open-closed duality,
carry also over to the heterotic string setting for $G=SU(N)$. 
More ambitiously, one would like to have an explicit relation
between the 4-fold periods and the holomorphic Chern Simons integral
also for a heterotic bundle with general structure group $G$. 
The approach of refs.~\refs{\KMV,\BMff} gives an explicit map from the the moduli 
of a $G$ bundle on $\Zb$ to a local mirror geometry $\Xabms$ for any $G$ 
and evaluation of the periods of $\Xabms$ gives the 4-fold side.
A computation on the heterotic side could proceed by a generalization of the arguments 
of sect.~2.3\yyy, e.g. by constructing the sections of the bundle from 
the more general approaches to $G$ bundles described in \refs{\FMW,\CurioBVA}. 
In sect.~8 we outline a possible alternative route, using a conjectural 
relation between two two-dimensional thories associated with the 3-fold and the 4-fold
compactification.

%%%%%%%%%%%%%%%%%%%%%%%%%%%%%%%%%%%
\newsec{Type II/heterotic duality in two space-time dimensions}
%%%%%%%%%%%%%%%%%%%%%%%%%%%%%%%%%%%
In the previous sections we demonstrated the chain of dualities in eq.~\netdualiia\
by matching the holomorphic superpotentials of the various
dual theories. In this section we further supplement this analysis by
relating the two-dimensional low energy effective theories of the
type IIA compactificatons on the 4-folds $\Xa$ and $\Xb$ with the
dual heterotic compactification on $T^2\times\Zb$. Many aspects of the
type II/heterotic duality on the level of the low energy effective action
are already examined in ref.~\HaackDI. We further extend this
discussion here. 

For the afore mentioned string compactifications the low energy effective
theory is described by two-dimensional $\cx N=(2,2)$ supergravity.\foot{Note
that these two-dimensional theories describe the effective space-time theory
and not the two-dimensional field theory of the underlying microscopic
string worldsheet.}
Chiral multiplets $\vphi$ and twisted chiral multiplets $\tx\vphi$ 
comprise the dynamical degrees of freedom of these supergravity
theories \refs{\GatesDU,\WittenYC}.
In a dimensional reduction of four-dimensional $\cx N=1$ theories å 
the two-dimensional chiral multiplets/twisted chiral multiplets arise
from four-dimensional chiral multiplets/vector multiplets, respectively.

The scalar potential of the two-dimensional $\cx N=(2,2)$ Lagrangian
arises from the holomorphic chiral and twisted chiral
superpotentials $W(\vphi)$ and ${\wtx W}(\tx\vphi)$, and the
kinetic terms are specified by the two-dimensional
K\"ahler potential\foot{This
splitting of the K\"ahler potential does not represent the most
general form. In fact in general the target space metric need not even
be K\"ahler \GatesDU. The given ansatz, however, suffices for our
purposes.}
\eqn\Ktwoi{ K^{(2)}(\vphi,\bb\vphi,\tx\vphi,\bb{\tx\vphi})\,=\, 
K^{(2)}(\vphi,\bb\vphi) + {\wtx K}^{(2)}(\tx\vphi,\bb{\tx\vphi})\ . }
Here $K^{(2)}$ and ${\wtx K}^{(2)}$ can be thought of individual
K\"ahler potentials for the chiral and twisted chiral sectors. In this
section we mainly focus on the K\"ahler potential~\Ktwoi\ to further
establish the type II/heterotic string duality of eq.~\netdualiia.

%%%%%%%%%%%%%%%%%%%%%%%%%s
\def\ZI{{\kappa}}\def\Dg{D}
\subsec{Type IIA on Calabi-Yau fourfolds}
The low energy degrees of freedom of type IIA compactifications on
the Calabi-Yau 4-fold $X$ are the twisted chiral multiplets 
$T^A, A=1,\ldots, h^{1,1}(X)$ and the chiral
multiplets $z^I, I=1,\ldots,h^{3,1}(X)$.\foot{In two dimensions
the graviton and the dilaton are not dynamical \deWitXR.}
They arise from the K\"ahler and the complex
structure moduli of the 4-fold $X$.\foot{For $h^{2,1}(X)\ne 0$
there are additional $h^{2,1}$ chiral multiplets, which we do not take
into account here. With these multiplets the simple splitting ansatz~\Ktwoi\
ceases to be sufficient \HaackDI.} Then the tree-level K\"ahler potential is given by \HaackDI
\eqn\KPotII{ K_{\rm IIA}^{(2)}\,=\,K_{\rm CS}^{(2)}(z,\bb z) + {\wtx K}_{\rm K}^{(2)}(T,\bb T)\,=\, 
-\ln Y_{\rm CS}^{\rm IIA}(z,\bb z) -\ln {\wtx Y}_{\rm K}^{\rm IIA}(T,\bb T) \ , }
where the exponential of the 
potential $K_{\rm CS}^{(2)}$ for the complex structure moduli is determined by 
\eqn\CSPot{ Y^{IIA}_{\rm CS}(z,\bb z)\,=\, \int_X \Omega(z)\wedge\bar\Omega(\bb z)\ , }
in terms of the holomorphic $(4,0)$ form $\Om$ of the Calabi-Yau $X$.
In the large radius regime the twisted potential ${\wtx K}_{\rm K}^{(2)}$ for
the K\"ahler moduli reads
\eqn\KPot{\wtx Y_{\rm K}^{\rm IIA}\,=\, å \fc{1}{4!}\int_X J^4 \,=\, 
\fc{1}{4!}\sum_{A,B,C,D} \I_{ABCD} (T^A-\bb T^A)(T^B-\bb T^B)(T^C-\bb T^C)(T^D-\bb T^D) \ , }
with $\I_{ABCD}$ the topological intersection numbers å of the 4-fold $X$. The K\"ahler moduli $T^A$
appear in the expansion of the complexified K\"ahler form $B+iJ=T^A \om_A,\, \om_A\in H^2(X,\IZ)$, where
$B$ and $J$ are the NS 2-form and the real K\"ahler form, respectively. Finally, in the presence of
background fluxes, we obtain the holomorphic superpotentials \refs{\GVW,\GukovIQ}
\eqn\WX{W(z) \,=\, \int_X \Omega \wedge F_{hor} \ , \qquad å \widetilde W(t) \,=\, \int_X e^{B+i J} \wedge F_{ver} \ . }
Here $F_{hor}\in H^4_{hor}(X)$ is a non-trivial horizontal RR 4-form
flux, whereas $F_{ver}\in H^{\rm ev}_{ver}(X)$ is a non-trivial even-dimensional
vertical RR flux.\foot{The 6- and 8-forms are the magnetic dual fluxes to
the RR 4- and 2-form fluxes in type IIA.} The twisted chiral
superpotential $\wtx W$ receives non-perturbative worldsheet corrections away
from the large radius point \refs{\LercheZB,\PMff}.

\subsec{Type IIA on the Calabi-Yau 4-folds $\Xa$ and $\Xb$}
We now turn to the type~IIA compactification on the special Calabi-Yau 4-fold $\Xa$.
As discussed in sect.~4.1.\yyy\ the 4-fold geometry $\Xa$ is a fibration over
the $\IP^1$ base, where the generic fiber is the Calabi-Yau 3-fold $\Za$.
Geometries of this type have been studied previously in \refs{\PMff,\HaackDI} and we extend
the discussion here to fibrations with singular fibers, which support the
brane/bundle degrees of freedom in the context of open-closed/heterotic duality.

For the divisor $\DS$ dual to the base this implies
\eqn\ZChi{ \int_{\DS} c_3(\Xa) = \chi(\Za) \ . }
Here $c_3(\Xa)$ is the third Chern class of the 4-fold $\Xa$ and $\chi(\Zb)$ is the
Euler characteristic of 3-fold $\Za$. Hence the divisor $\DS$ is homologous to the
generic (non-singular) fiber $\Za$.

For type~IIA compactified on the 4-fold $\Xa$ we are interested in the twisted chiral
sector, and hence in the twisted K\"ahler potential~\KPot. This means we need to obtain
the intersection numbers of the fibered 4-fold $\Xa$. We use similar arguments
as in ref.~\AspinwallVK, where the intersection numbers of $K3$-fibered
Calabi-Yau threefold are determined.

We denote by $S$ the (complexified) K\"ahler modulus that measures
the volume of the $\IP^1$ base, which is dual to the divisor $\DS$ representing
the generic fiber $\Za$. 
Consider now a divisor $\DH_a$ of the generic fiber $\Za$. As we move
this divisor about the base by mapping it to equivalent divisors in the
neighboring generic fibers, we define a divisor $\Dg_a$ in the Calabi-Yau
4-fold $\Xa$.\foot{Due to monodromies with respect to the degenerate
fibers, it may happen that two inequivalent
divisors $\DH_a$ and $\DH_b$ are identified globally, and hence
yield the same divisor $\Dg_a=\Dg_b$. Then we work on the 3-fold $\Za$
with monodromy-invariant (linear combinations of) divisors such that
only inequivalent divisors $\Dg_a$ are generated on the 4-fold $\Xa$.}
The remaining (inequivalent) divisors
of the 4-fold $\Xa$ are associated to singular fibers, and we denote them
by $\hD_{\hat a}$. 

The 2-forms $\om_S$, $\om_a$ and $\hx\om_{\hx a}$, which are dual to the
divisors $\DS$, $\Dg_a$ and $\hD_{\hx a}$, furnish now a basis of the
cohomology group $H^2(\Xa,\IZ)$, and we denote the corresponding
(complexified) K\"ahler moduli by $S$, $t_a$ and $\hx t_{\hx a}$. They measure
the volume of the $\IP^1$-base, the volume of the 2-cycles in the generic 3-fold
fiber $\Za$, and the volume of the remaining 2-cycles arising from the
degenerate fibers. 

From this analysis we can extract the structure of intersection numbers. 
Since $\DS$ is a homology representative of the generic fiber
it intersects only with the Calabi-Yau divisors $\Dg_a$ according to
the triple intersection numbers $\ZI_{abc}$ of the 3-fold $\Za$.
The intersection numbers for divisors, which do not involve $\DS$,
cannot be further specified by these general considerations. Therefore
we find
\eqn\intX{ \fc{1}{4!} \I_{ABCD} T^A\,T^B\,T^C\,T^D \,=\, 
\fc{1}{3!} \ZI_{abc} \,S\, t_a\,t_b\,t_c + \fc{1}{4!} \SI_{\al\be\ga\de} t'_\al\, t'_\be\, t'_\ga\, t'_\de å \ , }
where $t'_\al$ are the K\"ahler moduli $(t_a,\hx t_{\hx a})$ with their
quartic intersection numbers $\SI_{\al\be\ga\de}$. 
The twisted K\"ahler potential for the 4-fold $\Xa$ then reads
\eqn\KPotii{\eqalign{
\wtx Y_{\rm K}(\Xa)\,=\, 
&\fc{1}{3!}(S-\bb S) \sum\ZI_{abc}(t_a-\bb t_a)(t_b-\bb t_b)(t_c-\bb t_c) \cr
&\quad+\fc{1}{4!} \sum\SI_{\al\be\ga\de}(t'_\al-\bb t'_\al)(t'_\be-\bb t'_\be)(t'_\ga-\bb t'_\ga)(t'_\de-\bb t'_\de) \ . }}
The essential point here is that the leading term for large $S$
involves only the moduli $t_a$ associated with the bulk fields in the dual compactifications, whereas the 
brane/bundle degrees of freedom appear in the subleading term. In the decoupling limit $\Im S\to \infty$, the
kinetic terms derived from \KPotii\ factorize into the bulk and bundle sector of the dual theories as
$$
G_{A\bb B}(T^C)\ \p_\mu T^A\p^\mu \bb T^{\bb B} \ \to\ å G^{bulk}_{a\bb b}(t^c)\ \p_\mu t^a\p^\mu \bb t^{\bb b}
+\fc{1}{\Im S}\ G^{bundle}_{\al\bb \be}(t^c,t^\ga)\ \p_\mu t^\al\p^\mu \bb t^{\bb \be}\ ,
$$
illustrating the separation of the physical scales at which the fields in the two sectors fluctuate.
In this limit, the backreaction of the (dual) bulk fields to the (dual) bundle fields vanishes and
the latter fluctuate in the fixed background determined by the bulk fields. A more detailed
treatment of the heterotic dual will be given below.

Analogously to the three contributions to $H^{2}(\Xa,\IZ)$ distinguished above, 
we can decompose the even-dimensional fluxes $F_{\rm V}$
into three distinct classes 
\eqn\FVDec{ F_{\rm V} \,=\, f^{(1)} + f^{(2)} \wedge \om_S + f^{(3)} \ , }
where the components $f^{(1)}$ and $f^{(2)}$ pull back to even-forms in $H^{\rm ev}(\Zb)$,
while the fluxes $f^{(3)}$ vanish upon pullback to the regular 3-fold fiber $\Za$.
With the vertical fluxes~\FVDec\ the (semi-classical) twisted chiral superpotential $\wtx W(\Xa)$
simplifies to 
\eqn\WXii{ \wtx W(\Xa) \,=\, \int_{\Zb} e^{\sum_a t_a \om_a} 
\wedge (S f^{(1)} + f^{(2)}) + \int_{\Xa} e^{\sum_\al t'_\al\om'_\al} \wedge (f^{(1)}+f^{(3)}) \ , }
% å 
with the generators $(\om_a,\hx\om_{\hx a})$ collectively denoted by $\om'_\alpha$.

Next we turn to the chiral sector of type~IIA strings compactified on the mirror 4-fold $\Xb$. 
The K\"ahler potential~\CSPot\ is then expressed in terms of the periods
$\Pi^{\Sigma}\,=\,\int_{\ga_\Sigma} \Om^{(4,0)}$ of the Calabi-Yau 4-fold $\Xb$
\eqn\CSPotii{ Y_{\rm CS}(\Xb)\,=\,\sum_{\ga_\Sigma,\ga_\Lambda\in H_4(\Xb)}
\Pi^{\Sigma}(z)\eta_{\Sigma\Lambda}\bb\Pi^{\Lambda}(\bb z) \ , }
where $\eta_{\Sigma\Lambda}$ is the topological intersection paring on $H_4(\Xb)$. 
The horizontal background fluxes $F_{\rm H}$ induce the chiral
superpotential $W(\Xb)$ given in eq.~\wwii, where the quanta $\ux N_\Sigma$
correspond to the integral flux quanta of 4-form flux $F_{\rm H}$.

By 4-fold mirror symmetry the superpotential $\wtx W(\Xa)$ and $W(\Xb)$ are
equal on the quantum level. In comparing the semi-classical expression \WXii\ for the
twisted superpotential to the structure of the chiral superpotential \wwii\ in the stable degeneration
limit \delim, we observe that the vertical fluxes $f^{(1)}$, $f^{(2)}$ and $f^{(3)}$ give
rise to the flux quanta $M_\Sigma$, $N_\Sigma$ and $\hx N_\Sigma$, respectively.

\subsec{Heterotic string on $T^2\times\Zb$}
The low energy effective action of the heterotic string compactified on the 4-fold $T^2\times\Zb$
together with a (non-trivial) gauge bundle $V$ has in the large radius regime the structure \HaackDI
\eqn\KPhet{ K_{\rm het}^{(2)} \,=\, K_{\rm het}^{(4)}(\Phi,\bb\Phi)
+ {\wtx K}^{(2)}_{\rm het}(\tx\Phi,\bb{\tx\Phi}) \ . }
The chiral K\"ahler potential $K_{\rm het}^{(4)}$ coincides with the four-dimensional
K\"ahler potential of the heterotic string compactified on the Calabi-Yau 3-fold $\Zb$
with the gauge bundle $V$. Apart from the
heterotic dilaton, which is not a dynamic field in two dimensions \deWitXR, it comprises all the
kinetic terms for both the chiral multiplets of the K\"ahler/complex structure moduli of
the 3-fold $\Zb$ and the chiral multiplets from the gauge bundle $V$. 
The K\"ahler potential ${\wtx K}^{(2)}_{\rm het}$ of the twisted chiral multiplet
consists of the modes arising from the torus $T^2$ and
the gauge fields, which correspond to the vector multiplets in higher dimensions.

For heterotic Calabi-Yau compactifications with the standard embedding of the
spin connection the K\"ahler potential $K_{\rm het}^{(4)}$ splits further according to
$$
K_{\rm het}^{(4)} \,=\, K_{\rm CS}^{(4)}(z,\bb z) + K_{\rm K}^{(4)}(t,\bb t) + \ldots ,
$$ å å 
where $K_{\rm CS}^{(4)}$ and $K_{\rm K}^{(4)}$ are the K\"ahler potentials for the
chiral complex structure and K\"ahler moduli $z$ and $t$ of the Calabi-Yau $\Zb$. For a general
heterotic string compactification,
we do not know of any generic model independent properties of the K\"ahler potential. 
However, in the context of type IIA/heterotic duality~\netdualiia, we 
expect a special subsector associated with the kinetic terms of the complex structure
moduli $z^a$ of the 3-fold together with the specific moduli fields $\hx z^{\hx a}$ of the bundle 
captured by the dual 4-fold.

In order to infer some qualitative information about the relevant kinetic terms
of the moduli $z^a$ and $\hx z^{\hx a}$ we briefly discuss the general structure
of the bosonic part of the four-dimensional low-energy effective heterotic action
in the four-dimensional Einstein frame 
\eqn\LagHet{ S_{\rm het}^{(4)} \,=\, \fc{1}{2\kappa_4^2} 
\int d^4 x \sqrt{g_4} \left( R^{(4)} 
-\fc{1}{2} å \left( C_{a\bb b} \partial_\mu z^a \partial^\mu \bb z^{\bb b} \right)
-\fc{1}{2} \left( B_{\hat a \bb{\hat b}} \partial_\mu \hx z^{\hat a} \partial^\mu \bb{\hx z}^{\bb{\hat b}} \right) å + \ldots \right) \ . }
Here $R^{(4)}$ is the Einstein-Hilbert term, $\kappa_4$ is
the four-dimensional gravitational coupling
constant. $C_{a\bb b}$ and $B_{\hat a\bb{\hat b}}$ denote
the K\"ahler metrics of the chiral fields $z^a$ and $\hx z^{\hx a}$. For simplicity
cross terms among bulk and bundle moduli and
the kinetic terms of other moduli fields are omitted. Note that
the $\alpha'$ dependence of the bundle moduli
is absorbed into the K\"ahler metric $B_{\hat a\bb{\hat b}}$.

From a dimensional reduction point of view the bundle moduli $\hx z^{\hx a}$
arise from a Kaluza-Klein reduction of the ten-dimensional vector field $A^{(10)}$,
which in terms of four-dimensional coordinates $x$ and internal coordinates $y$
enjoys the expansion
$$
A^{(10)}(x,y) = 
A_\mu^{(4)}(x) dx^\mu + \sum_{\hx a} ({\hx z}^{\hat a}(x)\,v_{\hx a}(y) + {\rm c.c.} ) + \ldots \ . 
$$
The four-dimensional vector $A^{(4)}$ gives rise to the Yang-Mills kinetic term,
while the internal vectors fields $v_{\hx a}$ are integrated out in the dimensional
reduction process and yield the metric $B_{\hat a\bb{\hat b}}$
\eqn\BMet{ B_{\hat a\bb{\hat b}} \,=\, \fc{1}{V(\Zb)}
\int_{\Zb}d^6y \sqrt{g_6}\, \alpha' g^{i\bb\jmath} \,{\rm Tr}\left(v_{\hx a,i}\,\bb v_{\bb{\hx b},\bb\jmath}\right) \ . }
The volume factor $V(\Zb)$ arises due to the Weyl rescaling to the four-dimensional
Einstein frame, and it compensates the scaling of the (internal) measure
$d^6y\sqrt{g_6}$. Thus the dimensionless quantity $\alpha' \over \ell^2$, where $\ell$
is the length scale of the internal Calabi-Yau manifold $\Zb$, governs the magnitude
of the kinetic terms $B_{\hat a\bb{\hat b}}$.

As discussed in sect.~4.1.\yyy, the decoupling limit $\Im S\to \infty$ defined in ref.~\AlimBX\ 
is mapped on the heterotic side to the large fiber limit of the elliptically fibered Calabi-Yau 3-fold 
$\Zb\rightarrow B$. In order to work in at semi-classical regime, the volume $V(B)$ 
of the base $B$, common to the K3 fibration 
$\Xb \to B$ and the elliptic fibration $\Zb \to B$, has to be taken of large volume as well, due to the 
relations \HaackDI
$$
\la^{-2}_{II,2d}=\la^{-2}_{het,2d}\, ,\qquad å V_{het}(B) \cdot V_{II}(B)= \la_{II,2d}^{-4}\ ,
$$
which follow from the relations $\la_{II,6d}=\la_{het,6d}^{-1}$, 
$g_{het} =\la_{II,6d}^{-2}g_{II}$ in six dimensions \WittenEX. 
As we move away from the stable degeneration point in the dual type IIA
description, the volume of the elliptic fiber in the 3-fold $\Zb$ becomes finite
while we keep the volume of the base large
\eqn\Hscal{ 0 \ll \ell_F \ll \ell_B \ . }
Here $\ell_F$ is the length scale for the generic elliptic fiber
and $\ell_B$ is the length scale for the base. 

As a consequence, as we move away from the stable degeneration point,
the bundle components, which scale with the dimensionless quantity
$$
g_F\equiv\fc{\alpha'}{\ell_F^2}\ ,
$$
are the dominant contributions to the metric~\BMet. The moduli of the spectral cover
correspond on the (dual) elliptic fiber to vector fields $v_{\hx a}$, which
are contracted with the metric component scaling as $g_F$.
Therefore the bundle moduli ${\hx z}^{\hx a}$ associated to
the subbundle~$E$ of the spectral cover become relevant.

Thus for the heterotic string compactification on the 3-fold $\Zb$ with gauge
bundle the complex structure/bundle moduli space of the pair $(\Zb,E)$ 
is governed by the deformation problem of a family of Calabi-Yau 3-folds $\Zb$
together with a family of spectral covers $\Si_+$. As proposed in \wwii, this moduli dependence is
encoded in the relative periods $\ux \Pi^\Sigma(z,\hx z)$ of the relative three
forms $H^3(\Zb,\Si_+)$, and therefore in the semi-classical
regime the K\"ahler potential of the complex structure/bundle moduli
space $(\Zb,E)$ is expressed explicitly by \refs{\AlimBX,\JockersYJ}
\eqn\Khetdiv{
K_{{\rm CS},E}^{(4)}\,=\, -\ln Y_{{\rm CS},E}(\Zb,\Si_+) \ , \quad 
Y_{{\rm CS},E}(\Zb,\Si_+) \,= \!\!\!\!\!\!\!\!\!\sum_{\ga_\Sigma,\ga_\Lambda\in H_3(\Zb,\Si_+)}\!\!\!\!\!\!\!\!\!
\ux\Pi^{\Sigma}(z,\hx z)\ux\eta_{\Sigma\Lambda}\bb{\ux\Pi}^{\Lambda}(\bb z,\bb{\hx z}) \ . }
The topological metric $\eta_{\Sigma\Lambda}$ arises from the intersection
matrix of the relative cycles $\ga_\Sigma$. This intersection matrix has the form \AlimBX
$$
\left(\ux\eta \right) \,=\, \pmatrix{\eta_{\Zb} & 0 \cr 0 & i\,g_F\,\hx\eta_{\Si_+}} \ ,
$$
where $\eta_{\Zb}$ is the topological metric of the absolute
cohomology $H^3(\Zb)$ and $\hx\eta_{\Si_+}$ is the topological
metric of the variable cohomology sector $H^2_{\rm var}(\Si_+)$ of the
relative cohomology group $H^3(\Zb,\Si_+)$.

Note that the structure of the K\"ahler potential~\Khetdiv\ is also in agreement
with the mirror K\"ahler potential of type~IIA compactified on the 4-fold $\Xa$.
By the arguments of sect.~4\yyy, the K\"ahler modulus $S$ of the $\IP^1$ base of the 4-fold $\Xa$ is related to
the heterotic volume modulus of the elliptic fiber of the fibration $\Zb\to B$. In the large base limit
of $\Xa$/bundle decoupling limit of $(\Zb,V)$ the leading order terms are the
K\"ahler moduli of the 3-fold fiber $\Za$/complex structure moduli of the
3-fold $\Zb$. These moduli spaces are identified by mirror symmetry of the 3-fold
mirror pair $(\Za,\Zb)$.
The subleading terms for type~IIA on $\Xa$ in eq.~\KPotii\ should be compared
to the subleading bundle moduli terms in eq.~\Khetdiv\ on the heterotic side. 

Finally we remark that since the chiral sector of the heterotic string
compactification on $T^2\times\Zb$ and on $\Zb$ are equivalent (\cf eq.~\KPhet), 
the identification of the chiral K\"ahler potentials in the type IIA/heterotic duality in
two space-time dimensions carries over to the analog identification of K\"ahler
potentials in the F-theory/heterotic dual theories in four space-time dimensions discussed
in sect.~4.\yyy

\newsec{A heterotic bundle on the mirror of the quintic}
Our first example will be an $\cx N=1$ supersymmetric 
compactification
on the quintic in $\IP^4$ and its mirror.
This was the first compact manifold for which disc instanton corrected brane
superpotentials have been computed from 
open string mirror symmetry in \refs{\Wa,\MW}. This computation was confirmed
by an $A$ model computation in \PaWa. An off-shell version of the 
superpotential was later obtained in \refs{\JockersPE,\AlimRF,\AlimBX,\AganagicJQ},
both in the relative cohomology approach, eq.\spii,
as well as from open-closed duality, eq.\Wnc.

\subsec{Heterotic string on the threefold in the decoupling limit}
Here we follow the treatment in \refs{\AlimRF,\AlimBX}, 
In the framework of \Bat,
the mirror pair $(\Xa,\Xb)$ of toric hypersurfaces can be defined by a 
pair $(\D,\Ds)$ of toric polyhedra, given in app.~B.1\yyy\ for the concrete
example.
The $h^{1,1}=3$ K\"ahler moduli $t_a$, $a=1,2,3$, å of the fibration $\Za\to \Xa \to \IP^1$ describe the volume $t=t_1+t_2$ 
of the generic quintic fiber of the type $\Za$, the volume $S=t_3$ 
of the base $\IP^1$ and one additional K\"ahler volume $\that=t_2$ measuring the volume
of an exceptional divisor intersecting the singular fiber $\Za^0$. This 
divisor is å associated with the vertex $\nu_6\subset\D$ in eq.(B.1)\yyy\
and its K\"ahler modulus represents an open string deformation of a toric $A$ brane 
geometry $(\Za,L)$ of the class considered in \AV. 

The hypersurface equation for the mirror 4-fold $\Xb$ is given by the general expression
\eqn\ePBat{
P(\Xb)=\sum_{i=0}^N a_i \; \prod_{j=0}^M x_j^{\langle \nu_i,\nus_j\rangle+1}\ .
}
Here the sums for $i$ and $j$ run over the relevant integral points of the polyhedra $\D$ and 
$\Ds$, respectively, and $a_i$ are complex coefficients that determine
the complex structure of $\Xb$. A similar expression holds for the hypersurface equation 
of the mirror manifold $\Xa$,
with the roles of $\D$ and $\Ds$ exchanged.

Instead of writing the full expression, which 
would be too complicated due to the large number of relevant points of $\Ds$, we first write a 
simplified expression in local coordinates that displays the quintic fibration of the mirror:
\eqn\ePV{
P(\Xb)= p_0+v^{1}p_++v^{-1}p_-\ ,
}
with 
\eqn\ePi{
\eqalign{
p_0&=x_1^5+x_2^5+x_3^5+x_4^5+x_5^5-(z_1z_2)^{-1/5}\, x_1x_2x_3x_4x_5,\cr
p_+&=x_1^5+z_2\, (z_1z_2)^{-1/5}\, å x_1x_2x_3x_4x_5,\qquad 
p_-=z_3\, x_1^5\ .
}
}
Here $v$ is a local coordinate on $\IC^*$ and $z_a$ the three complex structure moduli
of $\Xb$ related to the afore mentioned K\"ahler moduli of $\Xa$
by the mirror map, $t_i=t_i(z_.)$. In the large volume limit the 
leading behavior is $t_i(z_.)=\fc{1}{2\pi i}\ln(z_i)+\cx O(z_.)$. 
The special combination $z_1z_2$ appearing above is mirror to the volume of the quintic fiber of 
$\pi:\, \Xa\to\IP^1$,
We refer to app.~B.1\yyy\ for further details of the parametrization used here and in the following.

Although the above expression for $P(\Xb)$ is oversimplified (most 
of the coordinates $x_j$ in \ePBat\ have been set to one), it
suffices to illustrate the general structure and
to sketch the effect of the decoupling limit, which, again simplifying, corresponds to 
setting $z_3=0$, removing the term $\sim p_-$ in \ePi.\foot{A more precise description
of this process as a local mirror limit is given in ref.~\BMff.}
This produces a hypersurface equation
of the promised form \ncffs.
In particular, $p_0(\Zb)=0$ defines the mirror of the quintic, which
has a single complex structure deformation parametrized by $z=z_1z_2$.
The hypersurface $\cxH$ for the relative cohomology space $H^3(\Zb,\cxH)$, which specifies å 
the Hodge variation problem, is defined by $p_+=0$, that is å 
\eqn\eHi{
\Zb\supset \cxH:\ x_1^5+z_2\, (z_1z_2)^{-1/5}\, å x_1x_2x_3x_4x_5=0 \ .
}
More precisely the component of \eHi\ relevant to the brane superpotential of refs.~\refs{\Wa,\AlimRF}
is in a patch with $x_i\neq 0\, \forall\, å i$ and passing to appropriate local coordinates for this patch,
the Hodge variation on $\cxH$ is equivalent to that on a quartic K3 surface in $\IP^3$ \AlimRF.

The F-theory content of the toric hypersurface $\Xb$ and its heterotic dual are 
exposed in different local coordinates 
on the ambient space, which put the hypersurface equation into the form studied in the 
context of F-theory/heterotic duality in \refs{\BMff}:
\eqn\eQi{\eqalign{
p_0&=Y^3+X^3+YXZ(stu+s^3+t^3) -z_1z_2\, Z^3(s^2t^2u^5)\ ,\cr
p_+&= X^3-z_2\, YXZ(stu)\ ,\hskip30pt
p_-= z_3\, X^3\ .\cr
}
}
Here $(Y,X,Z)$ are the coordinates on the elliptic fiber, a cubic in $\IP^2$. 
Again the zero set $p_0=0$ defines the 3-fold geometry $\Zb$, while the 
polynomials $p_\pm$ specify the components
$\cxS_\pm$ of the spectral cover of the heterotic bundle in the two $E_8$ factors.
While $p_-$ corresponds to the trivial spectral cover, $p_+$ describes a non-trivial component 
\eqn\sci{
\S_+:\ X^2-z_2\, YZ(stu)=0\ .
}
This equation can be seen to correspond to a bundle with structure group $SU(2)$ as follows.
The intersection of the equation~$\Si_+$ with the cubic elliptic equation gives six
zeros. However these zeros are identified by the Greene-Plesser
orbifold group $\IZ_3$, acting on the coordinates $\{Z,Y,X\}$ according to
\eqn\Korb{\{Z,Y,X\} \to \{ \rho^2\,Z, \rho\,Y, X \} \ , \qquad \rho^3=1 \ ,}
where $\rho$ is a third root of unity. Note that the spectral cover~$\Si_+$ represents the most general polynomial of
degree two invariant with respect to the orbifold group~\Korb. As a consequence
the six zeros become just two distinct zeros in the elliptic
fiber $E$, adding up to zero. Therefore the spectral cover describes a $SU(2)$ bundle
on the heterotic manifold $\Zb$.

Alternatively one may study the perturbative gauge symmetry of the heterotic compactification
from studying the singularities of the elliptic fibration $\Xb$. The
result of this procedure, described in detail in the appendix, is that the
bundle leads to the gauge symmetry breaking pattern
\eqn\gbpi{
\xymatrix{E_6\times E_6 \ar[r]& SU(6)\times E_6\cr}
}
in agreement with a new component of the bundle of structure group $SU(2)$. 

\break
\subsubsec{Flux superpotential in the decoupling limit}
To be more precise, the above discussion describes only the data of the bundle geometrized by
F-theory and ignores the 'non-geometric' part 
of the bundle arising from fluxes on the 7-branes, which may lead to a larger structure group 
of the bundle, and thus smaller gauge group of the compactification then the
one described above \refs{\BJPS}.

In particular, to compute the heterotic superpotential \wwiii, we have to specify
the class $\ga$ of sect.~2.2\yyy, which determines the flux number $\hx N_\Si$ in 
\wwii, and thus the superpotential as a linear combination of the 4-fold periods. 
This is the heterotic analogue of choosing the 5-brane flux on the type II brane \eHi.
Since eq.\idi\ identifies the type II open string brane modulus $z_2$ literally with the 
heterotic bundle modulus in the decoupling limit $\Im S \to \infty$, the 
relative cohomology space and the associated Hodge variation problem is identical to the one studied 
in the context of type II branes in \AlimBX.
Using the identification $\ga=\tx \ga$ between
the classes defined in \deftxgamma\ and \defgamma, the heterotic
superpotential
in the decoupling limit is identical to that for the type II brane
computed in sect.~5 of \AlimBX, see eq.~(5.3).
We now discuss the corrections
to this result for finite $\Im S$.

\subsec{F-theory superpotential on the four-fold $\Xb$}
According to the arguments of sect.~3\yyy, Hodge theory on the F-theory 4-fold
$\Xb$ computes further corrections to the superpotential of the type II/heterotic compactification for
finite $S$. We will now perform a detailed study of the periods of $\Xb$ using 
mirror symmetry of the 4-folds $(\Xa,\Xb)$.

Mirror symmetry is vital in two ways. Firstly, it allows to determine the geometric 
periods on $H_4(\Xb,\IZ)$, appearing as the coefficients of the flux numbers $N_\Si$
in \wwii, from an intersection computation on the mirror $\Xa$. Secondly, the mirror map $t(z)$ can be used to define 
preferred local coordinates on the complex structure moduli space $\cx M_{CS}(\Xb)$ near a large complex structure point. 
In the context of open-closed
string duality these two steps are central to extracting the large volume world-sheet instanton expansion
of the periods for the mirror $A$-model geometry $\Xa$, as they yield 
the disc instanton expansion of the superpotential for 
$A$-type brane geometry $(\Za,L)$ by open-closed duality \refs{\AlimRF,\AlimBX}. In the
present context we use this $A$ model expansion to describe the 
superpotential $W_F(\Xb)$ near a large complex structure limit of $\Xb$, which 
by the previous arguments describes the decoupling limit $\Im S \to \infty$ of 
the dual heterotic compactification $(\Zb,E)$ near large complex structure of $\Zb$.\foot{The fact that
the large complex structure limit of the 4-fold $\Xb$ implies a large structure limit of the dual heterotic 
3-fold $\Zb$ follows already from the hypersurface equation, eq.\eQi, and is explicit in
the monodromy weight filtration of the 4-fold periods discussed below.}

The methods of mirror symmetry for toric 4-fold hypersurfaces used in the following have been described 
in detail in \refs{\GMPh,\PMff,\KLRY}\
and we refer to these papers to avoid excessive repetitions. We work at the large complex structure point of $\Xb$ 
defined by the values å $z_a=0,\ a=1,2,3$ for the moduli in the hypersurface equation \ePV. This corresponds to a large volume
phase 
$t_a\sim \fc{1}{2\pi i}\ln(z_a)\to i \infty$ in the K\"ahler moduli of the mirror manifold $\Xa$ generated by the charge vectors
\eqn\cmori{\vbox{\offinterlineskip\halign{
\strut # 
&\hfil~$#$~=\hfil
&\hfil~$#$ &\hfil~$#$ &\hfil~$#$ &\hfil~$#$
&\hfil~$#$ &\hfil~$#$ &\hfil~$#$ &\hfil~$#$
&\hfil~$#$ &\hfil~$#$ &\hfil~$#$ &\hfil~$#$
\cr
&l^1&(&-4&0&1&1&1&1&-1&1&0)\ ,\cr
&l^2&(&-1&1&0&0&0&0&1&-1&0)\ ,\cr
&l^3&(&0&-2&0&0&0&0&0&1&1)\ .\cr
}}
}
The topological intersection data for this phase can be determined from toric geometry in the
standard way, see \refs{\PMff,\KLRY,\BMff,\GrimmEF} for examples. 
We refer to the appendix of \AlimBX\ for details on this particular 
example and restrict here to quote the quartic intersections 
\eqn\ffinters{\eqalign{
\cx F_4 &= \fc{1}{4!}\int_{X_c} J^4=\fc{1}{4!}\sum_{a,b,c,d} K_{\al\be\ga\de}t^\al t^\be t^\ga t^\de\cr
&=\fc{5}{6}(t_1+t_2)^3t_3+\fc{5}{12}(t_1+t_2)^4-\fc{1}{6}t_1^4
=\fc{5}{6}\tn_1^3\tn_3+\big(\fc{5}{12}\tn_1^4-\fc{1}{6}\tn_2^4\big) \ .
}}
Here $J=\sum_a t_a J_a=\sum_a \tn_a \check J_a$ denotes the K\"ahler form on $\Xa$, with
$J_a$, $a=1,2,3$ a basis of $H^{1,1}(\Xa)$ dual to the 
Mori cone defined by \cmori. In the above, we have introduced the 
linear combinations 
\eqn\cspv{
\tn_1=t=t_1+t_2,\qquad \tn_2=\that-t=-t_1,\qquad \tn_3=S=t_3\ ,
}
and the corresponding basis $\{\check J_a\}$ of $H^{1,1}(\Xa)$ 
to expose the simple dependence on the 
K\"ahler modulus $\tn_1=\Vol(\Za)$ of the generic quintic fiber of $\pi:\, \Za\to \Xa\to\IP^1$.

The leading terms of the period vector $\Pi_\Si=\int_{\ga_\Si}\Om$ for $\Xb$ in the limit $z_a\to 0$ 
can be computed from the classical volumes of even-dimensional algebraic cycles in $\Xa$
$$
\Pi_\Si(\Xb)=\int_{\ga_\Si}\Om(z) \sim \fc{1}{q!} \int_{\tx \ga_\Si} J^q,
$$
where $\ga_\Si\in H_4(\Xb,\IZ)$ refers to a basis of primitive 4-cycles 
in $\Xb$ and $\tx \ga_\Si$ a basis for the $2q$ dimensional algebraic cycles 
in $H_{2q}(\Xa),\ q=0,...,4$, related to the former by mirror symmetry.
Except for $q=2$, 
there are canonical basis elements for $H_{2q}(\Xa,\IZ)$, given
by the class of a point, the class of $\Xa$, the divisors dual to the 
generators $\check J_a$
and the curves dual to these divisors, respectively. 
On the subspace $q=2$ we choose as in \AlimBX\ the basis 
$
\ga_1=D_1\cap D_2,
\ga_2=D_2\cap D_8,
\ga_3=D_2\cap D_6.
$ 
Here the $D_i=\{x_i=0\},\ i=0,..,8$ are the toric divisors defined by the 
coordinates $x_i$ on the ambient space for $\Xa$
(cpw. eq.~\ePBat), which correspond to the vertices of the polyhedron $\D$ in (B.1)\yyy. The classical volumes of these basis elements computed from the intersections \ffinters\ are 
\eqn\epero{\eqalign{
\Pi_0=1,\ \Pi_{1,i}=\tn_i\ ,
\Pi_{2,1}=5\tn_1\tn_3\ ,
\Pi_{2,2}=\fc{5}{2}\tn_1^2\ ,
\Pi_{2,3}=2\tn_2^2\ ,\cr
\Pi_{3,1}=\fc{5}{2}\tn_1^2\tn_3+\fc{5}{3}\tn_1^3\ , 
\Pi_{3,2}=-\fc{2}{3}\tn_2^3\ ,
\Pi_{3,3}=\fc{5}{6}\tn_1^3\ ,
\Pi_{4}=\cx F_4\ ,
}}
where the first index $q$ on $\Pi_{q,.}$ denotes the complex dimension of the cycle.

The entries of the period vector $\Pi(\Xb)$ are
solutions of the Picard-Fuchs system for the mirror manifold $\Xb$ 
with the appropriate leading behavior 
\epero\ for $z_a\to 0$. The Picard-Fuchs operators can be derived from the toric GKZ system \refs{\PMff,\KLRY}
and are given in eq.~(A.6)\yyy\ in the appendix.

The Gauss-Manin system for the period matrix imposes certain integrability conditions
on the moduli dependence of the periods of a CY $n$-fold. For $n=2$ these conditions
imply that there are no instanton corrections on K3 and
for $n=3$ they imply the 
existence of a prepotential $\cx F$ for the periods. For $n=4$ the periods can
no longer be integrated to a prepotential, but still satisfy a set of integrability 
conditions discussed in ref.~\AlimBX.

Applying the integrability condition to the example
the leading behavior of $\Pi$ near $\tn_3=i \infty$, 
is captured by only seven functions denoted by $(1,\tn_1,\tn_2,\tx F_t,\tx W,\tx F_0,\tx T)$. 
The eleven solutions can be arranged into a period vector of the form
\eqn\eperi{
{\vbox{\offinterlineskip\halign{
\strut ~$#$~\hfil&~$#$~\hfil&~$#$~\hfil\cr
\hskip6pt \Pi_0=1\cr
\Pi_{1,1}=\tn_1\ ,
&\Pi_{1,2}=\tn_2\ ,
&\Pi_{1,3}=\tn_3 \ , 
\cr
\Pi_{2,1}=5\tn_1\tn_3+\pi_{2,1}\ ,\qquad
&\Pi_{2,2}=-\tx F_t\ ,\qquad
&\Pi_{2,3}=-\tx W,
\cr
\Pi_{3,1}=\tn_3\, \tx F_t+\pi_{3,1}\ ,
&\Pi_{3,2}=\tx T\ ,
&\Pi_{3,3}=-\tx F_0\ ,
\cr
\hskip7pt \Pi_{4}=\tn_3\,\tx F_0+\pi_4\ ,
\cr
}}}}
where the index $q$ on $\Pi_{q,.}$ now labels the monodromy weight 
filtration w.r.t. to the large volume monodromy $\tn_a\to \tn_a+1$.

Since the decoupling limit sends the compact 4-fold $\Xb$ to its 
non-compact open-closed dual $\XXb$, 
these functions should reproduce the relative 3-fold periods on $H^3(\Zb,\cxH)$
in virtue of eq.~\Wnc.
Indeed the four functions $(1,\tn_1,\tx F_t,\tx F_0)$ converge to the four periods on $H^3(\Zb)$
\eqn\elevenc{
\lim_{\tn_3\to i\infty}(1,\tn_1,\tx F_t,-\tx F_0) = (1,t,\p_t\cx F(t),-2\cx F(t)+t\p_t\cx F(t))\ ,
}
where $\cx F(t)=\fc{5}{6} t^3+\cx O(e^{2\pi i t})$ is the closed string prepotential 
on the mirror quintic.\foot{Here and in the following we neglect terms in 
the geometric periods from polynomials of lower degree in $\tn_i$.} 
The remaining three functions reproduce the three chain 
integrals on $H_3(\Zb,\cxH)$ with non-trivial $\p \ga\in H_2(\cxH)$:
\eqn\eleveno{
\lim_{\tn_3\to i\infty}(\tn_2,\tx W,\tx T) = (\that-t,W(t,\that),T(t,\that))\, ,
}
with classical terms $W(t,\that)=-2\tn_2^2+\cx O(e^{2\pi i \check t_k})$, $T(t,\that)=\fc{2}{3}\tn_2^3+\cx O(e^{2\pi i \check t_k})$, $k=1,2$.
In the context of open-closed duality, the double logarithmic solution 
$W(t,\that)$ of the 4-fold is conjectured \MayrXK\ to be the generating 
function of disc instantons in the type II mirror configuration $(\Za,L)$,
$$
W(t,\that)=-2\tn_2^2+\sum_{\be}\sum_{k=1}^\infty N_\be \fc{q^{k\be}}{k^2}\ ,
$$
similarly as $\cx F(t)$ is the generating function 
of closed string sphere instantons \refs{\CandelasRM}. In the above formula, $\be$ denotes the homology
class of the disc and the $N_\be$ are 
the integral Ooguri-Vafa disc invariants \OV.

Since the
closed string period vector \cspv\ appears twice in \eperi, with coefficients 
$1$ and $\tn_3=S$, respectively, the leading terms of the eleven periods on $\Xb$ 
are proportional to the seven relative periods on $H^3(\Zb,\cxH)$
$$
\lim_{\Im S \to \infty}\Pi_{q,.} \sim å \cases{(1,S)\times (1,t,\p_t\cx F,-2\cx F+t\p_t\cx F)\cr\cr(\that-t,W(t,\that),T(t,\that))} \ .
$$
A linear combination of these leading terms gives a large $S$ expansion for the 
superpotential of the form \wwii. 

\def\vr{\vrule height 10pt depth 4pt}
\subsec{Finite $S$ corrections: perturbative contributions}
There are two types of finite $S$ contributions in the 4-fold periods, which 
correct the 3-fold result: linear corrections 
$\sim S^{-1}$ and exponential corrections 
$\sim e^{2\pi i S}$. In the type II orientifold where $\Im S\sim 1/g_s$, the first å 
should correspond to perturbative corrections. 

These linear corrections are described 
by the three additional functions $\pi_{2,1},\pi_{3,1},\pi_4$ in \eperi\
with leading behavior 
\def\qn{{\check q}}
\eqn\elevenm{\eqalign{
\lim_{\tn_3\to i\infty}\, \pi_{2,1}&=f_{2,1}(\qn_1,\qn_2)\, , \cr
\lim_{\tn_3\to i\infty}\, \pi_{3,1}&=-\fc{5}{3}\tn_1^3 +f_{3,1}(\tn_1,\tn_2,\qn_1,\qn_2)\, \cr
\lim_{\tn_3\to i\infty}\, \pi_{4}&=\fc{5}{12}\tn_1^4-\fc{1}{6}\tn_2^4
+f_{4}(\tn_1,\tn_2,\qn_1,\qn_2)\, ,
}}
An immediate observation is that these terms seem to break the naive 
$S$-duality symmetry of the type II string (and the $T$-duality of the heterotic 
string) even in the large $S$ limit where one ignores the D-instanton
corrections $\sim e^{2\pi i S}$. The above
functions $f_{q,.}$ vanish exponentially in the $\qn_i=e^{2\pi i \tn_i}$ for $i=1,2$
near the large complex structure limit of $\Zb$, but contribute in the
interior of the complex structure moduli space of $\Zb$. 

E.g., the ratio of two periods corresponding to the 
central charges of an '$S$-dual' pair of BPS domain walls with 
classical tension $\sim \tx F_t$ is
$$
Z_2/Z_1 = \fc{S\tx F_t+\pi_{3,1}}{\tx F_t}=S+\fc{2}{3}t+\tx f(\tn_k,\qn_k)+\cx O(e^{-2\pi/g_s})\ .
$$
In principle there are various possibilities regarding the fate of $S$ duality. 
Firstly, there could be a complicated field redefinition which corrects the 
relation $\Im S=\fc{1}{g_s}$ away from the decoupling limit 
such that there is an $S$ duality for a redefined
field $\tx S$ including these corrections. 
Such a redefinition is known to be relevant in four-dimensional $\cx N=2$ 
compactifications of the heterotic string, where one may define a perturbatively 
modular invariant dilaton \refs{\refdil}. On the other hand,
duality transformations often originate from 
monodromies of the periods in the Calabi-Yau moduli space, which generate 
simple transformations at a boundary of the moduli space, such as $\Im S = \infty$, 
but correspond to complicated field transformations
away from this boundary. Again, such a 'deformation' of a duality transformation is 
known to happen in the heterotic string \refs{\AspinwallII}. At this point we can not decide
between these options, or a simple breaking of $S$-duality, without a detailed study 
of the monodromy transformations in the three-dimensional moduli space of the 
4-fold, which beyond the scope of this work.

\subsec{D-instanton corrections and Gromov--Witten invariants on the 4-fold}
There are further exponential corrections $\sim e^{2\pi i S}$ to the moduli dependent
functions in eqs.~\eperi.
Recall that we are considering here the classical periods of $\Xb$, 
which describe the complex structure moduli space of the 4-fold $\Xb$ and å 
complex deformations of the dual heterotic bundle compactification on $\Zb$. 
From the point the type IIA compactification on $\Xb$, obtained by compactifying
F-theory on $\Xb\times T^2$, these are $B$ model data and do not have 
an immediate instanton interpretation. 

However, according to the identification
of the decoupling limit in sect.~2\yyy, we expect these $B$ model data to 
describe D-instanton corrections $\sim e^{-2\pi/g_s}$ to the type II orientifold on the 3-fold,
see \netdualiia. Lacking a sufficient understanding of the afore mentioned issue of 
field redefinitions, we will express the expansion in exponentials $\sim e^{2\pi i S}$
in terms of Gromov--Witten invariants, or rather in terms of integral
invariants of Gopakumar--Vafa type, using the multi-cover formula for 4-folds given 
in \refs{\GMPh,\PMff}.
These invariants capture the world-sheet instanton expansion 
of the $A$-model on the mirror $\Xa$ of $\Xb$.
Note that if mirror pair $(X_A,X_B)$ supports a duality of the
type \hethet, then this expansion captures world-sheet and D-instanton
corrections computed by the twisted superpotential $\widetilde{W}(\Xa)$,
according to the arguments in sect~3.5.\yyy{} 
However, according to eq.~\dfib\ such a duality can only exist if
the mirror 4-fold $\Xa$ is given in terms of a suitable fibration structure,
which is not true for the quintic example of this section (since $\Xa$ is
neither elliptically nor K3 fibered), but for other examples considered
in sect.~7\yyy.

The integral $A$ model expansion of the 4-fold is defined by \refs{\GMPh,\PMff}\foot{The
fact that this multi-cover formula for spheres in a 4-fold is formally the same as
the multi-cover formula for discs in a 3-fold \OV\ is at the heart of the open-closed
duality of \refs{\MayrXK,\AlimRF,\AganagicJQ}.}
\eqn\mucoff{
\Pi_{2,\ga}=p^\ga_2(t_a)+\sum_\be\sum_{k>0} N^\ga_\be \fc{q^{\be \cdot k}}{k^2}\, ,
}
where $\Pi_{2,\ga}$ is one of the periods in the $q=2$ sector, å 
double logarithmic near the large complex structure limit $z_a=0$, and
$p_2$ a degree two polynomial in the coordinates $t_a$ defined by \cmori.
Moreover $\be$ is a label, which in the $A$ model on the mirror $\Xa$
specifies a homology class $\be \in H_2(\Xa,\IZ)$ with exponentiated 
K\"ahler volume $q^\be=\prod_a q_a^{n_a}$, $q_a=e^{2\pi i t_a}$. As discussed above,
these K\"ahler moduli of $\Xa$ map under mirror symmetry to coordinates
on the complex structure moduli space of the F-theory compactification on $\Xb$,\foot{The
$t_a$ are the distinguished flat coordinates of 
the Gauss-Manin connection.} and we use these coordinates to 
write an expansion for the $B$ model on $\Xb$.

We restrict here to discuss only the few leading coefficients $N^\ga_\be$ for the three 
linearly independent $q=2$ periods of $\Xb$. We label the 'class' $\be$ by
tree integers $(m,n,k)$, such that $N^\ga_\be$ is the coefficient of the exponential
$\exp(2\pi i (mt_1+nt_2+kt_3)$ in the basis \cmori. Thus $k$ is the exponent of 
$e^{2\pi i S}$ in the expansion.

\subsubsec{Deformation of the closed string prepotential $\cx F_t$}
The leading term of the period $\Pi_{2,2}$ is the closed string prepotential \eperi.
This period is mirror to a 4-cycle in the quintic fiber of $\Xa$ and depends
only on the closed string variable $t=t_1+t_2$ in the limit $\Im S\to\infty$. 
The leading terms in the expansion \mucoff\ of the 4-fold period are 
\eqn\quinstF{\vbox{
\halign{\hfil~#~\hfil&\hfil~#~\hfil\cr
\vbox{\offinterlineskip\halign{
\hfil~$#$~&\hfil~$#$~&\hfil~$#$~&\hfil~$#$~&\hfil~$#$~&\hfil~$#$~&\hfil~$#$~&\hfil~$#$~&\hfil~$#$~
&\hfil~$#$~&\hfil~$#$~&\hfil~$#$~&\hfil~$#$~&\hfil~$#$~&\hfil~$#$~&\hfil~$#$~\vrule\cr
%%%%
k=0\vr&0&1&2&3\cr
\noalign{\hrule}
0\ \vr&0&0&0&0\cr
1\ \vr&0&2875&0&0\cr
2\ \vr&0&0&1218500&0\cr
3\ \vr&0&0&0&951619125\cr
\noalign{\hrule}
%%%
}}
&
\vbox{\offinterlineskip\halign{
\hfil~$#$~&\hfil~$#$~&\hfil~$#$~&\hfil~$#$~&\hfil~$#$~&\hfil~$#$~&\hfil~$#$~&\hfil~$#$~&\hfil~$#$~
&\hfil~$#$~&\hfil~$#$~&\hfil~$#$~&\hfil~$#$~&\hfil~$#$~&\hfil~$#$~&\hfil~$#$~\vrule\cr
%%%%
k=1\vr&0&1&2&3\cr
\noalign{\hrule}
0\ \vr&5&20&0&0\cr
1\ \vr&0&8895&33700&600\cr
2\ \vr&0&19440&16721375&63071800\cr
3\ \vr&0&-1438720&49575600&32305559000\cr
\noalign{\hrule}
%%%
}}\cr&\cr
\vbox{\offinterlineskip\halign{
\hfil~$#$~&\hfil~$#$~&\hfil~$#$~&\hfil~$#$~&\hfil~$#$~&\hfil~$#$~&\hfil~$#$~&\hfil~$#$~&\hfil~$#$~
&\hfil~$#$~&\hfil~$#$~&\hfil~$#$~&\hfil~$#$~&\hfil~$#$~&\hfil~$#$~&\hfil~$#$~\vrule\cr
%%%%
k=2\, \vr&0&1&2&3\cr
\noalign{\hrule}
0\ \vr&0&0&0&0\cr
1\ \vr&0&0&3060&3750\cr
2\ \vr&0&0&5038070&98649500\cr
3\ \vr&0&0&19074160&47957485000\cr
\noalign{\hrule}
%%%
}}
&
\vbox{\offinterlineskip\halign{
\hfil~$#$~&\hfil~$#$~&\hfil~$#$~&\hfil~$#$~&\hfil~$#$~&\hfil~$#$~&\hfil~$#$~&\hfil~$#$~&\hfil~$#$~
&\hfil~$#$~&\hfil~$#$~&\hfil~$#$~&\hfil~$#$~&\hfil~$#$~&\hfil~$#$~&\hfil~$#$~\vrule\cr
%%%%
k=3\, \vr&0&1&2&3&4\cr
\noalign{\hrule}
0\ \vr&0&0&0&0&0\cr
1\ \vr&0&0&0&-2010&-1300\cr
2\ \vr&0&0&0&1710620&13806200\cr
3\ \vr&0&0&0&4610786345&243610412900\cr
\noalign{\hrule}
%%%
}}\cr}
}}
where the vertical (horizontal) directions corresponds to $m$ ($n$).
The $k=0$ expansion is a power series in the closed string exponential, which 
displays the independence of the closed string prepotential on the open string sector.
This independence is lost taking into account $e^{2\pi i S}$ corrections, as is expected from the 
backreaction of the closed string to the open string degrees of freedom at finite $g_s$.

The mixture between the closed and open string sector at finite $S$ is already visible in the 
definition of mirror map. In \refs{\AKV,\LMW} it had been observed, that
the definition of the flat closed string coordinate does {\it not} depend on the
open string moduli in the non-compact case, in other words, the mirror map $t=t(z)$ 
for the closed string modulus $t=t_1+t_2$ is the same as in the theory without branes,
with $z=z_1z_2$. This is no longer the case for finite $S$, as there are corrections to the mirror map
of the form $t(z_a)=t(z)+e^{2\pi i S}f(z,\zh)$.

\subsubsec{Deformation of disc superpotential $W(t,\that)$}
The leading term of the period $\Pi_{2,3}$ is the brane
superpotential of \AlimBX, which conjecturally computes the disc
instanton expansion of an $A$ type brane on the quintic. The leading terms 
in the expansion \mucoff\ of the 4-fold period 
with respect to the corrections $e^{2\pi i k S}$ are
\eqn\quinstW{
\vbox{\halign{\hfil~#~\hfil\cr
\vbox{\offinterlineskip\halign{
\hfil~$#$~&\hfil~$#$~&\hfil~$#$~&\hfil~$#$~&\hfil~$#$~&\hfil~$#$~&\hfil~$#$~&\hfil~$#$~&\hfil~$#$~
&\hfil~$#$~&\hfil~$#$~&\hfil~$#$~&\hfil~$#$~&\hfil~$#$~&\hfil~$#$~&\hfil~$#$~\vrule\cr
%%%%
k=0\, \vr&0&1&2&3&4&5\cr
\noalign{\hrule}
0\ \vr&0&20&0&0&0&0\cr
1\ \vr&-320&1600&2040&-1460&520&-80\cr
2\ \vr&13280&-116560&679600&1064180&-1497840&1561100\cr
3\ \vr&-1088960&12805120&-85115360&530848000&887761280&-1582620980\cr
\noalign{\hrule}
%%%
}}
\cr\cr
\vbox{\offinterlineskip\halign{
\hfil~$#$~&\hfil~$#$~&\hfil~$#$~&\hfil~$#$~&\hfil~$#$~&\hfil~$#$~&\hfil~$#$~&\hfil~$#$~&\hfil~$#$~
&\hfil~$#$~&\hfil~$#$~&\hfil~$#$~&\hfil~$#$~&\hfil~$#$~&\hfil~$#$~&\hfil~$#$~\vrule\cr
%%%%
k=1\, \vr&0&1&2&3&4&5\cr
\noalign{\hrule}
0\ \vr&0&20&0&0&0&0\cr
1\ \vr&0&1600&30640&3180&-1160&160\cr
2\ \vr&0&-116560&3772320&55277220&10018200&-6906880\cr
3\ \vr&0&12805120&-351282880&7862229440&104899190560&23999809580\cr
\noalign{\hrule}
%%%
}}
\cr\cr
\vbox{\offinterlineskip\halign{
\hfil~$#$~&\hfil~$#$~&\hfil~$#$~&\hfil~$#$~&\hfil~$#$~&\hfil~$#$~&\hfil~$#$~&\hfil~$#$~&\hfil~$#$~
&\hfil~$#$~&\hfil~$#$~&\hfil~$#$~&\hfil~$#$~&\hfil~$#$~&\hfil~$#$~&\hfil~$#$~\vrule\cr
%%%%
k=2\, \vr&0&1&2&3&4&5&6\cr
\noalign{\hrule}
0\ \vr&0&0&0&0&0&0&0\cr
1\ \vr&0&0&2040&3180&480&-40&0\cr
2\ \vr&0&0&679600&55277220&151559040&10282300&-4775320\cr
3\ \vr&0&0&-85115360&7862229440&333857152320&974522062840&92723257200\cr
\noalign{\hrule}
%%%
}}\cr}}}

\subsubsec{Deformation of $\Pi_{2,1}$}
As discussed in the previous subsections, the corrections to the 
third period å $\Pi_{2,1}$ contain $S^{-1}$ corrections and are in this sense
the most relevant. The leading terms of the expansion \mucoff\ are 
\eqn\quinstP{
\vbox{\halign{\hfil~#~\hfil\cr
\vbox{\offinterlineskip\halign{
\hfil~$#$~&\hfil~$#$~&\hfil~$#$~&\hfil~$#$~&\hfil~$#$~&\hfil~$#$~&\hfil~$#$~&\hfil~$#$~&\hfil~$#$~
&\hfil~$#$~&\hfil~$#$~&\hfil~$#$~&\hfil~$#$~&\hfil~$#$~&\hfil~$#$~&\hfil~$#$~\vrule\cr
%%%%
k=0\, \vr&0&1&2&3\cr
\noalign{\hrule}
0\ \vr&0&20&0&0\cr
1\ \vr&0&6020&3060&-2010\cr
2\ \vr&0&19440&3819570&1710620\cr
3\ \vr&0&-1438720&19074160&3659167220\cr
4\ \vr&0&148132440&-2365073280&20826366840\cr
\noalign{\hrule}
%%%
}}
\cr\cr
\vbox{\offinterlineskip\halign{
\hfil~$#$~&\hfil~$#$~&\hfil~$#$~&\hfil~$#$~&\hfil~$#$~&\hfil~$#$~&\hfil~$#$~&\hfil~$#$~&\hfil~$#$~
&\hfil~$#$~&\hfil~$#$~&\hfil~$#$~&\hfil~$#$~&\hfil~$#$~&\hfil~$#$~&\hfil~$#$~\vrule\cr
%%%%
k=1\, \vr&0&1&2&3\cr
\noalign{\hrule}
0\ \vr&-10&-20&0&0\cr
1\ \vr&0&-6020&0&3150\cr
2\ \vr&0&-19440&0&35577700\cr
3\ \vr&0&1438720&0&15651926000\cr
4\ \vr&0&-148132440&0&79135362000\cr
\noalign{\hrule}
%%%
}}\cr
}}}
\noi The $k=0$ corrections capture the linear corrections discussed in sect.~6.3.
These should arise from a one-loop effect on the brane; it would be interesting
to verify this by an independent computation.

\newsec{Heterotic five-branes and non-trivial Jacobians}
In this section we discuss a number of further examples
to illustrate the duality relations and the application 
of the method. The geometries are mostly taken
from \AlimRF, where the brane superpotential for $B$-type branes 
has been already computed. Since the superpotential \wwiii\
for the heterotic compactification on $\Zb$ with the appropriate bundle
$E$ agrees with the brane superpotential in the decoupling limit, 
the explicit heterotic superpotential in this limit can be read off 
from the results of \AlimRF. We have performed also a computation
of the finite $S$ corrections to the heterotic superpotential 
for the å examples below, by the methods described in detail the
previous section. The results are of a similar general structure as in the quintic case.
Detailed expressions for the examples are available upon request.

The main focus of this section will be to describe some 
additional aspects arising from the point of F-theory and
the heterotic compactification on $\Zb$. Let us recall the 
following basic result 
on F-theory/heterotic duality which will help
to understand the different outcomes in the following examples. 
The elements of the Hodge group $H_{1,1}(\Xb)$ of the 4-fold can be 
roughly divided into the following sets w.r.t. their 
meaning in the dual heterotic compactification on 
the CY 3-fold $\Zb$ with bundle $E$ (see \refs{\MV,\BershadskyNH,\BJPS,}):
\vskip 20pt

\vbox{
\noi{\it Generic classes:}\br
\noi The first set arises from the two 
generic classes from the K3 fiber $Y$ of the K3 fibration $\Xb\to B_2$:

\item{1.} The class $E$ of the fiber of the elliptic fibration 
$Y\to \IP^1$, which is also the elliptic fiber of $\Xb$. This curve
shrinks in the 4d F-theory limit and does not lead to a field in four dimensions;
\item{2.} The class $F$ of the section of the elliptic 
fibration $Y\to \IP^1$, which provides the universal 
tensor multiplet associated with the heterotic dilaton.
}
\vskip 10pt

\vbox{
\noi{\it Geometry of $\Zb$:}
\item{3.} $h^{1,1}(B_2)$ classes of the base of the K3 fibration $\Xb\to B_2$
with K3 fiber $Y$.
\item{4.} $h^{1,1}(\Zb)-h^{1,1}(B_2)-1$ classes associated with singular
fibers of the elliptic fibration $\Zb\to B_2$.
}
\vskip 10pt

\vbox{
\noi{\it Gauge fields \& å 5-branes:}
\item{5.} $h^{1,1}(Y)-2={\rm rank}\, G_{pert}$ classes from singular
fibers of the elliptic fibration $Y\to \IP^1$, corresponding to the
Cartan subgroup of the perturbative gauge group $G_{pert}$.
\item{6.} $h^{1,1}(B_3)-h^{1,1}(B_2)-1$ classes arising from blow ups of 
the $\IP^1$ bundle $B_3\to B_2$ with fiber of class $F$.
These blow ups correspond
to heterotic 5-branes wrapping a curve $C\in B_2$.
\item{7.} The remaining ${\rm rank}\, G_{non-pert}$
classes of $\Xb$ arise from extra singularities of the elliptic fibration,
which correspond to the 
Cartan subgroup of a non-perturbative gauge group $G_{non-pert}$.
}

\noi Fixing the heterotic 3-fold $\Zb$, one can still vary the 
4-fold data in the last group, to choose a bundle $E$. In the framework 
of toric geometry, this step can be made very explicit by using 
local mirror symmetry of bundles \KMV.
Starting from the toric 3-fold polyhedron for $\Zb$ one may 
to 'geometrically engineer' the bundle in terms of a 4-fold 
polyhedron, by appropriately adding or removing exceptional divisors, 
as described in great detail in \refs{\BMff,\RajeshIK}. By the type II/heterotic map \idi, this is the 
complement of adding singular fibers to the mirror fibration
$\Xa\to \IP^1$ in \epiL, to define a toric $A$ type brane on 
the 3-fold mirror $\Za$ \AlimRF.

The items 5.-7. in the above describe, how an å element 
of $H^{1,1}(\Xb)$ added in the engineering of the bundle falls into one of the 
three classes in the last set, depending on the relative location 
of the exceptional divisor w.r.t. the fibration structure. 
It follows that the $B$-type branes in the type II compactification
may map to quite different heterotic degrees of freedom under the 
type II/heterotic map \idi: perturbative gauge fields, 
heterotic five-branes and non-perturbative gauge fields.
This variety can be seen already in the examples of \AlimRF, 
as discussed below.

\subsec{Structure group $SU(1)$: Heterotic five-branes}
As seen in the previous section, 
the quintic example of \refs{\Wa,\JockersPE,\AlimRF} corresponds to a
perturbative heterotic bundle with structure group $SU(2)$.
Another example of a brane compactification taken from ref.~\AlimRF\ turns out to 
have a quite different interpretation. 
In this case, the brane deformation of the type II string does not
translate to a bundle modulus on the heterotic side under the
type II/heterotic map \idi, but rather to a brane 
modulus. On the heterotic side, this is a 5-brane representing 
a small instanton \Witsi.

Let us first recall the brane geometry on the type II side, which
is defined in \AlimRF\ as a compactification of a non-compact brane in 
the non-compact CY $\opt$, i.e. the anti-canonical bundle of $\IP^2$. 
This example has been very well studied in the context of open string mirror symmetry 
in \refs{\AKV,\LM,\GraberDW}. The non-compact CY can be thought of as the 
large fiber limit of an elliptic fibration $\Za\to\IP^2 $ which gives
the interesting possibility to check the result obtained from the 
compact 4-fold against the disc instanton corrected 3-fold superpotential 
computed by different methods in \refs{\AKV,\LM,\GraberDW}. Indeed it was shown in \AlimRF,
that 4-fold mirror symmetry reproduces the known results for the non-compact
brane in the large fiber limit, including the normalization 
computed from the intersections of the 4-fold $\Xa$. The result for the local 
result is corrected by instanton corrections for finite fiber 
volume.\foot{Note that this is a large 
fiber limit in the type IIA theory compactified on $\Za$, not the previously discussed 
large fiber limit of the heterotic string compactified on $\Zb$.}

Two different 3-fold compactifications of $\opt$ were considered in \AlimRF, with a different model 
for the elliptic fiber.\foot{A cubic in $\IP^2$ for the degree 9 and a sixtic 
in $\IP^2(1,2,3)$ for the degree 18 hypersurface.}
As the two examples produce very similar results, we discuss here the
degree 18 case of \AlimRF\ in some detail and only briefly comment on
the difference for the degree 9 hypersurface, below.

The $B$-type brane is defined in \AlimRF\ by adding 
a new vertex 
\eqn\newvert{
\nu_8=(-1,0,2,3,-1)
}
in the base of 
the 'enhanced' toric polyhedron $\D$. The Hodge numbers of the space $\Xb$ 
obtained in this way are
$
\Xb:\ h^{1,3}=4,\ h^{1,2}=0,\ h^{1,1}=2796,\ \chi=16848(=0 {\ \rm mod\ } 24)\ .
$
We refer the interested reader again to app.~B for 
the details on the toric geometry and the
parametrizations used in the following and continue with a non-technical
discussion å of the geometry. The addition of the vertex $\nu_8$
corresponds to the blow up of a divisor in the singular central fiber of 
the 4-fold fibration $\Xa\to \IP^1$. The new element
in $H^{1,1}(\Xa)$ is identified as the deformation parameter
of the $A$-brane on the 3-fold $\Za$, via open-closed duality.

On the mirror side, the blow up modulus corresponds to 
a new complex structure deformation parametrizing a holomorphic 
divisor in $\Zb$. As will be explained now, this deformation maps 
in the heterotic compactification to a modulus moving 
a heterotic 5-brane that wraps a curve $C$ in the base $B_2$ of the 3-fold 
$\Zb$.

In appropriate local coordinates, the form \ePV\ of the hypersurface equation, 
exposing the elliptic fibration of both, $\Zb$ and $\Xb$, is
\eqn\ePii{
\eqalign{
p_0&=Y^2+X^3+(z_1^3z_2z_3)^{-1/18}\, YXZ\, stu +Z^6\ ((z_2z_3)^{-1/3}\, (stu)^6 + s^{18}+ t^{18}+ u^{18}),\cr
p_+&=Z^6\, ((stu)^6+\zh s^{18})\, ,\qquad 
p_-=Z^6\, (stu)^6\ .
}
}
The brane geometry in $\Zb$, reducing to the mirror of the non-compact
brane in $\opt$ of \AKV, is defined by the hypersurface $\cxH:\, p_+=0$ within 
$\Zb$ defined by $p_0=0$ \AlimRF.

The hypersurface constraint \ePii\ is already in the form to which the 
methods of \BMff\ can be applied. 
The relevant component of $p_+$ deforming with the modulus $\zh$ lies
in a patch with $s,t,u\neq 0$ and is given by 
\eqn\scii{
\Si_+:\ Z^6\, (t^6u^6+\zh s^{12})=0\ .
}
Here the deformation $\zh$ does not involve the coordinates of
the elliptic fiber, and therefore it does {\it not} correspond to a bundle modulus.
Instead this F-theory geometry describes heterotic 
5-branes wrapping a curve $C$ in the base $B_2$ of the heterotic compactification. 
As described in detail in \refs{\MV,\AMsd,\BJPS} (see also ref.~\DonagiXE),
F-theory describes these heterotic 5-branes by a blow ups of the the $\IP^1$ bundle 
$B_3\to B_2$.

The toric 4-fold singularities associated with heterotic 
five-branes of type \ePii\ were also studied in great detail in \refs{\BMff,\RajeshIK}. 
In the present case, the 5-branes wrap a set of curves $C$
in the elliptic fibration $\Zb\to B_2$, defined by 
the zero of the function $f(s,t,u)=s^6(t^6u^6+\zh s^{12})$.
The deformation $\zh$ moves the branes on the second component, similarly å 
as it moves the type II brane in the dual type II compactification on $\Zb$.

By the F-theory/heterotic dictionary developed in å refs.~\refs{\MV,\AMsd,\BJPS}, 
the above singularity describes a small $E_8$ instanton, which
can be viewed as an M-theory/type IIA 5-brane \Witsi. 
Note that there are also exceptional blow up divisors in $\Xb$ associated with the 5-brane
wrapping, which support the elements in $H^{1,1}(\Xb)$ dual to the world-volume tensor 
fields on the 5-branes \refs{\MV,\AMsd,\BJPS}. However, these K\"ahler blow ups 
are not relevant for the purpose of computing the superpotential $W(\Xb)$.

The above conclusions may again be cross-checked by 
analyzing the perturbative gauge symmetry of the heterotic compactification, 
which does not changes in this case for $\zh\neq 0$
\eqn\gbpiia{
\xymatrix{E_8\times E_8 \ar[r]& E_8\times E_8 }\, ,
}
as is expected from the trivial structure group of the bundle, with the
anomaly cancelled entirely by 5-branes. 

The compactification of the non-compact brane in $\opt$ in the degree 9 hypersurface
leads to similar results. The 4-fold considered in \AlimRF\ has the
Hodge numbers
$$
\Xb:\ h^{1,3}=6(2),\ h^{1,2}=0,\ h^{1,1}=586,\ \chi=3600(=0 {\ \rm mod\ } 24) \ .
$$
and describes a heterotic compactification with 5-branes wrapping a 
curve given by the equation
\eqn\sciib{
\Si_+:\ Z^3\, s^3\, (t^3u^3+\zh s^{6})=0\ .
}
The further discussion is as above, except for the gauge symmetry 
breaking pattern, which is in this case
$E_6\times E_6 \to å E_6\times E_6$.

In the decoupling limit $\Im S\to \infty$ limit, the heterotic superpotential for the 5-branes
in these two cases agrees with the type II brane superpotential computed in sect.~3.2 and app.~B\ of \AlimRF, 
respectively. See also sect.~5 of \GrimmEF\ for a reconsideration of the first case,
with an identical result (Tab.3a/5.2).

\subsec{Non-trivial Jacobians: $SU(2)$ bundle on a degree 9 hypersurface}
A new aspect of another example of \AlimRF\ is the appearance of a 
non-trivial Jacobian $J(\Si)$ of the spectral surface, corresponding to non-zero
$h^{1,2}$ \FMW. In this case there are additional massless fields 
associated with the Jacobian $J(\Xb)=H^3(\Xb,\IR)/H^3(\Xb,\IZ)$ in the
F-theory compactification, and the non-trivial Jacobian of $\Si$ in the heterotic 
dual \refs{\FMW,\CurioBVA,\AspinwallBW}.

The present example has been considered in sect.~3.3 of \AlimRF\ and
describes a brane compactification on the same degree nine hypersurface $\Za$
as in the previous section, but with a different gauge background. $\Za$ is defined 
as a hypersurface in 
the weighted projective space $\IP^4(1,1,1,3,3)$ with
hodge numbers and Euler number 
\eqn\hdnin{
\Za:\, h^{1,1}=4(2),\ h^{1,2}=112,\ \chi=-216\ ,
}
The numbers in brackets denote the non-toric deformations of $\Za$, which 
are å unavailable in the given hypersurface representation. 

As familiar by now, the technical details on toric geometry 
are relegated to app.~B\yyy. The Hodge numbers of the dual F-theory 4-fold
$\Xb$ are
$$
\Xb:\, h^{1,3}=4,\ h^{1,2}=3,\ h^{1,1}=246(11),\ \chi=1530=18 {\ \rm mod\ } 24\ .
$$
The local form \ePV\ of the hypersurface equation for $\Xb$, 
exposing the elliptic fibration and the hypersurface $\Zb$ is 
\eqn\ePiii{
\eqalign{
p_0&=a_1 Y^3+a_2X^3+Z^3\ (a_3 (stu)^3 + a_4 s^{9}+ a_5 t^{9}+ a_6 u^{9})
+a_0 YXZ\, stu,\cr
p_+&=Y\, (a_8 Y^2+a_7 XZstu)\, ,\qquad 
p_-=a_9 Y^3.
}
}
Again the zero set $p_0=0$ defines the 3-fold geometry $\Zb$ for the compactification
of the type II/heterotic string, while the brane geometry considered in \AlimRF\
is defined by the hypersurface å $\cxH:\, p_+=0$. By
the type II/heterotic map \idi, we reinterprete these equations in terms of a 
heterotic bundle on $\Zb$. While $p_-$ corresponds to the trivial spectral cover, 
$p_+$ describes a component with non-trivial dependence on a single modulus
$\zh$:
\eqn\sciii{
\Si_+:\ Y^2+\zh XZstu=0\ ,
}
where $\zh$ is the brane/bundle deformation. 
As in the quintic case, $\Si_+$ may be identified with a component with 
structure group $SU(2)$. This is confirmed by a study of the perturbative 
gauge symmetry of the heterotic compactification, which changes for $\zh\neq 0$ as
\eqn\gbpiii{
\xymatrix{E_6\times E_6 \ar[r]& SU(6)\times E_6\cr} \ .
}
The $\Im S\to\infty$ limit of the heterotic superpotential for this bundle coincides with the
type II result computed in \AlimRF. 

\newsec{ADE Singularities, Kazama-Suzuki models and matrix factorizations}
In the above we have described how 4-fold mirror symmetry computes quantum corrections to
the superpotential and the K\"ahler potential of supersymmetric compactifications 
to four and lower dimensions with four supercharges. Specifically, these corrections are 
expected to correspond to D$(-1)$, D1, and D3 instanton contributions in the type II orientifold 
compactification on $\Zb$ and to world-sheet and space-time instanton corrections to 
a $(0,2)$ heterotic string compactification on the same manifold. At present, it is
hard to concretely verify these predictions by an independent computation. 
A particularly neat way to find further evidence for our proposal (in the $\cx N=2$
supersymmetric situation) would be to establish a connection with refs.~\refhm. 
In these works, considerable progress has been made in understanding corrections to the 
hyper-multiplet moduli, especially the interaction with mirror symmetry. It would be very 
interesting to study the overlap with the non-perturbative corrections discussed in the 
present paper. In this section, we discuss a different application of heterotic/F-theory 
duality which might be viewed as an interesting corroboration of our main statements, and 
is also of independent interest.

\subsubsec{$\cx N=2$ supersymmetry}
It is best again to begin with 8 supercharges. Consider a heterotic string compactification 
on a K3 manifold near an ADE singularity with a trivial gauge bundle on the blown up 2-spheres. 
The hypermultiplet moduli space of this heterotic compactification is corrected by $\al'$ corrections 
from perturbative and world-sheet instanton
effects. It has been shown in \WittenFQ\ that for an $A_1$ singularity,
the heterotic moduli space space in the hyperk\"ahler limit is given by the Atiyah-Hitchin 
manifold, which is also the moduli space of three-dimensional $\cx N=4$ $SU(2)$ Yang-Mills theory. 
This relation between the moduli space of the heterotic string on a singular K3 and the moduli space
of a three-dimensional gauge theory can be derived and generalized
by studying the stable degeneration limit of the dual type IIA/F-theory
3-fold. Specifically it is shown in refs.~\refs{\AspinwallXS,\MayrBK}, that the
3-fold $\Xb$ dual to the heterotic string on an ADE singularity of type $G$
and with a certain local behavior of the gauge bundle $V$ develops a singularity, 
which 'geometrically engineers' a three-dimensional
gauge theory of gauge group and matter content depending on $G$ and $V$, 
see ref.~\refs{\HoriZJ}. In connection with the $\cx N=2$ version of the decoupling limit
$\Im S\to \infty$, eq.\tfcor, this leads to a very concrete relation between the 3-fold period and the
world-sheet instanton corrections to the heterotic hypermultiplet space in the hyperk\"ahler 
limit. This could be explicitly checked against the known result, at least 
in the case dual to 3d $SU(2)$ SYM theory.

\subsubsec{$\cx N=1$ supersymmetry}
The above situation has an interesting $\cx N=1$ counter part. Namely, it has been 
conjectured in \refs{\MayrBK} that one may use the heterotic string on a certain 3-fold
singularity to geometrically engineer (the moduli space of) interesting 2-dimensional
field theories. The 3-fold singularities are of the type
\eqn\tfsings{
y^2+H(x_k)=0\, ,
}
where $H(x_k)$ describes an ADE surface singularity.
The idea is the obvious generalization of the above, by first applying
heterotic/F-theory duality and then exploiting the relation of ref.\ \GVW\ between 
similar 4-fold singularities and Kazama-Suzuki models. We here make this correspondence more precise.

Recall that the identification of \GVW\ proceeded through the comparison of the vacuum 
and soliton structure of a type IIA compactification on Calabi-Yau four-fold with its 
superpotential from four-form flux, and the Landau-Ginzburg description \LVW\ of the 
deformed Kazama-Suzuki coset models \KaSu. The four-folds relevant for this connection 
are local manifolds that are fibered by singular 2-dimensional ALE spaces and their
deformations. The ADE type of the singularity in the fiber determines the
numerator $G$ of the $\cx N=2$ coset $G/H$, while the flux determines the denominator $H$ 
and the level. More precisely, the fluxes studied in \GVW\ are the minimal fluxes 
corresponding to a minuscule weight of $G$. These give rise to the so-called
SLOHSS models (simply-laced, level one, Hermitian symmetric space), which is the subset of
Kazama-Suzuki models admitting a Landau-Ginzburg description. This identification 
was checked for the $A$-series in ref.~\GVW\ and worked out in detail for $D$ and $E$ in 
ref.~\EguchiFM. It has remained an interesting question to identify the theories for non-minimal 
flux, see e.g., the conclusions of \EguchiFM.

An important clue to address this question has come from the study of matrix factorizations
and their deformation theory. In particular, it was observed in ref.~\KnOm, see also ref.~\HerbstJP, that
the superpotential resulting from the deformation theory of certain matrix factorization in 
$\cx N=2$ minimal models coincides with the Landau-Ginzburg potential of a corresponding SLOHSS 
model. More precisely, the matrix factorizations are associated with the fundamental weights 
of ADE simple Lie algebras via the standard McKay correspondence, and the relevant subset are
those matrix factorizations corresponding to the minuscule weights. We argue that this 
coincidence of superpotentials can be explained via heterotic/F-theory/type II duality. 

The missing link is provided by ref.\ \MoCu. Among the results of this work is that the 
matrix factorizations of ADE minimal models can be used to describe bundles on partial 
resolutions (Grassmann blowups) of the threefold singularities of ADE type \tfsings\ that appear 
in the above-mentioned conjecture of ref.~\MayrBK. The bundles have support only on the
smooth part of the partial blowup, which is important to apply the arguments of ref.~\WittenFQ.

The combination of the last three paragraphs suggests that we should couple the heterotic
worldsheet to the matrix factorizations of ref.~\MoCu! This can be implemented by using the
$(0,2)$ linear sigma model \WittenYC\ resp.\ $(0,2)$ Landau-Ginzburg models \DistlerMK,
along the lines of \GovindarajanJS, in combination with an appropriate non-compact 
Landau-Ginzburg model to describe the fibration structure. The resulting strongly
coupled heterotic worldsheet theories are conjectured to be dual to those 2-d field theories that
are engineered on the four-fold side. The ADE type of the minimal model is that of the fiber
of the four-fold, while the fundamental weight specifies the choice of four-form flux.

As formulated, the above conjecture makes sense for all, fundamental weights. The main testable 
prediction is thus the coincidence of the deformation superpotentials of the higher rank
matrix factorizations corresponding to non-minuscule fundamental weights with the appropriate 
periods of the four-folds of refs.~\refs{\GVW,\EguchiFM}. Note that the Kazama-Suzuki models only 
appear for the minuscule weights, and that we have not covered the case of fluxes 
corresponding to non-fundamental weights. We plan to return to these questions in the
near future.

%%%%%%%%%%%%%%%%%%%%%%%%%%%%%%%%%
%\GovindarajanKR
\lref\GovindarajanKR{
S.~Govindarajan and T.~Jayaraman,
``Boundary fermions, coherent sheaves and D-branes on Calabi-Yau
manifolds,''
Nucl.\ Phys.\ å B {\bf 618}, 50 (2001)
[arXiv:hep-th/0104126].
%%CITATION = NUPHA,B618,50;%%
}
%\GrimmDQ
\lref\GrimmDQ{
T.~W.~Grimm, T.~W.~Ha, A.~Klemm and D.~Klevers,
``The D5-brane effective action and superpotential in N=1
compactifications,''
Nucl.\ Phys.\ B {\bf 816}, 139 (2009)
[arXiv:0811.2996 [hep-th]].
%%CITATION = NUPHA,B816,139;%%
}
%\JockersMN
\lref\JockersMN{
H.~Jockers and M.~Soroush,
``Relative periods and open-string integer invariants for a compact
Calabi-Yau hypersurface,''
Nucl.\ Phys.\ B {\bf 821}, 535 (2009)
[arXiv:0904.4674 [hep-th]].
%%CITATION = NUPHA,B821,535;%%
}
\lref\Lukas{
L.~B.~Anderson, J.~Gray, D.~Grayson, Y.~H.~He and A.~Lukas,
``Yukawa Couplings in Heterotic Compactification,''
arXiv:0904.2186 [hep-th];\br
Y.~H.~He, S.~J.~Lee and A.~Lukas,
``Heterotic Models from Vector Bundles on Toric Calabi-Yau Manifolds,''
arXiv:0911.0865 [hep-th];\br
L.~B.~Anderson, J.~Gray, Y.~H.~He and A.~Lukas,
``Exploring Positive Monad Bundles And A New Heterotic Standard Model,''
arXiv:0911.1569 [hep-th].
}
%%%%%%%%%%%%%%%%%%%%%%%%%%%%%%%%%

\newsec{Conclusions}
In this note we study the variation of Hodge structure of the complex
structure moduli space of certain Calabi-Yau 4-folds. These
moduli spaces capture certain effective couplings of the $\cx N=1$
supergravity theory arising from the associated F-theory 4-fold compactification.
Furthermore, through a chain of dualities we relate such F-theory
scenarios to heterotic compactifications with non-trivial gauge bundle and small instanton 5-branes
and to type~II compactifications with branes.  

The connection to the heterotic string is made through the
stable degeneration limit of the F-theory 4-fold \refs{\MV,\FMW,\AMsd}. Taking this limit 
specifies the corresponding heterotic geometry. Due to
the employed F-theory/heterotic duality the resulting
heterotic geometry is given in terms of elliptically fibered
Calabi-Yau 3-folds. Furthermore, in the simplest cases, the geometric
bundle moduli are described in terms of the spectral cover, which
is also encoded in the 4-fold geometry in the stable degeneration limit \FMW. 
Alternatively, depending on the details of the F-theory 4-fold, we
describe the moduli space of heterotic 5-branes instead of bundle
moduli. On the other hand the link to the open-closed type~II string
theories is achieved through the weak coupling limit \refs{\AlimBX},
and it realizes the open-closed duality introduced in ref.~\refs{\MayrXK,\LM,
\AganagicJQ}.

We argue that the two distinct limits to the heterotic string and
to the open-closed string map the variation of Hodge structure
of the F-theory Calabi-Yau 4-fold to the variation of
mixed Hodge structure of the corresponding Calabi-Yau 3-fold 
relative to a certain divisor.
For the heterotic string this divisor is either identified with the
spectral cover of the heterotic bundle or with the embedding of
small instantons. In the context of open-closed type II geometries
the divisor encodes a certain class of brane deformations as studied in
refs.~\refs{\LMW,\JockersPE,\AlimRF,\AganagicJQ,\AlimBX,\GrimmEF,\Yau,\JockersMN,\GrimmDQ}.

Starting from the F-theory 4-fold geometry we discuss in detail 
non-trivial background fluxes and compute the
$\cx N=1$ superpotential, which couples to the moduli fields described by
the variation of Hodge structure.
We trace these F-terms along the chain of dualities to
the open-closed and heterotic string compactifications.
For the heterotic string we find that, depending on the characteristics
of the 4-fold flux quanta, these fluxes either deform the bulk
geometry of the heterotic string to generalized Calabi-Yau
manifolds \refs{\DasguptaSS,\FuSM,\BeckerET},
or they give rise to superpotential terms for the bundle/five-brane
moduli fields. The
superpotentials associated to the flux quanta encode obstructions
to deformations of the spectral cover. Furthermore, we
show that in the stable degeneration limit the holomorphic Chern-Simons functional of the
heterotic gauge bundle gives rise to these F-terms for the geometric
bundle moduli.

The underlying 4-fold description of the heterotic and the type II strings
allows us to extract (non-perturbative) corrections to the stable degeneration
limit and the weak coupling limit respectively. We discuss the nature of these
corrections, and we find that they encode world sheet instanton, D-instanton
and space-time instanton corrections depending on the specific theory
in the analyzed web of dualities. In order to exhibit the origin of these
corrections we compare our analysis with the analog $\cx N=2$ scenarios,
which have been studied in detail in refs.~\refs{\BMff,\BMof}.

Apart from these F-term couplings we demonstrate that our techniques
are also suitable to extract the K\"ahler potentials for the metrics of the
studied moduli spaces in appropriate semi-classical regimes. In ref.~\AlimBX\
the connection to the open-closed K\"ahler potential for 3-fold compactifications with
7-branes has been developed. Here, starting from the K\"ahler\
potential of the complex structure moduli space of the Calabi-Yau 4-fold,
we also extract the corresponding K\"ahler potential associated to the combined
moduli space of the complex structure and certain moduli of the heterotic
gauge bundle. In leading order these K\"ahler potentials are in agreement
with the results obtained by dimensional reduction of higher dimensional
supergravity theories \refs{\HaackDI,\JockersYJ}. In addition our calculation predicts
subleading corrections.

Thus, the used duality relations together with the presented computational
techniques offer novel tools to extract (non-perturbative) corrections to
$\cx N=1$ string compactifications arising from F-theory, 
from heterotic strings or from type~II strings in the presence of branes.
It would be interesting to confirm the anticipated quantum corrections
by independent computations and to understand in greater detail the
physics of various (non-perturbative) corrections discussed here.
In particular, our analysis suggests a connection to the 
quantum corrections in the hypermultiplet sector of $\cx N=2$ compactifications
analyzed in refs.~\refhm. 

Our techniques should also be useful to address phenomenological
interesting questions in the context of F-theory, type II or heterotic
string compactifications. 
As discussed in sects.~5,6, the finite $S$ corrections to the 
superpotential capture the backreaction of the geometric moduli to the bundle moduli.
Such corrections are a new and important ingredient in fixing the bundle moduli
in phenomenological applications, as emphasized, e.g., in ref.~\DonagiCA.
Thus the calculated (quantum
corrected) superpotentials provide a starting point to
investigate moduli stabilization and/or supersymmetry breaking
for the class of models discussed here. 
In the context of the heterotic string it seems plausible that our
approach can be extended to more general heterotic bundle configurations,
which can be described in terms of monad constructions \refs{\DistlerMK,\GovindarajanKR}. 
Such an extension is not only
interesting from a conceptual point of view, but in addition it also
gives a handle on analyzing the effective theory of phenomenologically 
appealing heterotic bundle configurations as discussed, for instance,
in ref.~\Lukas.

In section 8, we propose an explanation, and conjecture an extension of, an observation 
originally made by Warner, which relates the deformation superpotential of matrix 
factorizations of minimal models to the flux superpotential of local four-folds near 
an ADE singularity. One of the results of this connection is the 
suggestion that (higher rank) matrix factorizations should also play a role in 
constructing the $(0,2)$ worldsheet theories of heterotic strings.

%%%%%%%%%%%%%%%%%%%%%%%%%%%%%%%%%%%%%%
%\AshokXQ
\lref\AshokXQ{
 S.~K.~Ashok, E.~Dell'Aquila, D.~E.~Diaconescu and B.~Florea,
 ``Obstructed D-branes in Landau-Ginzburg orbifolds,''
 Adv.\ Theor.\ Math.\ Phys.\  {\bf 8}, 427 (2004)
 [arXiv:hep-th/0404167].
 %%CITATION = 00203,8,427;%%
}
%\AspinwallBS
\lref\AspinwallBS{
P.~S.~Aspinwall and S.~H.~Katz,
``Computation of superpotentials for D-Branes,''
Commun.\ Math.\ Phys.\  {\bf 264}, 227 (2006)
[arXiv:hep-th/0412209].
%%CITATION = CMPHA,264,227;%%
}
%\HoriJA
\lref\HoriJA{
 K.~Hori and J.~Walcher,
 ``F-term equations near Gepner points,''
 JHEP {\bf 0501}, 008 (2005)
 [arXiv:hep-th/0404196].
 %%CITATION = JHEPA,0501,008;%%
}
%\KnappTV
\lref\KnappTV{
 J.~Knapp and E.~Scheidegger,
 ``Matrix Factorizations, Massey Products and F-Terms for Two-Parameter
 Calabi-Yau Hypersurfaces,''
 arXiv:0812.2429 [hep-th].
 %%CITATION = ARXIV:0812.2429;%%
}
%\JockersNG
\lref\JockersNG{
 H.~Jockers and W.~Lerche,
 ``Matrix Factorizations, D-Branes and their Deformations,''
 Nucl.\ Phys.\ Proc.\ Suppl.\  {\bf 171}, 196 (2007)
 [arXiv:0708.0157 [hep-th]].
 %%CITATION = NUPHZ,171,196;%%
}
%\BaumgartlAN
\lref\BaumgartlAN{
 M.~Baumgartl, I.~Brunner and M.~R.~Gaberdiel,
 ``D-brane superpotentials and RG flows on the quintic,''
 JHEP {\bf 0707}, 061 (2007)
 [arXiv:0704.2666 [hep-th]].
 %%CITATION = JHEPA,0707,061;%%
}
The presented approach to calculate deformation superpotentials by studying
adequate Hodge problems is ultimately linked to the derivation of effective 
obstruction superpotentials with matrix factorization or, more generally, worldsheet
techniques \refs{\HoriJA,\AshokXQ,\AspinwallBS,\BaumgartlAN,
\JockersNG,\KnappTV}. While the latter approach
leads to effective superpotentials up to field redefinitions, our computations
give rise to effective superpotentials in terms of flat coordinates due to the
underlying integrability of the associated Hodge problem. It would be
interesting to explore the physical origin and the necessary
conditions for the emergence of such a flat structure in the context of
the deformation spaces studied in this note.
%%%%%%%%%%%%%%%%%%%%%%%%%%%%%%%%%%%%%%%%

\vskip2cm
\noi {\bf Acknowledgements:}\break 
We would like to thank 
Mina Aganagic,
Ilka Brunner,
Shamit Kachru,
Wolfgang Lerche,
Jan Louis,
Dieter L\"ust,
Masoud Soroush
and
Nick Warner
for discussions and correspondence.
We would also like to thank Albrecht Klemm for coordinating the submission of related work.
The work of H.J. is supported by the Stanford Institute of Theoretical Physics
and by the NSF Grant 0244728. 
The work of P.M. is supported by the program
``Origin and Structure of the Universe'' of the German Excellence Initiative.

\appendix{A}{Some toric data for the examples}
\def\vvrule{ \phantom{\vrule} }

\subsec{The quintic in $P^4(1,1,1,1,1)$}
\vskip-0.6cm

\subsubsec{Parametrization of the hypersurface constraints}
\noi The toric polyhedra for the example considered in sect.~6\yyy\
are defined as the convex hull of the vertices

\vbox{
\eqn\tpi{
\vbox{\offinterlineskip\halign{
# $\phantom{\Si}$ %height 12pt depth 4pt
&~$#$\hfil
&~$#$\hfil&\hfil\quad$#$\quad\hfil\vvrule
&~$#$\hfil&~$#$\hfil&\hfil\quad$#$\quad\hfil\vvrule
\cr
%\noalign{\hrule}
\vvrule&\D\hskip10pt
&	\nu_0=& (\- 0,\- 0,\- 0,\- 0,\- 0)	&\Ds\hskip10pt
&\nus_0=& (\- 0,\- 0,\- 0,\- 0,\- 0)\cr
\vvrule&&\nu_1=& ( - 1,\- 0,\- 0,\- 0,\- 0)	
&&\nus_1=& (\- 1,- 4,\- 1,\- 1,\- 0)\cr
\vvrule&&\nu_2=& (\- 0, å -1,\- 0,\- 0,\- 0)	
&&\nus_2=& (\- 1,\- 1,- 4,\- 1,\- 0)\cr
\vvrule&&\nu_3=& (\- 0,\- 0, å -1,\- 0,\- 0)	
&&\nus_3=& (\- 1,\- 1,\- 1,- 4,\- 0)\cr
\vvrule&&\nu_4=& (\- 0,\- 0,\- 0, å -1,\- 0)	
&&\nus_4=& (\- 1,\- 1,\- 1,\- 1,\- 0)\cr
\vvrule&&\nu_5=& (\- 1,\- 1,\- 1,\- 1,\- 0)	
&&\nus_5=& (- 4,\- 1,\- 1,\- 1,\- 1)\cr
\vvrule&&\nu_6=& (\- 0,\- 0,\- 0,\- 0,- 1)	
&&\nus_6=& (- 4,\- 1,\- 1,\- 1,- 5)\cr
\vvrule&&\nu_7=& (- 1,\- 0,\- 0,\- 0,- 1)	
&&\nus_7=& (\- 0,- 3,\- 1,\- 1,\- 1)\cr
\vvrule&&\nu_8=& (- 1,\- 0,\- 0,\- 0,\- 1)	
&&\nus_8=& (\- 0,\- 1,- 3,\- 1,\- 1)\cr
\vvrule&&& 
&&\nus_9=& (\- 0,\- 1,\- 1,- 3,\- 1)\cr
\vvrule&&& 
&&\nus_{10}=& (\- 0,\- 1,\- 1,\- 1,\- 1)\cr
}}
}} 
\vskip-15pt
\noi The local coordinates in the expressions \ePi\ and \eQi\ are defined by the following selections
$\Xi_1$ and $\Xi_2$ of points of $\Ds$, respectively: 
\eqn\XiI{\vbox{\offinterlineskip\halign{
\hfil$#\phantom{\int}\!\!\!\!\hfil$&\hfil$#$&\hfil$#$&\hfil$#$&\hfil$#$&\hfil$#$&\hfil$#$\cr
&&&\Xi_1\cr
x_1'&-4&1&1&1&0\cr x_2&1&-4&1&1&0\cr x_3&1&1&-4&1&0\cr
x_4&1&1&1&-4&0\cr x_5&1&1&1&1&0\cr a&-4&1&1&1&-1\cr
b&-4&1&1&1&1\cr\cr
}}
\hskip2cm
\vbox{\offinterlineskip\halign{
\hfil$#\phantom{\int}\!\!\!\!\hfil$&\hfil$#$&\hfil$#$&\hfil$#$&\hfil$#$&\hfil$#$&\hfil$#$\cr
&&&\Xi_2\cr
Z&1&1&0&0&0\cr Y&1&-2&0&0&0\cr X'&-2&1&0&0&0\cr s&1&1&-2&1&0\cr t&1&1&1&-2&0\cr u&1&1&1&1&0\cr
a&-2&1&0&0&1\cr b&-2&1&0&0&-1\cr
}}}
As described in sect.~6, the local coordinates $\{x_i\}$ and $\{Z,Y,X',s,t,u\}$ 
may be associated with the 'heterotic' manifold $\Zb$ encoded in the F-theory 4-fold $\Xb$.
In the example, $\Zb$ is the mirror quintic, which is embedded in a toric ambient space with 
a large number $h^{1,1}=101$ of K\"ahler classes, resulting in 101 coordinates $x_k$ in 
the hypersurface constraint $\ePBat$. $\{x_i\}$ and $\{Z,Y,X',s,t,u\}$ are special selections
of these 101 coordinates, where the latter display (one of) the elliptic
fibration(s) of $\Zb$.

On the other hand $(a,b)$ are coordinates inherent to the 4-fold $\Xb$, 
parametrizing a special $\IP^1$, $F$, which plays the
central role in the stable degeneration limit of \refs{\FMW,\AMsd} and the local mirror limit of 
\refs{\KMV,\BMff}. $F$ is the base of the elliptic fibration of a K3 $Y$, which in turn
is the fiber of the K3 fibration of $\Xb$:
$$
Y\to F,\qquad Y\to \Xb \to B_2\ .
$$
In the above example, $B_2$ can be thought of as a blow up of $\IP^2$.
The stable degeneration limit of the toric hypersurface can be defined as a local mirror limit in the complex structure
moduli of $\Xb$, where one passes to new coordinates \BMff
$$
\ePi\, :\ x_1=x'_1ab,\ v=a/b\ ,\qquad \eQi\, :\ X=X'ab,\ v=a/b\ .
$$
The distinguished local coordinate $v=a/b$ on $\IC^*$ parametrizes 
a patch near the local singularity associated with the bundle/brane data
for a Lie group $G$ \KMV.
For $G=SU(n)$, $v$ appears linearly, which leads to a substantial simplification of 
the Hodge variation problem, as described in the appendices of refs.~\refs{\MayrXK,\AganagicJQ}.

%\break
\subsubsec{Perturbative gauge symmetry of the heterotic string}
The perturbative gauge symmetry of the dual heterotic string is determined by
the singularities in the elliptic fibration of the K3 fiber $Y$ \MV. 
There is a simple technique to read off fibration structures for the 
CY 4-fold $\Xb$ from the toric polyhedra described in refs.~\Avram. Namely 
a fibration of $\Xb\to B_{4-n}$ with fibers a Calabi-Yau $n$-fold $Y_n$
corresponds to the existence of a hypersurface $H$ of codimension $4-n$,
such that the integral points in the set $H\cap \Ds$ define the toric polyhedron of $Y_n$. 

In the present case, the toric polyhedron $\Ds_{K3}$ for the K3 fiber $Y$ is given as the
convex hull of the points in $\Ds$ lying on the hypersurface $H:\, \{x_3=x_4=0\}$:
\vskip5pt
\vbox{
\eqn\tpK{
\vbox{\offinterlineskip\halign{
# $\phantom{\Si}$ %height 12pt depth 4pt
&~$#$\hfil
&~$#$\hfil&\hfil\quad$#$\quad\hfil\vvrule
&~$#$\hfil&~$#$\hfil&\hfil\quad$#$\quad\hfil\vvrule
\cr
%\noalign{\hrule}
\vvrule&\D_{K3}\hskip10pt
&	\mu_0=& (\- 0,\- 0,\- 0)	&\Ds_{K3}\hskip10pt &\mus_0=& (\- 0,\- 0,\- 0)\cr
\vvrule&&\mu_1=& (\- 0,- 1,\- 0) &&\mus_1=& (-2,\- 1,- 3)\cr
\vvrule&&\mu_2=& (\- 1, \-1,\- 0) &&\mus_2=& (- 2,\- 1,\- 1)\cr
\vvrule&&\mu_3=& (\- 0,\- 0,- 1) &&\mus_3=& (\- 0,- 1,\- 1)\cr
\vvrule&&\mu_4=& (- 1,\- 0, å -1) å &&\mus_4=& (\- 0,\- 1,\- 1)\cr
\vvrule&&\mu_5=& (- 1,\- 0, \- 1) å &&\mus_5=& (\- 1,- 2,\- 0)\cr
\vvrule&&& && \mus_6=& (\- 1,\- 1,\- 0)\cr }}
}}
\vskip-5pt
\noi where the zero entries at the 3rd and 4th position have been deleted and
$\D_{K3}$ is the dual polyhedron of $\Ds_{K3}$. The elliptic fibration of $Y$ is visible as the 
polyhedron $\Ds_E=\Ds_{K3}\cap\{x_5=0\}$ of the elliptic curve
$$
\D_E = {\rm Conv}\left\{ (-1,0),(0,-1),(1,1) \right\},\quad 
\Ds_E={\rm Conv}\left\{ (-2, 1), (1, -2), (1, 1)\right\}\ .
$$
Since the model for the elliptic fiber is not of the standard form, 
but the cubic in $\IP^2$ orbifolded by the action \Korb, the application of the standard
methods to determine the singularity of the elliptic fibration and thus the 
perturbative heterotic gauge group should be reconsidered carefully. The 
singularity of the elliptic fibration can be determined directly from the hypersurface equation of $X$ 
of the elliptically fibered K3 polynomial
\eqn\KPi{ p(K3)\,=\,Z^3+ Y^3+X'^3(a^2b^4+a^4b^2+a^3b^3) +Z Y X' (a b+b^2)\ , }
which is associated to the toric data~\tpK.
The $\IZ_3$ orbifold singularity is captured
by $r^3=p\,q$ in terms of the invariant monomials $p={Y^3\over X'^3}$,
$q={Z^3\over X'^3}$ and $r={Z Y\over X'^2}$. Then, to leading order,
the singularities of the elliptic fiber $E$ 
in the vicinity $a=0$ and in the vicinity $b=0$ are respectively given by
$$
p_{a\rightarrow 0}(K3)\,=\, a^2 q + q^2 + q r + r^3 \ , \qquad
p_{b\rightarrow 0}(K3)\,=\, b^2 q + q^2 + b q r + r^3 \ .
$$
Using a computer algebra system, such as ref.~{\SING}, it is straight forward to check that the polynomials
$p_{a\rightarrow 0}(K3)$ and $p_{b\rightarrow 0}(K3)$ correspond to the ADE singularities
$SU(6)$ and $E_6$.

In fact it turns out, that the same answer is obtained by naively applying the method developed in 
refs.~\refs{\CandF,\CandelasEH} for the standard model of the elliptic fiber, which implements the 
Kodaira classification of singular elliptic fibers in the language of toric polyhedra, such that the
orbifold group is taken into account automatically. 
The polyhedron $\Ds_{K3}$ splits into a top and bottom piece $\Xi_+$ and $\Xi_-$ with the points
$$
\vbox{\offinterlineskip\halign{
\hfil$#\vphantom{\int}$&\hfil$#$&\hfil$#$\cr
\phantom{\Xi_{+}}&\Xi_{+}&\phantom{\Xi_{+}}\cr
-2&1&1\cr
-1&0&1\cr
-1&1&1\cr
0&-1&1\cr
0&0&1\cr
0&1&1\cr
&&\cr
}}
\hskip2cm
\vbox{\offinterlineskip\halign{
\hfil$#\vphantom{\int}$&\hfil$#$&\hfil$#$\cr
\phantom{\Xi_{-}}&\Xi_{-}&\
\phantom{\Xi_{-}}\cr
-2&1&-3\cr
-2&1&-2\cr
-2&1&-1\cr
-1&0&-2\cr
-1&1&-2\cr
0&-1&-1\cr
0&1&-1\cr}}
$$
which build up the affine Dynkin diagrams of $SU(6)$ and $E_6$, respectively.
As asserted in \refs{\BershadskyNH,\CandF,\CandelasEH}, these toric vertices å corresponds to two ADE singularities 
of the same type, in agreement with the direct computation. Moreover, deleting the vertex 
$\nu_7\in\D$ which is associated with the exceptional toric divisor that described the brane/bundle
modulus $\zh$, the same analysis produces a K3 fiber with two ADE singularities of type $E_6$,
leading to the pattern \gbpi.

\subsubsec{Moduli and Picard-Fuchs system}
The moduli $z_a$ are related to the parameters
$a_i$ in \ePBat\ by 
\eqn\defmod{
z_a=(-)^{l^a_0} \prod_i a_i^{l^a_i}\ ,
}
where $l^a_i$ are the charge vectors that define the 
phase of the linear sigma model for the mirror $\Xa$. For the phase considered in \refs{\AlimRF,\AlimBX},
these are given in \cmori. 
The complex structure modulus $z\sim e^{2\pi i t}$
mirror to the volume of the generic quintic fiber, the brane/bundle modulus $\zh\sim e^{2\pi i \that}$ and the
distinguished modulus $z_S\sim e^{2\pi i S}$ capturing the decoupling limit are given by 
$$
z=z_1z_2=-\fc{a_1a_2a_3a_4a_5}{a_0^5}\, ,\qquad 
\zh=z_2=-\fc{a_1a_6}{a_0a_7}\, , \qquad 
z_S=z_3=\fc{a_7a_8}{a_1^2}\, .
$$
The GKZ system for CY 4-folds has been discussed in the context of mirror symmetry
e.g. in \refs{\PMff,\KLRY,\AlimRF}. A straightforward manipulation of it leads to the
following system of Picard-Fuchs operators for the above example: 
\eqn\pfi{\eqalign{
\cx L_1 &=
\th_1^4 (\th_1+\th_3-\th_2)-z_1 å (-\th_1+\th_2) (4 \th_1+1+\th_2) (4 \th_1+2+\th_2) (4 \th_1+3+\th_2) (4 \th_1+4+\th_2)
\ ,\cr
\cx L_2 &=
(\th_1+\th_3-\th_2) \th_3-z_3 å (2 \th_3-\th_2) (2 \th_3+1-\th_2)
\ ,\cr
\cx L_3 &=
-(2 \th_3-\th_2) (-\th_1+\th_2)-z_2 å (\th_1+\th_3-\th_2) (4 \th_1+1+\th_2)
\ ,\cr
\cx L_4 &=
(-\th_1+\th_2) \th_3+z_2 å z_3 å (2 \th_3-\th_2) (4 \th_1+1+\th_2)
\ ,\cr
\cx L_5 &=
-(2 \th_3-\th_2) \th_1^4-z_1 å z_2 å (4 \th_1+1+\th_2) (4 \th_1+2+\th_2) (4 \th_1+3+\th_2) (4 \th_1+4+\th_2) (4 \th_1+5+\th_2)
\ ,\cr
\cx L_6 &=
-(2 \th_3-\th_2) \th_1^3-5 z_1 å z_2 å (4 \th_1+1+\th_2) (4 \th_1+2+\th_2) (4 \th_1+3+\th_2) (4 \th_1+4+\th_2)\cr&\quad -z_2 å \th_1^3 (\th_1+\th_3-\th_2)
\ .
}}
Here $\th_a=z_a\fc{\p}{z_a}$ are the logarithmic derivatives in 
the coordinates $z_a,\, a=1,2,3$.

\subsec{Heterotic 5-branes}
\vskip-0.6cm
\subsubsec{Degree 18 hypersurface in $\IP^4(1,1,1,6,9)$}
The polyhedra for the mirror pair $(\Xa,\Xb)$ of 4-folds dual to the 3-fold 
compactifications on $(\Za,\Zb)$ are defined as the convex hull of the points:
\eqn\tpiia{
\vbox{\offinterlineskip\halign{
\hfil$#\phantom{\int}\!\!\!\!$&\hfil$#$&\hfil$#$&\hfil$#$&\hfil$#$&\hfil$#$&\hfil$#$\cr
&&&\D\cr
\nu_0&0&0&0&0&\phantom{-}0\cr \nu_1&0&0&0&-1&0\cr \nu_2&0&0&-1&0&0\cr \nu_3&0&0&2&3&0\cr \nu_4&-1&0&2&3&0\cr \nu_5&0&-1&2&3&0\cr \nu_6&1&1&2&3&0\cr \nu_7& å 0&0&2&3&-1\cr \nu_8&-1&0&2&3&-1\cr \nu_9&0&0&2&3&1\cr
}}\hskip 1cm 
\vbox{\offinterlineskip\halign{
\hfil$#\phantom{\int}\!\!\!\!$&\hfil$#$&\hfil$#$&\hfil$#$&\hfil$#$&\hfil$#$\cr
&&\D^*\cr
6& 6& 1& 1& 0\cr 6& 6& 1& 1& -6\cr 6& -12& 1& 1& 0\cr 6& -12& 1& 1& -6\cr 0&
6& 1& 1& 6\cr 0& 0& 1& -1& 0\cr 0& 0& -2& 1& 0\cr 0& -6& 1& 1& 6\cr -12& 
6& 1& 1& 6\cr -12& 6& 1& 1& -6\cr
}}
\hskip 1cm 
\vbox{\offinterlineskip\halign{
\hfil$#\phantom{\int}\!\!\!\!$&\hfil$#$&\hfil$#$&\hfil$#$&\hfil$#$&\hfil$#$&\hfil$#$\cr
x_i&&&\Xi\cr
Y&0&0&1&-1&0\cr X&0&0&-2&1&0\cr Z'&0&0&1&1&0\cr
s&-12&6&1&1&0\cr t&6&-12&1&1&0\cr u&6&6&1&1&0\cr
a&0&0&1&1&-1\cr b&0&0&1&1&1\cr
\cr\cr
}}}
$\D$ is the enhanced polyhedron for $\XXa$ in Table~2 of \AlimRF, with the point $\nu_9$
added in the compactification $\Xa$ of $\XXa$. The polyhedron $\D_3$ 
for the 3-fold $\Za$ defined as a degree 18 hypersurface in $\IP^4(1,1,1,6,9)$ 
is given by the points on the hypersurface $\nu_{i,5}=0$, 
with the last entry deleted. The vertices of the 
dual polyhedron $\Ds_3$ of $\D_3$ are given by the points of $\Ds$ 
with $\nus_{i,5}=0$ and on extra vertex $(-12,6,1,1)$. On the r.h.s we have given the 
selection $\Xi$ of points in $\Ds$ used to define local coordinates in \ePii. 
The relation to the coordinates used there is 
$
Z=Z'ab,\ v=a/b\ .
$

The relevant phase of the K\"ahler cone considered in \refs{\AlimRF,\GrimmEF} is
\eqn\mvII{\vbox{\offinterlineskip\halign{
\strut # 
&\hfil~$#$~=\hfil
&\hfil~$#$ &\hfil~$#$ &\hfil~$#$ &\hfil~$#$
&\hfil~$#$ &\hfil~$#$ &\hfil~$#$ &\hfil~$#$
&\hfil~$#$ &\hfil~$#$ &\hfil~$#$ &\hfil~$#$
\cr
&l^1&(&-6&3&2&1&0&0&0&0&0&0)\ ,\cr
&l^2&(&0&0&0&-2&0&1&1&-1&1&0)\ ,\cr
&l^3&(&0&0&0&-1&1&0&0&1&-1&0)\ .\cr
&l^4&(&0&0&0&-1&-1&0&0&0&1&1)\ .\cr
}}
}
In the coordinates \defmod, the brane modulus in \sci\ is given by $\zh=z_2^{1/3}z_3^{-2/3}$.

\subsubsec{Degree 9 hypersurface in $\IP^4(1,1,1,3,3)$}
The polyhedra for the mirror pair $(\Xa,\Xb)$ of 4-folds dual to the 3-fold 
compactifications on $(\Za,\Zb)$ are defined as the convex hull of the points:
\eqn\tpiib{
\vbox{\offinterlineskip\halign{
\hfil$#\phantom{\int}\!\!\!\!$&\hfil$#$&\hfil$#$&\hfil$#$&\hfil$#$&\hfil$#$&\hfil$#$\cr
&&&\D\cr
\nu_0&0&0&0&0&\phantom{-}0\cr \nu_1&0&0&0&-1&0\cr \nu_2&0&0&-1&0&0\cr \nu_3&0&0&1&1&0\cr \nu_4&-1&0&1&1&0\cr \nu_5&0&-1&1&1&0\cr \nu_6&1&1&1&1&0\cr \nu_7& å 0&0&1&1&-1\cr \nu_8&-1&0&1&1&-1\cr \nu_9&0&0&1&1&1\cr
}}\hskip 1cm 
\vbox{\offinterlineskip\halign{
\hfil$#\phantom{\int}\!\!\!\!$&\hfil$#$&\hfil$#$&\hfil$#$&\hfil$#$&\hfil$#$\cr
&&\D^*\cr
-6& 3& 1& 1& 3\cr 0& 3& 1& 1& 3\cr 0& -3& 1& 1& 3\cr 3& 3& 1& 1& -3\cr -6& 3&
1& 1& -3\cr 3& -6& 1& 1& -3\cr 3& 3& 1& 1& 0\cr 3& -6& 1& 1& 0\cr 0& 
0& -2& 1& 0\cr 0& 0& 1& -2& 0\cr
}}
\hskip 1cm 
\vbox{\offinterlineskip\halign{
\hfil$#\phantom{\int}\!\!\!\!$&\hfil$#$&\hfil$#$&\hfil$#$&\hfil$#$&\hfil$#$&\hfil$#$\cr
x_i&&&\Xi\cr
Y&0&0&1&-2&0\cr X&0&0&-2&1&0\cr Z'&0&0&1&1&0\cr
s&-6&3&1&1&0\cr t&3&-6&1&1&0\cr u&3&3&1&1&0\cr
a&0&0&1&1&-1\cr b&0&0&1&1&1\cr
\cr\cr
}}}
The polyhedron $\D_3$ 
for the 3-fold $\Za$ defined as a degree 9 hypersurface in $\IP^4(1,1,1,3,3)$ 
is again given by the points on the hypersurface $\nu_{i,5}=0$. On the r.h.s we have given the 
selection $\Xi$ of points in $\Ds$ used in \sciib, with the redefinitions $Z=Z'ab$, $v=a/b$.
The phase of the K\"ahler cone considered in \AlimRF\ is
\eqn\mvII{\vbox{\offinterlineskip\halign{
\strut # 
&\hfil~$#$~=\hfil
&\hfil~$#$ &\hfil~$#$ &\hfil~$#$ &\hfil~$#$
&\hfil~$#$ &\hfil~$#$ &\hfil~$#$ &\hfil~$#$
&\hfil~$#$ &\hfil~$#$ &\hfil~$#$ &\hfil~$#$
\cr
&l^1&(&-3&1&1&1&0&0&0&0&0&0)\ ,\cr
&l^2&(&0&0&0&-2&0&1&1&-1&1&0)\ ,\cr
&l^3&(&0&0&0&-1&1&0&0&1&-1&0)\ .\cr
&l^4&(&0&0&0&-1&-1&0&0&0&1&1)\ .\cr
}}
}
In the coordinates \defmod, the brane modulus in \sciib\ is given by $\zh=z_2^{1/3}z_3^{-2/3}$.

\subsec{$SU(2)$ bundle of the degree 9 hypersurface in $\IP^4(1,1,1,3,3)$}
The polyhedra for the mirror pair $(\Xa,\Xb)$ of 4-folds dual to the 3-fold 
compactifications on $(\Za,\Zb)$ are defined as the convex hull of the points:
\eqn\tpiiia{
\vbox{\offinterlineskip\halign{
\hfil$#\phantom{\int}\!\!\!\!$&\hfil$#$&\hfil$#$&\hfil$#$&\hfil$#$&\hfil$#$&\hfil$#$\cr
&&&\D\cr
\nu_0&0&0&0&0&\phantom{-}0\cr \nu_1&0&0&0&-1&0\cr \nu_2&0&0&-1&0&0\cr \nu_3&0&0&1&1&0\cr \nu_4&-1&0&1&1&0\cr \nu_5&0&-1&1&1&0\cr \nu_6&1&1&1&1&0\cr \nu_7& å 0&0&0&0&-1\cr \nu_8&0&0&0&-1&-1\cr \nu_9&0&0&0&-1&1\cr
}}\hskip 1cm 
\vbox{\offinterlineskip\halign{
\hfil$#\phantom{\int}\!\!\!\!$&\hfil$#$&\hfil$#$&\hfil$#$&\hfil$#$&\hfil$#$\cr
&&\D^*\cr
3& 3& 1& 1& 0\cr 3& -6& 1& 1& 0\cr 2& 2& 1& 0& 1\cr 2& -4& 1& 0& 1\cr 0& 0& 
1& -2& 1\cr 0& 0& 1& -2& -3\cr 0& 0& -1& 0& 1\cr 0& 0& -2& 1& 0\cr -4& 2& 
1& 0& 1\cr -6& 3& 1& 1& 0\cr
}}
\hskip 1cm 
\vbox{\offinterlineskip\halign{
\hfil$#\phantom{\int}\!\!\!\!$&\hfil$#$&\hfil$#$&\hfil$#$&\hfil$#$&\hfil$#$&\hfil$#$\cr
x_i&&&\Xi\cr
Y'&0&0&1&-2&0\cr X&0&0&-2&1&0\cr Z&0&0&1&1&0\cr
s&-6&3&1&1&0\cr t&3&-6&1&1&0\cr u&3&3&1&1&0\cr
a&0&0&1&-2&-1\cr b&0&0&1&-2&1\cr
\cr\cr
}}}
$\D$ is the enhanced polyhedron for $\XXa$ in Table 4 of \AlimRF, with the point $\nu_9$
added in the compactification $\Xa$ of $\XXa$. The polyhedron $\D_3$ 
for the 3-fold fiber $\Za$ of the fibration $\Xa\to\IP^1$ is given by the points on the hypersurface $\nu_{i,5}=0$, 
with the last entry deleted \AlimRF. The vertices of the 
dual polyhedron $\Ds_3$ of $\D_3$ are given by the points of $\Ds$ 
with $\nus_{i,5}=0$ and one extra vertex $(0,0,1,-2)$ (which is 
a point, but no vertex, in $\Ds$). On the r.h.s we have given the 
selection $\Xi$ of points in $\Ds$ used to define local coordinates in \ePiii. 
The relation to the coordinates used there is 
$
Y=Y'ab,\ v=a/b\ .
$
The charge vectors for the phase of the linear sigma model considered in \AlimRF\ is
\eqn\mviii{\vbox{\offinterlineskip\halign{
\strut # 
&\hfil~$#$~=\hfil
&\hfil~$#$ &\hfil~$#$ &\hfil~$#$ &\hfil~$#$
&\hfil~$#$ &\hfil~$#$ &\hfil~$#$ &\hfil~$#$
&\hfil~$#$ &\hfil~$#$ &\hfil~$#$ &\hfil~$#$
\cr
&l^1&(&-2&0&1&1&0&0&0&-1&1&0)\ ,\cr
&l^2&(&0&0&0&-3&1&1&1&0&0&0)\ ,\cr
&l^3&(&-1&1&0&0&0&0&0&1&-1&0)\ .\cr
&l^4&(&0&-2&0&0&0&0&0&0&1&1)\ .\cr
}}
}
In the coordinates \defmod, the brane modulus in \sciii\ is given by $\zh=z_3\, (z_1^3z_2z_3^3)^{-1/9}$.
The combination $z_1^3z_2z_3^3$ of complex structure parameters is mirror to the overall volume of $\Za$.

Explicit expressions for the superpotential in the decoupling limit can be found 
in sect.~3.3 and app.~B. of \AlimRF.

\listrefs
\end